\documentclass[twocolumn]{aastex631}
\usepackage{CLASSY_Preamble}
\newcommand{\mpfit}{MPFIT}
\newcommand{\ciii}{\ion{C}{3}]}
\newcommand{\fciii}{[\ion{C}{3}]}
\newcommand{\civ}{\ion{C}{4}}
\newcommand{\lya}{Ly$\alpha$}
\newcommand{\heii}{\ion{He}{2}}
\newcommand{\oii}{[\ion{O}{2}]}
\newcommand{\oiii}{[\ion{O}{3}]} %[O\,\textsc{iii}]}
\newcommand{\oiiiuv}{\ion{O}{3}]}

\newcommand{\hii}{\ion{H}{2}}

\newcommand{\ha}{H$\alpha$}
\newcommand{\hb}{H$\beta$}
\newcommand{\hg}{H$\gamma$}
\newcommand{\hd}{H$\delta$}
\newcommand{\siii}{[\ion{S}{3}]}
\newcommand{\fex}{[\ion{Fe}{10}]}
\newcommand{\fevii}{[\ion{Fe}{7}]}
\newcommand{\fevi}{[\ion{Fe}{6}]}
\newcommand{\fexiv}{[\ion{Fe}{14}]}
\newcommand{\nii}{[\ion{N}{2}]}
\newcommand{\niii}{\ion{N}{3}]}
\newcommand{\niv}{\ion{N}{4}]}
\newcommand{\fniv}{[\ion{N}{4}]}
\newcommand{\sii}{[\ion{S}{2}]}

\newcommand{\silii}{\ion{Si}{2}}     
\newcommand{\siiii}{\ion{Si}{3}]}
\newcommand{\oi}{[\ion{O}{1}]}
\newcommand{\mgii}{\ion{Mg}{2}}

\newcommand{\nv}{\ion{N}{5}}

\newcommand{\pyneb}{\texttt{PyNeb}}

\usepackage{lipsum}
\usepackage{multirow}

\begin{document}

%\submitjournal{AASJournal ApJ}
\accepted{for publication in ApJ}
\shortauthors{Mingozzi et al.}
\shorttitle{CLASSY~VIII}

\title{CLASSY VIII: Exploring the Source of Ionization with UV ISM diagnostics in local High-$z$ Analogs \footnote{
Based on observations made with the NASA/ESA Hubble Space Telescope,
obtained from the Data Archive at the Space Telescope Science Institute, which
is operated by the Association of Universities for Research in Astronomy, Inc.,
under NASA contract NAS 5-26555.}}

\author[0000-0003-2589-762X]{Matilde Mingozzi}
\affiliation{Space Telescope Science Institute, 3700 San Martin Drive, Baltimore, MD 21218, USA}
\affiliation{AURA for ESA, Space Telescope Science Institute, 3700 San Martin Drive, Baltimore, MD 21218, USA}

\author[0000-0003-4372-2006]{Bethan L. James}
\affiliation{AURA for ESA, Space Telescope Science Institute, 3700 San Martin Drive, Baltimore, MD 21218, USA}

\author[0000-0002-4153-053X]{Danielle A. Berg}
\affiliation{Department of Astronomy, The University of Texas at Austin, 2515 Speedway, Stop C1400, Austin, TX 78712, USA}

\author[0000-0002-2644-3518]{Karla Z. Arellano-C\'{o}rdova}
\affiliation{Institute for Astronomy, University of Edinburgh, Royal Observatory, Edinburgh, EH9 3HJ, UK}

\author[0000-0003-0390-0656]{Adele Plat}
\affiliation{Institute for Physics, Laboratory for Galaxy Evolution, EPFL, Observatoire de Sauverny, Chemin Pegasi 51, 1290 Versoix, Switzerland}

\author[0000-0002-9136-8876]{Claudia Scarlata}
\affiliation{Minnesota Institute for Astrophysics, University of Minnesota, 116 Church Street SE, Minneapolis, MN 55455, USA}

\author[0000-0003-4137-882X]{Alessandra Aloisi}
\affiliation{Space Telescope Science Institute, 3700 San Martin Drive, Baltimore, MD 21218, USA}

\author[0000-0001-5758-1000]{Ricardo O. Amor\'{i}n}
\affiliation{ARAID Foundation. Centro de Estudios de F\'{\i}sica del Cosmos de Arag\'{o}n (CEFCA), Unidad Asociada al CSIC, Plaza San Juan 1, E--44001 Teruel, Spain}
\affiliation{Departamento de Astronom\'{i}a, Universidad de La Serena, Av. Juan Cisternas 1200 Norte, La Serena 1720236, Chile}

% % \author[0000-0003-1074-4807]{Matthew Bayliss}
% % \affiliation{Department of Physics, University of Cincinnati, Cincinnati, OH 45221, USA}

\author[0000-0003-4359-8797]{Jarle Brinchmann}
\affiliation{Instituto de Astrof\'{i]}sica e Ci\^{e}ncias do Espa\c{c}o, Universidade do Porto, CAUP, Rua das Estrelas, PT4150-762 Porto, Portugal}

\author[0000-0003-3458-2275]{St\'{e}phane Charlot}
\affiliation{Sorbonne Universit\'{e}, CNRS, UMR7095, Institut d'Astrophysique de Paris, F-75014, Paris, France}

% % \author[0000-0002-7636-0534]{Jacopo Chevallard}
% % \affiliation{Sorbonne Universit\'{e}, UPMC-CNRS, UMR7095, Institut d'Astrophysique de Paris, F-75014, Paris, France}

\author[0000-0002-0302-2577]{John Chisholm}
\affiliation{Department of Astronomy, The University of Texas at Austin, 2515 Speedway, Stop C1400, Austin, TX 78712, USA}

% % \author[0000-0000-0000-0000]{Ilyse Clark}
% % \affiliation{Department of Astronomy, The University of Texas at Austin, 2515 Speedway, Stop C1400, Austin, TX 78712, USA}

% % \author[0000-0001-9714-2758]{Dawn K. Erb}
% % \affiliation{Center for Gravitation, Cosmology and Astrophysics, Department of Physics, University of Wisconsin Milwaukee, 3135 N Maryland Ave., Milwaukee, WI 53211, USA}

\author[0000-0001-6865-2871]{Anna Feltre}
\affiliation{INAF - Osservatorio di Astrofisica e Scienza dello Spazio di Bologna, Via P. Gobetti 93/3, 40129 Bologna, Italy}

\author[0000-0002-5659-4974]{Simon Gazagnes}
\affiliation{Department of Astronomy, The University of Texas at Austin, 2515 Speedway, Stop C1400, Austin, TX 78712, USA}

\author[0000-0001-8587-218X]{Matthew Hayes}
\affiliation{Stockholm University, Department of Astronomy and Oskar Klein Centre for Cosmoparticle Physics, AlbaNova University Centre, SE-10691, Stockholm, Sweden}

\author[0000-0003-1127-7497]{Timothy Heckman}
\affiliation{Center for Astrophysical Sciences, Department of Physics \& Astronomy, Johns Hopkins University, Baltimore, MD 21218, USA}

% \author[0000-0002-6586-4446]{Alaina Henry}
% \affiliation{Space Telescope Science Institute, 3700 San Martin Drive, Baltimore, MD 21218, USA}
% \affiliation{Center for Astrophysical Sciences, Department of Physics \& Astronomy, Johns Hopkins University, Baltimore, MD 21218, USA}

\author[0000-0003-4857-8699]{Svea Hernandez}
\affiliation{AURA for ESA, Space Telescope Science Institute, 3700 San Martin Drive, Baltimore, MD 21218, USA}

% % \author[0000-0002-6790-5125]{Anne Jaskot}
% % \affiliation{Department of Astronomy, Williams College, USA}

\author[0000-0001-8152-3943]{Lisa J. Kewley}
\affiliation{Research School of Astronomy and Astrophysics, Australian National University, Cotter Road, Weston Creek, ACT 2611, Australia; ARC Centre of Excellence for All Sky Astrophysics in 3 Dimensions (ASTRO 3D), Canberra, ACT 2611, Australia}

\author[0000-0002-5320-2568]{Nimisha Kumari}
\affiliation{AURA for ESA, Space Telescope Science Institute, 3700 San Martin Drive, Baltimore, MD 21218, USA}

\author[0000-0003-2685-4488]{Claus Leitherer}
\affiliation{Space Telescope Science Institute, 3700 San Martin Drive, Baltimore, MD 21218, USA}

% \author[0000-0003-1354-4296]{Mario Llerena}
% \affiliation{Instituto de Investigaci\'{o}n Multidisciplinar en Ciencia y Tecnolog\'{i}a, Universidad de La Serena, Raul Bitr\'{a}n 1305, La Serena 2204000, Chile}
% \affiliation{Departamento de Astronom\'{i}a, Universidad de La Serena, Av. Juan Cisternas 1200 Norte, La Serena 1720236, Chile}

\author[0000-0001-9189-7818]{Crystal L. Martin}
\affiliation{Department of Physics, University of California, Santa Barbara, Santa Barbara, CA 93106, USA}

\author[0000-0003-4359-8797]{Michael Maseda}
\affiliation{Department of Astronomy, University of Wisconsin-Madison, 475 N Charter Street, Madison, WI 53706 USA}

\author[0000-0003-2804-0648]{Themiya Nanayakkara}
\affiliation{Swinburne University of Technology, Melbourne, Victoria, AU}

% % \author[0000-0002-1049-6658]{Masami Ouchi}
% % \affiliation{National Astronomical Observatory of Japan, 2-21-1 Osawa, Mitaka, Tokyo 181-8588, Japan}
% % \affiliation{Institute for Cosmic Ray Research, The University of Tokyo, Kashiwa-no-ha, Kashiwa 277-8582, Japan}
% % \affiliation{Kavli Institute for the Physics and Mathematics of the Universe (WPI), University of Tokyo, Kashiwa, Chiba 277-8583, Japan}

\author[0000-0002-5269-6527]{Swara Ravindranath}
\affiliation{Observational Cosmology Lab, Code 665, NASA Goddard Space Flight Center, 8800 Greenbelt Rd, Greenbelt, MD 20771, USA}

\author[0000-0002-7627-6551]{Jane R. Rigby}
\affiliation{Observational Cosmology Lab, Code 665, NASA Goddard Space Flight Center, 8800 Greenbelt Rd, Greenbelt, MD 20771, USA}

\author[0000-0002-9132-6561]{Peter Senchyna}
\affiliation{Carnegie Observatories, 813 Santa Barbara Street, Pasadena, CA 91101, USA}

\author[0000-0003-0605-8732]{Evan D. Skillman}
\affiliation{Minnesota Institute for Astrophysics, University of Minnesota, 116 Church Street SE, Minneapolis, MN 55455, USA}

% % \author[0000-0002-6172-733X]{Dan P. Stark}
% % \affiliation{Steward Observatory, The University of Arizona, 933 N Cherry Ave, Tucson, AZ, 85721, USA}

% % \author[0000-0002-4834-7260]{Charles C. Steidel}
% % \affiliation{Cahill Center for Astronomy and Astrophysics, California Institute of Technology, MC249-17, Pasadena, CA 91125, USA}

% % \author[0000-0001-6369-1636]{Allison L. Strom}
% % \affiliation{Department of Astrophysical Sciences, 4 Ivy Lane, Princeton University, Princeton, NJ 08544, USA}

\author[0000-0001-6958-7856]{Yuma Sugahara}
\affiliation{Institute for Cosmic Ray Research, The University of Tokyo, Kashiwa-no-ha, Kashiwa 277-8582, Japan}
\affiliation{National Astronomical Observatory of Japan, 2-21-1 Osawa, Mitaka, Tokyo 181-8588, Japan}
\affiliation{Waseda Research Institute for Science and Engineering, Faculty of Science and Engineering, Waseda University, 3-4-1, Okubo, Shinjuku, Tokyo 169-8555, Japan}

\author[0000-0003-3903-6935]{Stephen M. Wilkins}
\affiliation{Astronomy Centre, University of Sussex, Falmer, Brighton BN1 9QH, UK}

\author[0000-0001-8289-3428]{Aida Wofford}
\affiliation{Instituto de Astronom\'{i}a, Universidad Nacional Aut\'{o}noma de M\'{e}xico, Unidad Acad\'{e}mica en Ensenada, Km 103 Carr. Tijuana-Ensenada, Ensenada 22860, M\'{e}xico}

\author[0000-0002-9217-7051]{Xinfeng Xu}
\affiliation{Center for Astrophysical Sciences, Department of Physics \& Astronomy, Johns Hopkins University, Baltimore, MD 21218, USA}

% \linenumbers

% \suppressAffiliations
\correspondingauthor{Matilde Mingozzi} 
\email{mmingozzi@stsci.edu}

%-----------------------------------------------------------------------------------------

\begin{abstract} %check 250 word limit
In the current JWST era, rest-frame UV spectra play a crucial role in enhancing our understanding of the interstellar medium (ISM) and stellar properties of the first galaxies in the epoch of reionization (EoR, $z>6$).
Here, we compare well-known and reliable optical diagrams sensitive to the main ionization source (i.e., star formation, SF; active galactic nuclei, AGN; shocks) to UV counterparts proposed in the literature - the so-called ``UV-BPT diagrams'' - using the HST COS Legacy Archive Spectroscopic SurveY (CLASSY), the largest high-quality, high-resolution and broad-wavelength range atlas of far-UV spectra for 45 local star-forming galaxies.
In particular, we explore where CLASSY UV line ratios are located in the different UV diagnostic plots, taking into account state-of-the-art photoionization and shock models and, for the first time, the measured ISM and stellar properties (e.g., gas-phase metallicity, ionization parameter, carbon abundance, stellar age). 
We find that the combination of \ciii$\lambda\lambda$1907,9 \heii$\lambda1640$ and \oiiiuv$\lambda$1666 can be a powerful tool to separate between SF, shocks and AGN at sub-solar metallicities. 
We also confirm that alternative diagrams without \oiiiuv$\lambda$1666 still allow us to define a SF-locus with some caveats. Diagrams including \civ$\lambda\lambda$1548,51 should be taken with caution given the complexity of this doublet profile.
Finally, we present a discussion detailing the ISM conditions required to detect UV emission lines, 
visible only in low gas-phase metallicity (12+log(O/H)$\lesssim8.3$) and high ionization parameter (log($U$)$\gtrsim-2.5$) environments.
Overall, CLASSY and our UV toolkit will be crucial in interpreting the spectra of the earliest galaxies that JWST is currently revealing.

% Rest-frame UV spectra play a key role in the understanding of massive stellar populations, chemical evolution, feedback processes, and reionization. Indeed, in the current JWST era, the UV spectroscopic frontier has been pushed to higher redshifts than ever before, to finally reveal the first galaxies in the distant Universe. In this context, the HST COS Legacy Archive Spectroscopic SurveY (CLASSY) provides the first high-quality, high-resolution and broad-wavelength range catalogue of 45 local star-forming galaxies (SFGs) in the rest-frame UV (1150 −2000 Å) to investigate their stellar and gas properties. The sample is representative of SFGs across all redshifts, including extremely metal-poor objects similar to reionization-era systems. Hence, CLASSY provides the ideal UV atlas with which we can tailor a complete UV diagnostic toolkit to explore the interstellar medium (ISM) properties (i.e., density, temperature, gas-phase metallicity, ionization parameter). In this talk I will present such a toolkit, obtained from the analysis of the main emission lines of CLASSY spectra and the comparison with well-known optical diagnostics. We also compared our measurements with state-of-the-art photoionization models, to provide the best diagnostics plots to identify the main source of ionization. Overall, CLASSY and our UV toolkit can be crucial to interpret the earliest galaxies revealed with JWST.

\end{abstract} 

\keywords{Dwarf galaxies (416), Ultraviolet astronomy (1736), Galaxy chemical evolution (580), 
Galaxy spectroscopy (2171), High-redshift galaxies (734), Emission line galaxies (459)}

%-----------------------------------------------------------------------------------------
%-----------------------------------------------------------------------------------------

% SECTION 1
\section{Introduction}\label{sec:intro}
% \color{blue}
% \noindent{ Notes for CLASSY co-Is:} 
% Welcome CLASSY co-Is to the overleaf draft of the CLASSY VIII paper!
% I am grateful that you made it here and are ready to leave comments, suggestions, 
% and questions for me. 
% To help you do so, here are a few important notes:
% \begin{enumerate}
%     \item To leave comments, click on the "Review" button in the upper right
%     hand corner. 
%     This should cause a pale blue-gray column to appear next to the actual text.
%     Then just use your cursor to click and highlight any text of interest to make an "Add comment" button show up in the blue-gray column.
%     Click the "Add comment" button, type your comment, and hit enter. 
%     \item Note the comments of other collaborators - no need to duplicate identical concerns, but please feel free to comment or respond to existing comments.
%     \item {\it Please} add comments for citations, especially to your own work (!), that I missed.
%     \item {\it Please} feel free to correct typos if you see them (I read the paper several times, but I think there are still a lot!).
% \end{enumerate}
% \color{black}
Emission lines provide precious information about the conditions and ionizing sources of the interstellar medium (ISM) in galaxies across cosmic time. 
Several classification methods have been proposed to identify the main ionization sources: hot and young massive stars, tracing recent star formation (SF); post-asymptotic giant branch (post-AGB) stars, tracing older stellar populations; Active Galactic Nuclei (AGN), and thus the energy of the accretion disc surrounding a central supermassive black hole; low-ionization (nuclear) emission-line regions (LI(N)ERs); shocks, due to supernovae and stellar winds, outflows from starbursts/AGN activity and mergers, or from random motions of interstellar clouds; accretion on compact objects (e.g., High-Mass X-ray Binaries, HMXB); or a mixture of them (e.g., \citealt{heckman80, izotov99, nagao06, kewley06, allen08, stasinska15,belfiore16}). 

Historically, the main diagnostics to discriminate between different ISM ionization mechanisms are the classical optical diagnostic diagrams, first presented in \citet{baldwin81} (and usually referred as BPT diagrams) and then progressively updated by \citet{keel83, veilleux87,kauffmann03,kewley06}.
These diagrams are constructed on intensity ratios of strong emission lines close in wavelength, namely \nii$~\lambda$6584/\ha, \sii$~\lambda\lambda$6717,31/\ha\ and \oi$~\lambda6300$/\ha\ versus \oiii$~\lambda$5007/\hb\ (i.e., \nii, \sii\ and \oi\ BPT diagrams), minimizing the effects of differential reddening by dust and flux calibration issues. They have been found useful to discriminate between \hii-like sources and objects photoionised by a harder radiation field (e.g., power-law continuum by an AGN or shock excitation). However, existing BPT diagrams can be less effective in discriminating mechanisms other than SF in metal-poor galaxies (e.g., \citealt{groves06,reines20,polimera22}), since, on the one hand, the involved line ratios can be dependent on metallicity and, on the other, the hot metal-poor stellar populations have a hard spectrum to irradiate the gas (e.g., \citealt{feltre16,byler18,xiao18,wofford21}). Also, these diagnostics are not optimal in discriminating shocks from other ionization sources. 

Exploiting the large statistics of the SDSS DR7 \citep{abazajian09}, \citet{shirazi12} proposed a less metallicity-dependent diagnostic diagram to discriminate the ionization source: \nii/\ha\ versus \heii$~\lambda$4686/\hb. \heii$~\lambda$4686 is produced via recombination and thus indicates the existence of sources of hard ionizing radiation capable of ionizing the helium (ionization potential of 54.4~eV).  
Such a high radiation field can be produced by young stellar populations ({massive and very massive stars}), including O-type stars (e.g., \citealt{sixtos23}) and some types of Wolf–Rayet (WR) stars (e.g., WC, WO, and WNE, \citealt{schaerer98}; see also \citealt{schmutz92}), but also other mechanisms such as AGN \citep{shirazi12}, X-ray binaries \citep{garnett91,mayya23} or shocks \citep{thuan05,dopita96,stasinska15,alarie19} can play a role. 
Recently, \citealt{tozzi23} investigated the \citet{shirazi12} diagnostic diagram with integral field spectroscopy (IFS) observations from the SDSS-IV MaNGA (Mapping Nearby Galaxies at APO; \citealt{bundy15}) survey, confirming its power in revealing AGN activity undetected from the classical BPT diagrams.
% \heii\ lines, however, can be particularly faint and difficult to observe (see Sec.~\ref{sec:shirazi}). 
Finally, to better disentangle the presence of shocks, \citet{kewley19} presented an overview of alternative optical diagrams, including \oiii$\lambda5007$, \oii$\lambda$3727 and \oi$\lambda6300$ emission lines (see also \citealt{heckman80, allen08}) as well as the importance of correlations between line ratios and the velocity dispersion of the gas. %\oi$\lambda6300$ lines, however, can be very faint and also difficult to reproduce with models.
Overall, the combination of (some of) these optical criteria provides a reliable powerful tool to investigate the ISM ionization sources in galaxies.

However, the well-studied optical emission lines (from \oii$\lambda\lambda$3727 to \siii$\lambda\lambda9069,9532$) on which the diagnostics described in the previous paragraphs rely are not easily accessible to investigate the early phases of galaxy evolution in the epoch of reionization (EoR, $z>6$) in the current JWST and future extremely large ground-based telescopes (ELTs) era. 
For example, JWST/NIRspec does not cover anymore \ha\ at $z\gtrsim6.5$ and \oiii~$\lambda\lambda$4959,5007 at $z\gtrsim10$, while JWST/MIRI is less sensitive and does not have multi-object slit capabilities.
Hence, rest-frame UV emission lines have started to play a critical role in understanding EoR objects, thus making a reliable UV toolkit of ISM diagnostics essential. 
EoR galaxies are expected to be increasingly compact, metal-poor, with low-masses and large specific star formation rates (e.g., \citealt{wise14, madau15, robertson15, stanway16, stark16}) and characterized by prominent high-ionization nebular UV emission lines (e.g., \heii$\lambda1640$, \oiiiuv$\lambda\lambda1661,6$, \ciii$\lambda\lambda$1907,9, \civ$\lambda\lambda$1548,51; \citealt{stark15,mainali17,mainali18}). 
The spectra of some of these objects have already started to be revealed with JWST/NIRSpec (e.g., \citealt{curtijwst22,arellano-cordova22b,curtis-lake23,matthee22,trump23,arrabal-haro23}), with one of the furthest ($z\sim10.603$) galaxies ever spectroscopically observed characterized by plenty of UV (from \lya\ to \mgii$\lambda\lambda2795,2802$) and optical (from \oii$\lambda\lambda3727$ to \oiii$\lambda$4363) emission lines \citep{bunker23}. 
In the foreseeable future, further JWST observations including large surveys such as GLASS (e.g., \citealt{treu22,roberts-borsani22,castellano22}), CEERS (e.g., \citealt{finkelstein22,arrabal-haro23}) and JADES (e.g., \citealt{curtis-lake23,bunker23}) will provide more and more rest-frame UV spectra of $z>6$ systems, exploring in depth the EoR. 

Several recent works proposed UV alternatives to the optical diagnostics, either using photoionization models from SF and AGN \citep{feltre16,jaskot16,nakajima18,dors18,byler20} and shock models \citep{jaskot16}, or from the comparison of either models and simulations \citep{hirschmann19,hirschmann22}.
For instance, \citetalias{feltre16} proposed various UV diagnostic plots, including the \ciii$\lambda\lambda$1907,9/\heii$\lambda1640$ versus \civ$\lambda\lambda$1548,51/\heii$\lambda1640$ to discriminate between SF and AGN models (see also \citealt{dors18}). 
\citet{jaskot16} used the same diagnostic diagram to separate SF from shocks models, while \citet{nakajima18} proposed alternative models and diagrams based on \civ$\lambda\lambda$1548,51/\ciii$\lambda\lambda$1907,9 and their equivalent widths (EWs). 
Finally, \citet{hirschmann19,hirschmann22} coupled models and cosmological zoom-in simulations, exploring synthetic optical and UV emission-line diagnostic diagrams, listing the most promising in Table~1 of \citet{hirschmann22}.
However, UV diagnostics have yet to be compared against the optical ones in a systematic way, because it is not trivial to observe a full suite of optical and UV lines capable of probing the different ISM properties (e.g., density, temperature, metallicity, ionization parameter), which is fundamental for interpreting the results. 

The analysis of local galaxies can provide the tools to interpret high-$z$ galaxies, given the high signal-to-noise ($S/N$) of their spectra and the possibility of getting multi-wavelength coverage, which can help in the data interpretation.
In \citet{mingozzi22} (\citetalias{mingozzi22} hereafter) we have started to create such a UV toolkit using the
Cosmic Origins Spectrograph (COS) Legacy Spectroscopic SurveY (CLASSY\footnote{\href{https://archive.stsci.edu/hlsp/classy}{https://archive.stsci.edu/hlsp/classy}}; \citealt{berg22,james22}, 
\citetalias{berg22} and \citetalias{james22} hereafter). 
CLASSY represents the first high-quality, high-
resolution far-UV (FUV, $1150-2000$~\AA) Treasury of 45 nearby ($0.002 < z < 0.182$) star-forming galaxies, with properties that make them representative at all redshifts (see \citetalias{berg22} Fig.~8), but generally characterized by more extreme ionization fields, lower stellar masses, and higher star formation rates (SFRs) than $z\sim0$ objects, as typically observed in the EoR (e.g., \citealt{wise14, madau15, robertson15, stanway16, stark16}). 
Specifically, in \citetalias{mingozzi22} we provided detailed measurements of dust attenuation ($E(B-V)$), electron density ($n_e$), electron temperature ($T_e$), gas-phase metallicity (12+log(O/H)) and ionization parameter (log($U$)), using both UV and optical direct diagnostics, taking into account the different ionization zones of the ISM. Then, from the comparison of the derived properties, we provided a set of equations to estimate ISM properties only from UV emission lines.
This paper presents the second part of the UV-based toolkit introduced in \citetalias{mingozzi22} and aims at exploring the well-known optical emission-line diagnostics tracing the different ionization sources and the recently proposed UV-based counterparts, to check their consistency and reliability to discriminate SF from AGN and shock ionization. 

In Sec.~\ref{sec:sample} we describe the CLASSY sample, summarizing the characteristics of the UV and optical data, the main steps of the data analysis performed in \citetalias{mingozzi22}, while in Sec.~\ref{sec:methods}, we give an overview of the models used in this work to interpret the optical and UV diagnostics. 
In Sec.~\ref{sec:results-diagnostics-opt} we present the main optical diagnostic diagrams sensitive to the ionization source to demonstrate that CLASSY galaxies are indeed dominated by SF.
% Then, in Sec.~\ref{sec:results-diagnostics-uv} we show where these galaxies locate in the UV diagnostics diagram counterparts 
CLASSY represents the local reference sample of metal-poor star-forming galaxies that we can use to explore the boundaries of the star formation locus in UV diagnostic diagrams, as we show in Sec.~\ref{sec:results-diagnostics-uv}, interpreting the results taking into account ISM properties and state-of-the-art photoionization and shock models.
{Then, in Sec.~\ref{sec:discussion} we discuss the conditions needed to observe UV emission lines, possible caveats in using them as diagnostics and which of them are currently observed with JWST at high-z.} Finally, in Sec.~\ref{sec:conclusions} we summarize our main findings. 
Throughout this work we assume a flat $\Lambda$CDM cosmology (H$_0 = 70$~km/s/Mpc, $\Omega=0.3$) and 12+log(O/H)$_\odot = 8.69$ \citep{asplund09}.

%---------------------------------------------------------------------------------------------------------------
%---------------------------------------------------------------------------------------------------------------

\section{The CLASSY survey} \label{sec:sample}
The 45 local galaxies from the CLASSY survey span a wide range of stellar masses ($6.2 < $~log($M_\star$/M$_\odot$)~$< 10.1$), SFRs ($-2<~$log($SFR$/M$_\odot$yr$^{-1}$)~$<+1.6$), oxygen abundances ($7 < $~12+log(O/H)~$< 8.8$), electron densities ($10 < n_e$(\sii$\lambda\lambda$6717,31)/cm$^{-3}$~$< 1120$), degree of ionization ($0.54 < O3O2 < 38.0$, with $O3O2 =$~\oiii~$~\lambda\lambda$5007/\oii~$~\lambda\lambda$3727,9), and reddening values ($0.02 < E(B-V) < 0.67$).
In \citetalias{berg22}, we present our sample, explaining in detail the selection criteria along with an extensive overview of the HST/COS and archival optical spectra. {Also, \citetalias{berg22} broadly compares the global properties of the CLASSY galaxies with local and high-$z$ samples (see \citetalias{berg22} Sec.~5, Fig.~8), showing that these objects are characterized by similar low masses and metallicity to the dwarf galaxies of the local volume, but are characterized by higher SFRs and sSFRs, consistent with high-$z$ systems. Also, CLASSY objects follow the same trend as other $z\sim0$ star-forming galaxies in the mass-metallicity relation but with a larger scatter. Recent JWST studies are further confirming that such characteristics are typical of the high-$z$ Universe (see e.g., Fig~4 and 5 of \citealt{curti23}; \citealt{atek23,simmonds23}).}
The data reduction is presented in \citetalias{james22}, including spectra extraction, co-addition, wavelength calibration, and vignetting. 
In \citetalias{mingozzi22}, we explained in detail the data analysis, from the stellar continuum modeling to the emission line fitting, providing measurements for the UV redshifts ($z_{UV}$), the main UV and optical emission line fluxes and equivalent widths, and ISM properties. In the following subsections, we summarize the most important characteristics of the sample and the main steps of the data analysis performed in these previous works.

\subsection{The sample}\label{sec:thesample}
CLASSY spectra span from 1150~\AA\ to 2100--2500~\AA, combining the G130M, G160M, G185M, G225M and G140L \HST/COS gratings, from 135 orbits of new COS data (PID: 15840, PI: Berg) and 177 orbits of archival COS data, for a total of 312 orbits. 
Both this paper and \citetalias{mingozzi22} focus on the analysis of all the emission lines (except for \lya) in the range $1150-2000$~\AA, using the so-called \emph{High Resolution} (HR: G130M$+$G160M; $R\sim10000-24000$) and
\emph{Moderate Resolution} (MR: G130M$+$G160M$+$G185M$/$G225M; $R\sim10000-20000$) co-added spectra\footnote{The instrumental broadening at the wavelengths of the emission lines taken into account in the HR and MR spectra is, respectively, $\sigma_{\rm int} \sim 15-30$~km/s.}, presented in detail in \citet{berg22}\footnote{Only for J1044+0353 and J1418+2102, we used the so-called \emph{Low Resolution} (LR: G130M$+$G160M$+$G140L or G130M$+$G160M$+$G140L; $R\sim1500-4000$, $\sigma_{\rm int} \sim 80$~km/s) coadds because of the lack of higher resolution data.}. 
%with a dispersion of 12.23 m\AA/pixel, and a resolution of 0.073 \AA\ per resolution element (\AA/resel, where 1 resel equates to 6 native COS pixels\footnote{This corresponds to an instrumental broadening $\sigma_{\rm int} \sim 25-45$~km/s at the wavelengths of the emission lines taken into account in the HR and MR spectra, respectively.}), and 33 m\AA/pixel and 0.200 \AA/resel, respectively. 
% Only for J1044+0353 and J1418+2102, we used the so-called \emph{Low Resolution} (LR: G130M$+$G160M$+$G140L or G130M$+$G160M$+$G140L; $R\sim1500-4000$) coadds, with a nominal point source resolution of 80.3 m\AA/pixel or 0.498 \AA/resel, since the G185M data are missing. %Additionally, for J1112+5503 we do not have UV emission line measurements, because the COS G185M and G225M observations for the J1112+5503 galaxy were impacted by guide star failures. 
We collected data for the entire CLASSY sample also in the optical wavelength regime, gathering SDSS spectra if available, integral field spectroscopy data (VLT/VIMOS, MUSE, Keck/KCWI), or long-slit spectra (MMT; LBT/MODS, see \citealt{arellano-cordova22}, CLASSY~\citetalias{arellano-cordova22}). 
A full description of the CLASSY optical dataset can be found in \citetalias{mingozzi22}\footnote{The instrumental broadening of the optical data is in the range $\sigma_{\rm int} \sim35-120$~km/s.}.
%with a 3.0" aperture are available for 38 of the CLASSY galaxies \citep{abazajian09}, while DR13 BOSS spectrograph data with a 2.0" aperture exist for J0144+0453 \citep{albareti17,guseva17}. 
% For the remaining galaxies of the sample (J0036-3333, J0127-0619, J0337-0502, J0405-3648, J0934+5514, and J0144+0453), we used integral field spectroscopy data (VLT/VIMOS, MUSE, Keck/KCWI), when available, or long-slit spectroscopy (LBT/MODS, MMT; see \citealt{arellano-cordova22}, CLASSY~\citetalias{arellano-cordova22}), instead of SDSS. A full description of the optical data utilized can be found in \citetalias{mingozzi22}\footnote{The instrumental broadening of the optical data is in the range $\sigma_{\rm int} \sim35-120$~km/s.}.
%%{add paragraph about WISE data? If we use them: Spitzer Space Telescope and the Wide-field Infrared Survey Explorer (WISE)}
% --introduce WISE bands at 3.4, 4.6, 12 and 22~$\mu$m (W1, W2, W3 and W4 bands, respectively; \citealt{wright10}) ---
% WISE observed the whole sky in four mid-infrared bands, 3.4, 4.6, 12, and 22
% from Sartori15: For the Stern selection, we required the objects to be detected in the W1 and W2 bands (in the WISE catalogue this means S/N > 2)

In \citetalias{mingozzi22} and throughout this work, we take into account the properties of the CLASSY galaxies in terms of redshift, stellar mass $M_\star$, $SFR$, and 12+log(O/H), estimated and described in \citetalias{berg22} (see Sec.~4.5 and 4.7).
Specifically, 12+log(O/H) measurements are based on the direct $T_e$ method, using \sii~$~\lambda$6717/$~\lambda$6731 and \oiii~$~\lambda$4363/$~\lambda$5007 as electron density and temperature tracers, respectively. We also take into account the ionization parameter log($U$) from \oiii$\lambda$5007/\oii$\lambda\lambda$3727 and the stellar age of our targets, determined from the analysis presented in \citetalias{mingozzi22}.
For this paper, we also measured the carbon-to-oxygen abundance C/O, which we calculated following the prescriptions of \citet{berg16}. 
Specifically, we calculated the ionic abundances of C$^{++}$ and O$^{++}$ from \ciii$~\lambda$1907,9/\hb\ and \oiii$~\lambda$1666/\hb\, using \pyneb\ \citep{luridiana15} and $T_e$(\oiii). Then, we divided one by the other multiplying per the ionization correction factor (ICF), provided by \citet{berg16}, taking into account log($U$). 
We find (C/O)/(C/O)$_\odot$ to vary in the range 0.1--0.6, with (C/O)$_\odot = 0.44$. 
A detailed discussion about CLASSY C/O will be given by Berg et al. (in preparation).

\subsection{Data analysis}\label{sec:data-analysis}
Here, we briefly summarize the main steps of the data analysis of UV and optical spectra performed in \citetalias{mingozzi22} (see their Sec.~3). 
In particular, we used a set of customized Python scripts in order to first fit and subtract the stellar component and then fit the main emission lines with multiple Gaussian components where needed.
\begin{figure*}
\begin{center}
    \includegraphics[width=0.5\textwidth]{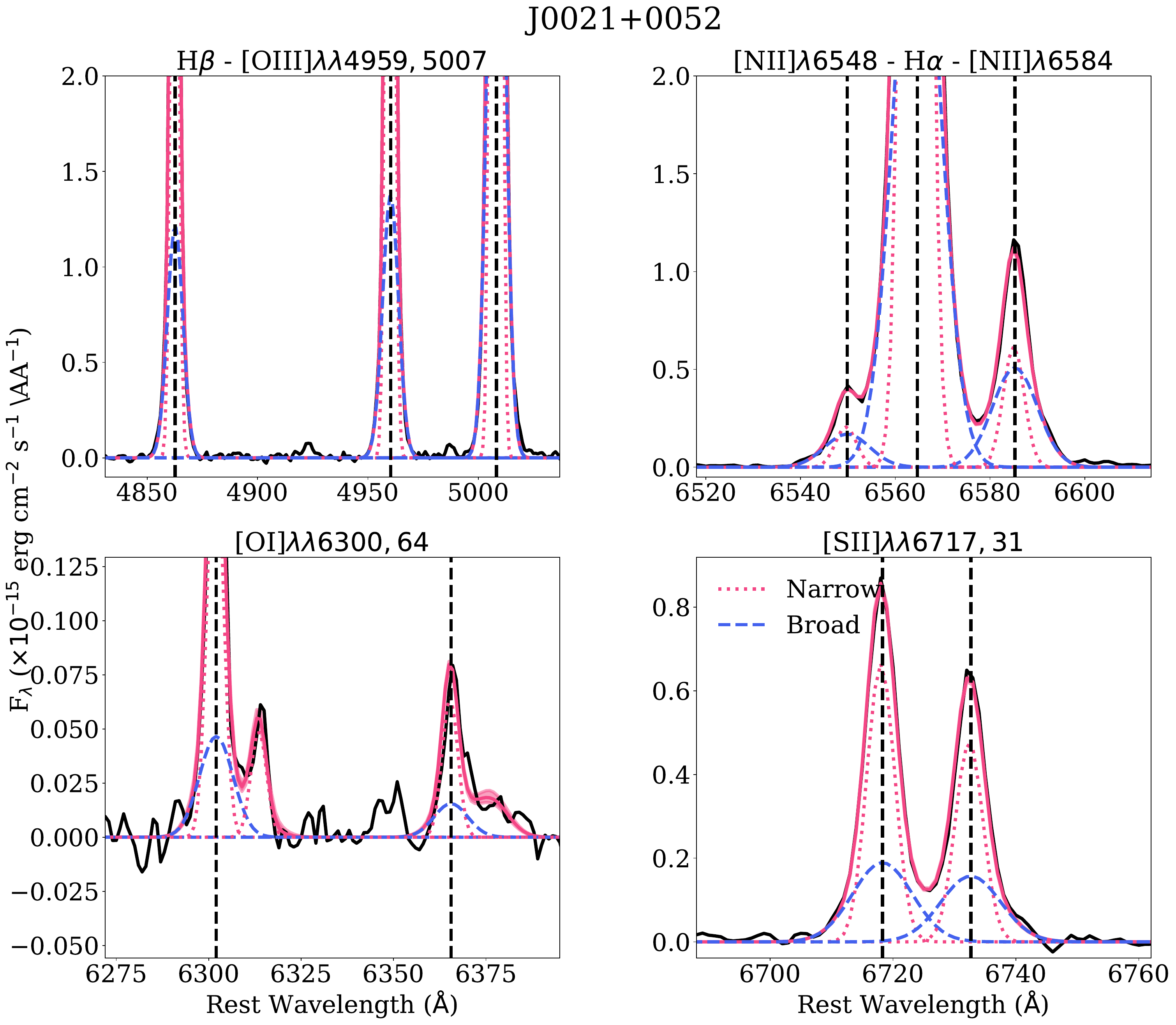}
    \includegraphics[width=0.45\textwidth]{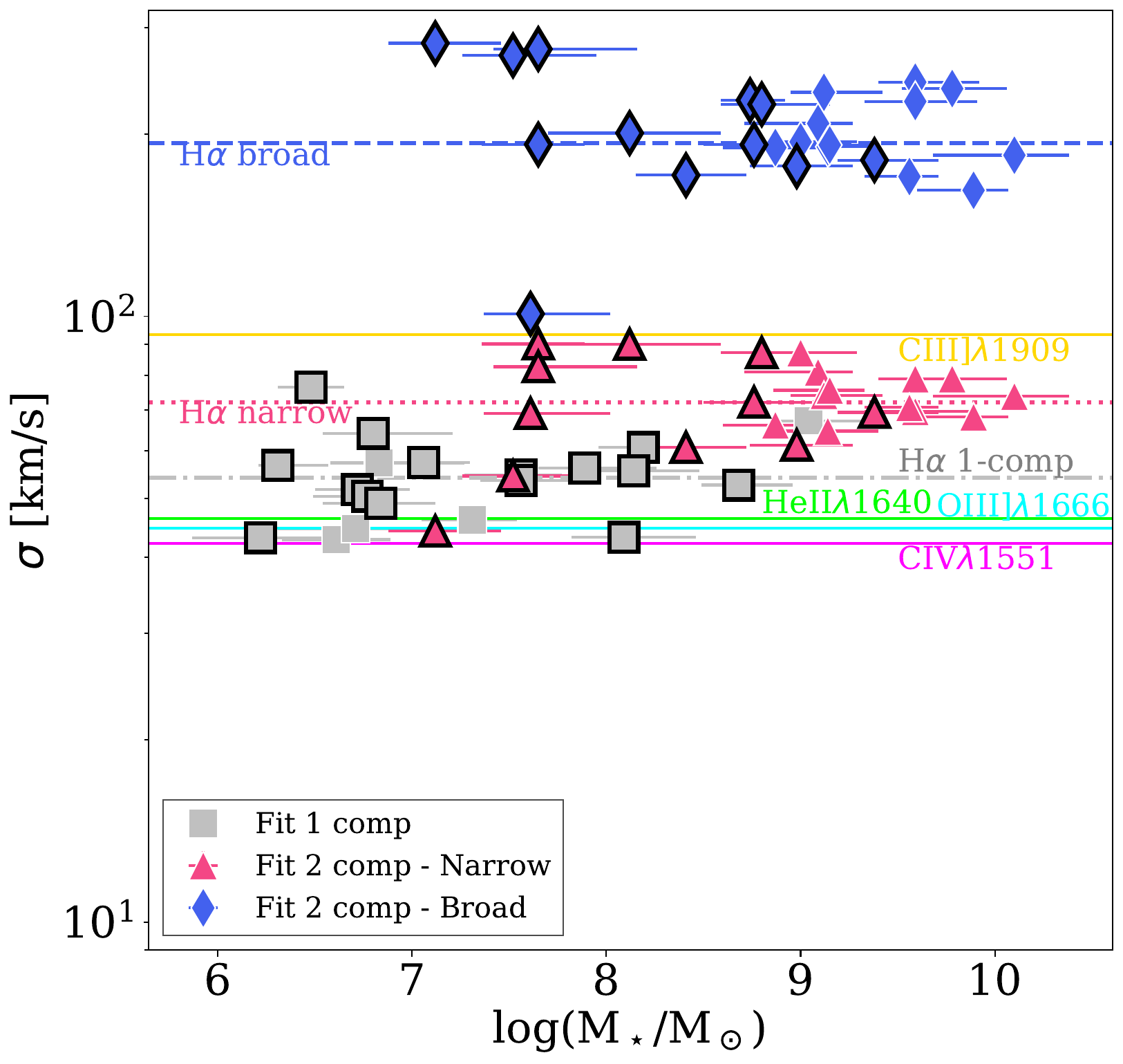}
\end{center}
\caption{Left panel: optical emission lines best-fit with narrow and broad Gaussian components (dotted red and dashed blue lines, respectively; total fit shown by red solid line) for one of the CLASSY galaxies, J0021+0052. Here we show the lines needed to create the classical optical BPT diagrams (\hb, \oiii$~\lambda\lambda$4959,5007, \nii~$~\lambda\lambda$6548,84, \ha, \oi~$~\lambda\lambda$6300,64, \sii$~\lambda\lambda$6717,31) to discriminate among different ionization sources (see Sec.~\ref{sec:results-diagnostics-opt}). 
The emission line next to \oi$~\lambda$6300 is \siii~$~\lambda$6312, while the one next to \oi$~\lambda$6364 is \silii~$~\lambda$6371.
Right panel: intrinsic velocity dispersion $\sigma$ (y-axis in log scale) for the 1-component (grey squares) and, when needed, the 2-component best-fit (narrow and broad components shown as red triangles and blue diamonds, respectively) of optical emission lines, as a function of the stellar mass. The broad component $\sigma$ (median value $\sim 190$~km/s, dashed blue line) is generally larger than the narrow ($\sim 70$~km/s, dotted pink line) and single ($\sim 55$~km/s, dash-dotted grey line) components. 
Galaxies with at least one $S/N>3$ UV emission line (i.e., \ciii$\lambda\lambda$1907,9; 12+log(O/H)~$\lesssim8.3$) are highlighted by a black thick edge.
The UV lines intrinsic $\sigma$ median values are generally consistent with the optical, as shown by the solid horizontal lines: \civ$\lambda\lambda$1548,51 (magenta), \heii$\lambda$1640 (green), \oiiiuv$\lambda\lambda$1661,6 (cyan), \ciii$\lambda\lambda$1907,9 (gold).}
\label{fig:fitcomps}
\end{figure*}

As a preliminary step, both UV and optical CLASSY spectra were corrected for the total Galactic foreground reddening along the line of sight of their coordinates with the \citet{cardelli89} reddening law (see \citetalias{berg22}).
Then, we fitted and subtracted the UV and optical stellar components, adopting the latest version of the \citet{bruzual03} models, assuming the \citet{chabrier03} initial mass function (IMF), as described in detail in \citetalias{mingozzi22} Sec.~3.1. 
% To model the UV stellar component, we fitted the observed high-resolution COS spectra, rebinned by 15 pixels, adopting the latest version of the \citet{bruzual03} models \citep[S.\ Charlot \& G.\ Bruzual, in-preparation, hereafter C\&B; see also][]{gutkin16,vidal-garcia17,plat19}, assuming the initial mass function (IMF) of \citet{chabrier03}. 
%Our fiducial model includes also the nebular continuum, which is a crucial constituent of the UV light in star-forming galaxies, assuming a closed geometry \citep{gutkin16,plat19} and a volume-averaged log($U$)~$=-2.5$ \citep[defined as in][]{gutkin16}.
% This step is important to accurately measure the UV and optical emission lines in the wavelength ranges [$1150-2000$]~\AA, and [$3700-9100$]~\AA, respectively. 
Finally, we performed the fit of the main UV and optical emission lines separately. In particular, we simultaneously fitted each spectrum taking into account a set of UV (from \fniv~$\lambda$1483, \niv~$\lambda$1487, to \fciii~$\lambda$1907, \ciii~$\lambda$1909) and optical emission lines (from \oii$~\lambda\lambda$3727 to \siii$~\lambda$9069) with a linear baseline centered on zero and a single Gaussian, making use of the code \mpfit\ \citep{markwardt09}, which performs a robust non-linear least squares curve fitting (see \citetalias{mingozzi22} Sec.~3.2 for more details). 
%In our procedure we masked the main Milky Way absorption lines, we performed the spectral fitting in windows of 3000~\kms\ centered around each emission line.
% We allowed the line flux to vary, tying the velocity (i.e., the line center) and the velocity dispersion for all the emission lines that , to better constrain weak or blended features\footnote{There are few exceptions for the emission-line fitting procedure described in detail in Sec.~3.2.1 of \citetalias{mingozzi22}}. %We made an exception for the line center of the UV emission-lines \ciii\ and \civ, because they can be significantly shifted in velocity with respect to the others due to their origin, as we will discuss in Sec.~({cite}).

We noted that the \ha\ line, in particular, shows a broad component in many CLASSY galaxies (as shown in the top panel of Fig.~\ref{fig:fitcomps} for J0021+0052). 
Therefore, we performed 2-component Gaussian fits to the main optical emission lines (i.e., \oii~$~\lambda\lambda$3727,9, \hd, \hg, \hb, \oiii~$~\lambda\lambda$4959,5007, \oi~$~\lambda\lambda$6300,74, \nii~$~\lambda\lambda$6548,84, \ha, \sii~$~\lambda\lambda$6717,31). 
In particular, we used a narrow ($\sigma < 200$~km/s) and a broad ($\sigma <1000$~km/s) Gaussian component, tying the velocity and velocity dispersion of each component to be the same for all the emission lines. 
We applied a reduced chi-square $\tilde\chi^2$ selection in order to state if a fit with a second component better reproduces the observed spectral profiles (see Sec.~3.2.2 of \citetalias{mingozzi22} for more details). 
Overall, a second broad component is needed in 24 out of 45 galaxies of our sample. This component tends to be slightly blue-shifted ($\sim 50-80$~km/s) at increasing $M_\star$ and accounts for $\lesssim 30$\% of the total emission (\citetalias{mingozzi22}). 

The right panel of Fig.~\ref{fig:fitcomps} shows the intrinsic velocity dispersion $\sigma$ that we calculate from the optical emission lines as a function of the galaxy stellar mass for our sample\footnote{The emission line velocity dispersion $\sigma$ refers to the \ha\ line for the optical and is corrected for the instrumental broadening $\sigma_{int}$ of the different instruments taken into account, assuming $\sigma = \sqrt{\sigma_{obs}^2 - \sigma_{int}^2}$. }. 
In particular, we show either the single Gaussian $\sigma$ (grey squares) or the narrow (red triangles) and broad component (blue diamonds) $\sigma$, according to the best-fit selected for each galaxy. 
The median values of the single, narrow, and broad components are $\sim 55 $~km/s, $\sim 70 $~km/s and $\sim 190 $~km/s, respectively, as indicated by the dash-dotted grey, dotted red and dashed blue horizontal lines. 
The single Gaussian $\sigma$ for the objects that need a 2-component fit is not shown, but its median value would be $\sim 100 $~km/s, and thus much higher than the value calculated for the narrow component.
Overall, the broad emission component can trace gas at different velocities along the line of sight due to different mechanisms, including stellar winds, galactic outflows or turbulence, linked to different ionization sources (e.g., stellar or AGN photoionization and/or shocks; \citealt{izotov07,james09,amorin12,bosch19,komarova21,hogarth20}).
Hence, in Sec.~\ref{sec:results-diagnostics-opt} we use $\sigma$ to explore whether correlations exist between gas kinematics and the optical diagnostics tracing the level of ionization of the gas (e.g., \citealt{kewley19}). 

In Fig.~\ref{fig:fitcomps}, the CLASSY targets with at least one UV emission line (i.e., \ciii~$\lambda\lambda$1907,9) are highlighted by a black thick edge: they all have generally low stellar masses and 12+log(O/H)~$\lesssim8.3$. 
\ciii$~\lambda\lambda$1907,9 is the most common UV emission line in CLASSY galaxies (observed in 28 objects) without considering \lya\ (see \citealt{hu23} or \citetalias{hu23}), as it is usually seen in low-mass SFGs (e.g., \citealt{rigby15,maseda17,du17}). Specifically, all the CLASSY galaxies showing either \heii$~\lambda$1640 (19) or \oiiiuv$~\lambda1666$ (22) have also \ciii\ emission (see \citetalias{mingozzi22}). Also, these three emission lines are all visible in the 9 CLASSY galaxies characterized by \civ~$\lambda\lambda$1548,51 in pure emission. 
The colored solid horizontal lines in Fig.~\ref{fig:fitcomps} indicate their median intrinsic velocity dispersions, which we calculated, taking into account COS instrumental resolution as well as the extension of the sources derived from the COS NUV acquisition images (see \citetalias{james22} Sec.~2 and Sec 3.2; \citealt{james14} Sec.~3.2).
In general, \heii$~\lambda$1640, \oiiiuv$~\lambda1661,6$ and \civ~$\lambda\lambda$1548,51 share very similar values ($\sigma \sim 50$~km/s), consistent with optical emission lines at similar stellar masses. We will further comment on this in Sec.~\ref{sec:discussion}.
We emphasize that the UV emission-line $S/N$ is typically not high enough to detect faint broad components in the emission lines, apart from a few exceptions {that show broad emission in \civ~$\lambda\lambda$1548,51, \heii~$\lambda$1640 and \oiii~$\lambda\lambda$1661,6 (but not in \ciii)} (\citetalias{mingozzi22}, {Sec.~3.3.}). 
Also, UV lines are not seen in the most massive targets in which the optical line broad components are particularly enhanced.
However, the fact that $\sigma$(\ciii) is systematically higher than other UV lines ($\sigma$(\ciii)~$\sim90$~km/s) could indicate the presence of a hidden undetected broad component. 
A further discussion of the optical and UV kinematical properties of the sample can be found in App.~\ref{app:uv-opt-kins}.

%---------------------------------------------------------------------------------------------------------------
%---------------------------------------------------------------------------------------------------------------

\section{Diagnostic Models Used within this Work}\label{sec:methods}
This work aims to understand how state-of-the-art models from the literature compare to observed line ratios of the CLASSY survey, using well-studied optical diagnostics and less well-constrained UV diagnostics. 
This section provides an overview of each model considered for the main sources of ionization found within galaxies: photoionization, AGN, and shocks. We provide a summary of the models taken into account in Tab.~\ref{tab:models}. In the next sections of the paper, we investigate and comment on optical and UV diagnostics to distinguish between these different ionizing mechanisms. 

Concerning stellar photoionization grid models (referred to as SF models hereafter), we take into account emission from \hii\ regions around young stars, using nebular-emission models of SFGs from \citet{gutkin16} (\citetalias{gutkin16} hereafter) based on two limiting cases: one instantaneous burst (single stellar population; SSP models) or constant activity with time (CST models). 
In particular, single-burst models, in which the emission is strongest at young ages (i.e., from 3-4~Myr up to 5-7~Myr if stellar rotation is included) and disappears as the massive star population evolves, are generally used for young stellar populations and very recent starbursts, and thus more consistent with the CLASSY galaxies and EoR systems. %
\citetalias{gutkin16} models do not take into account either stellar rotation or stellar multiplicity, which would imply a longer duration of ionizing photon production (e.g., MIST models from \citealt{byler17} models) and, in the latter case, a harder radiation field (e.g., {Binary Population and Spectral Synthesis,} `BPASS' models from \citealt{xiao18}).
Other possible differences between \citetalias{gutkin16} and other models in the literature concern the choice of stellar libraries, IMF or the way to scale gas-phase metallicity for elements such as nitrogen and carbon\footnote{The former has secondary nucleosynthetic production at high metallicity, while part of the latter returns to the ISM through metallicity-dependent processes such as stellar winds (see e.g., \citealt{byler18} for more details).}.
For the purpose of this study, {we take into account \citetalias{gutkin16} models (with constant and burst of SF), BPASS models from \citet{xiao18}\footnote{\url{http://www.bpass.auckland.ac.nz}} (\citetalias{xiao18} hereafter) and \citet{byler17} (\citetalias{byler17} hereafter) models. \citetalias{gutkin16} model parameters (see Tab.~\ref{tab:models}) are well matched with the AGN and shock grids that we present below, allowing us to make a consistent comparison between optical and UV diagnostic diagrams. On the other hand, \citetalias{xiao18} and \citetalias{byler17} models allow us to explore the differences caused by the inclusion of interacting binary stars in the stellar population and stellar rotation, respectively (and thus a harder radiation field), as we show in Sec.~\ref{sec:results-diagnostics-uv} and App.~\ref{app:uv-diag-models}.}
%The comparison between different model prescriptions is beyond the scope of this paper.
% here we want to explore if UV diagnostic diagram classification can be consistent with the optical and thus identify the best-observed UV diagnostics to discriminate SF from AGN and shock ionization.

As shown in \citet{feltre16} Fig.~1, the ionizing spectrum powered by a lower metallicity stellar population is harder with respect to a more metal-enriched one, producing a different spectral energy distribution at photon energies greater than $\sim20$~eV. 
However, it generally cannot account for ionic species with ionization energies above $\sim50$~eV. Hence, the presence of emission lines requiring such ionization energy could be a good indicator of the presence of an additional ionization mechanism than SF (e.g., AGN or shocks). 
In this work, we take into account the AGN models from \citet{feltre16} (\citetalias{feltre16} hereafter). 
\begin{longrotatetable}
\begin{table*}
\begin{center}
\caption{Summary of the models used in this work: constant (CST) and single-burst (single stellar population; SSP) \citet{gutkin16} grids (\citetalias{gutkin16}) {and \citet{xiao18} BPASS models (\citetalias{xiao18})} for SF,
\citet{feltre16} ({\citetalias{feltre16}}) for AGN narrow-line region (NLR), \citet{alarie19} (\citetalias{alarie19}) %and \citet{allen08} (\citetalias{allen08}) 
for shocks. The parameters include the ionizing spectrum, gas-phase metallicity (Z$_{ISM}$), ionization parameter log($U_S$), dust-to-gas ratio ($\xi_d$), carbon-to-oxygen abundance (C/O) and initial mass function (IMF). $\xi_d$ accounts for the depletion of metals onto dust grains in the ionized gas. %The depletion factors in \citetalias{byler17} models are taken from \citet{dopita13} (see also \citealt{byler18} Table~1). 
We refer to Sec.~4 of \citet{plat19} to see the impact of the variation of the parameters that we fix in this work.}
\label{tab:models}
\begin{tabular}{cccc} 
\hline
\hline  
Model & Parameters & Sampled Values & Notes \\
\hline
\hline
\citetalias{gutkin16} & Ionizing spectrum & CST SF; age~$=100$~Myr & \multirow{9}{10cm}{\citet{bruzual03} stellar pop. synthesis models + Cloudy \citep{ferland13}. The values of gas-phase metallicity correspond to 12+log(O/H)~$=6.64, 7.64, 8.55, 8.88$ and log(Z$_{ISM}$/[Z$_\odot$])~$-2.2$-0.04. We tested all values of $n_H$, $\xi_d$ and (C/O), but here we show only $n_H=100$~cm$^{-3}$, $\xi_d = 0.3$ and (C/O)/[(C/O)$_\odot$]~$=0.1, 0.72$. Special prescriptions for C and N because scale non-linearly with O (G16 Sec.~2.3.1; see \citealt{henry00,berg16}).} \\
    & & & \\
    & Z$_{ISM}$     & 0.0001,0.001,0.008,0.017; Z$_\odot = 0.01524$ & \\
    & log($n_H$/[cm$^{-3}$])       & 1, 2, 3, 4 &  \\
    & log($U_S$)    & -3.5, -3. , -2.5, -2. , -1.5, -1. & \\
    & $\xi_d$  & 0.1, 0.3, 0.5  &  \\
    & (C/O)/[(C/O)$_\odot$]     & 0.10, 0.14, 0.20, 0.27, 0.38, 0.52, 0.72, 1.00, 1.40; (C/O)$_\odot =0.44$ & \\%0.10, 0.14, 0.20, 0.27, 0.38, 0.52, 0.72, 1.00, 1.40
    & IMF      & \citet{chabrier03} ($m_{min} =$~0.1~M$_\odot$; $m_{up} = 100., 300.$~M$_\odot$) & \\
    % & Others      & \multicolumn{2}{}{} \\
\hline
\citetalias{gutkin16} & Ionizing spectrum & SSP; age~$=1-10$~Myr & \multirow{4}{10cm}{Same as in CST \citetalias{gutkin16} grids. Given the absence of stellar rotation and multiplicity, the ionizing radiation drops at older ages than 5~Myr. All the parameters are as in CST models. In this work, we show only grids with $t=3,5,10$~Myr.}   \\
    & & & \\
    % & Z$_{ISM}$     & Same as in CST \citetalias{gutkin16} & \\
    % & $n_H$/[cm$^{-3}$]         & Same as in CST \citetalias{gutkin16} &  \\
    % & log($U_S$)    & Same as in CST \citetalias{gutkin16} & \\
    % & $\xi_d$  & Same as in CST \citetalias{gutkin16}  &  \\
    % & (C/O)/[(C/O)$_\odot$]         & Same as in CST \citetalias{gutkin16} & \\%0.10, 0.14, 0.20, 0.27, 0.38, 0.52, 0.72, 1.00, 1.40
    % & IMF      & Same as in CST \citetalias{gutkin16} & \\
    % & Others      & Same as in CST \citetalias{gutkin16} &  \\
    & & & \\ 
    & & & \\ 
    % & & & \\ 
    % & & & \\ (i.e., Z$_{ISM}$, $n_H$/[cm$^{-3}$], log($U_S$), $\xi_d$ , (C/O)/[(C/O)$_\odot$], IMF) 
\hline
{\citetalias{xiao18} }  & Ionizing spectrum & SSP; age~$=1-100$~Myr & \multirow{4}{10cm}{BPASS V2.1 population models \citep{stanway16,eldridge17} + Cloudy \citep{ferland13}. The gas nebula is assumed without dust and a constant $n_H$ with 12+log(O/H)~$=6.6, 7.61, 8.52, 8.93$ (matched to Z$_{\star}$). In this work, we show only grids with $n_H=100$~cm$^{-3}$ and $t=3,10$~Myr. C, N and other elements fractions are scaled according to the population metallicity Z (see Table 2, \citealt{xiao18}). }  \\
    & & & \\
    & Z$_{ISM}$     & 0.0001,0.001,0.008,0.020; Z$_\odot = 0.020$ & \\
    & log($n_H$/[cm$^{-3}$])         & $0-3$ in 0.5 dex intervals &  \\
    & log($U_S$)    & -3.5, -3. , -2.5, -2. , -1.5 & \\
    & & & \\ 
    & & & \\ 
    % & & & \\ 
\hline
{\citetalias{byler17}} & Ionizing spectrum & Bursts of SF; age~$t =0.5, 1, 2, 3, 4, 5, 6, 7, 10$~Myr &  \multirow{4}{10cm}{MESA Isochrones\&Stellar Tracks + Cloudy + Flexible Stellar Population Synthesis code \citep{byler-fsps18}. Stellar rotation and nebular line and continuum emission included. Ionizing radiation is strong at young ages (up to 5-7~Myr) and disappears as massive stars evolve. Larger metallicity range than \citetalias{gutkin16} %but starting from Z$_{ISM} \sim0.001$
(12+log(O/H)~$\sim6.93-8.68$). In this work, we show only $t=3,5,10$~Myr.} \\
    & log(Z$_{ISM}$/[Z$_\odot$])   & −1.5, −1.0, −0.6, −0.4, −0.3, −0.2,−0.1, 0.0, 0.1, 0.2; Z$_\odot = 0.0142$ &  \\
    & $n_H$/[cm$^{-3}$]         & 100. & \\
    & log($U_S$)         & Same as G16 & \\
    & IMF & \citet{kroupa01} ($m_{min} =$~0.1~M$_\odot$; $m_{up} = 120.$~M$_\odot$) & \\
    & Others & N and C treated with equations reported in \citet{byler17}. & \\
    % & & & \\
    % & & & \\
    % & & & \\
    % & & & \\
\hline
\hline
\citetalias{feltre16} & Ionizing spectrum     & AGN NLR; $\alpha = −1.2, −1.4, −1.7, −2.0$ & \multirow{8}{10cm}{Cloudy models with open geometry, luminosity $L_{AGN}=10^{45}$~erg/s, inner radius from the NLR of 300~pc. Ionizing spectrum parameterized with the power-law index $\alpha$ of the spectral energy distribution of the incident ionizing radiation from the AGN accretion disc at UV and optical wavelengths in AGN models. All heavy
elements except for nitrogen (see \citetalias{feltre16} Eq.~6) are assumed to scale linearly with oxygen abundance. } \\
    & Z$_{ISM}$ & Same as G16    & \\
    & $n_H$/[cm$^{-3}$]         & Same as G16 &  \\
    & log($U_S$)    & Same as G16 & \\
    & $\xi_d$  & Same as G16  &  \\
    & & & \\
    & & & \\
    % & & & \\
\hline
\hline
% \citetalias{allen08} & Ionizing spectrum & Fast shocks ($v_s>100$~km/s) & Mappings-V \citep{sutherland17} and a plane-parallel geometry.   & \\
%     & Z$_{ISM}$ & 0.04   & Solar metallicity \\
%     & $n_H$/[cm$^{-3}$]         & 1. &  \\
%     & $v_s$/[km/s]    &  100, 125, ..., 1000 & \\
%     & $B_0$/[$\mu$G]     &  10$^{-4}$, 0.5, 1.0, 2.0, 3.23, 4.0, 5.0, 10 &  \\
% \hline
\citetalias{alarie19} & Ionizing spectrum &  Fast shocks ($v_s>100$~km/s)   & \multirow{9}{10cm}{Mappings-V \citep{sutherland17} and a plane-parallel geometry. Models created using same prescriptions of the well-known \citet{allen08} shock models by \citetalias{alarie19}, but using the metallicity values of \citetalias{gutkin16}, to study variation with metallicity. We tested all values, but in this paper we show only grids with $n_H = 1$~cm$^{-3}$, (C/O)/(C/O)$_\odot = 0.26$ for Z$_{ISM} = 0.0001,0.001$ and (C/O)/(C/O)$_\odot = 1.00$ for Z$_{ISM} = 0.017$ (trans-solar metallicity grid).} \\
    & Z$_{ISM}$ & 0.0001, 0.001, 0.008, 0.08, 0.017 &  \\ %, 0.03, 0.04    & \\
    & $n_H$/[cm$^{-3}$]         & 1., 10., 100., 1000., 10000. &   \\
    & $v_s$/[km/s]    &  100, 125, ..., 1000 & \\
    & (C/O)/[(C/O)$_\odot$]         & 0.26, 1.00 &  \\    
    & $B_0$/[$\mu$G]     &  10$^{-4}$, 0.5, 1.0, 2.0, 3.23, 4.0, 5.0, 10 &  \\
    & & & \\
    % & & & \\ 
    & & & \\ 
\hline
\hline
%\multicolumn{5}{l}{Input Parameters}\\ 
% \multirow{3}[]{100}{We selected the same values of Z$_{ISM}$, log($U_S$) and used for G16.}
\end{tabular}
\end{center}
\end{table*}
% \tablecomments{}
\end{longrotatetable}
The main parameters in {F16 models} are the slope $\alpha$ of the ionizing spectrum\footnote{{The luminosity per unit frequency S$_\nu$ of the AGN accretion disc is usually approximated by a broken power law, with S$_\nu \sim \nu^{\alpha}$ at wavelengths $0.001 \leq \lambda/\mu \leq 0.25$, as shown in Eq.~5 of \citetalias{feltre16}}}, as well as the other quantities shown in detail in Tab.~\ref{tab:models}.
Collisional excitation and photoionization from slow ($v\lesssim150$~km/s) and fast shocks ($v\sim150-1000$~km/s), respectively, can also produce a rich spectrum of UV and 
optical emission lines, since shocked regions have high electron temperature and ionization state \citep{dopita03}.  
Here we take into account the shock models from the 3MdBs\footnote{\url{http://3mdb.astro.unam.mx/}} database (\citealt{alarie19}; \citetalias{alarie19}). 
These model grids span a broad range of metallicity, matched with \citetalias{gutkin16} grid values (see Tab.~\ref{tab:models}), with respect to the well-known MAPPINGS models from \citet{allen08}\footnote{\citetalias{alarie19} models are made with MAPPINGS-V and agree with \citet{allen08} predictions at LMC metallicity (see \citetalias{alarie19} for more details).}, and thus are the best suited for the CLASSY sample. 
The emission line spectra primarily depend on the shock velocity $v_s$ and the magnetic field parameter $B_0$, which regulate the shock ionization spectrum and the effective ionization parameter. 
An increase in the density of the material preceding the shock ionization front (preshock density $n_H$) can play a role as well because it enforces the collisional de-excitation of forbidden lines.
Generally, two sets of models are taken into account: shocks and shocks+precursor. 
The precursor emission becomes important at increasing velocities of the shock front ($v\gtrsim170$~km/s), where the ionization front velocity exceeds $v_s$. 
Indeed, the high velocity causes the photoionization front to detach from the shock and form a ``precursor'', which can contribute to or even dominate the shock emission. 
In general, we noticed that an increase in density from $n_H\sim1$~cm$^{-3}$ to $n_H\sim1000$~cm$^{-3}$ does not lead to dramatic changes in the optical and UV line ratios discussed in the following sections unless noted in the text. 
Hence, in this work, we show only grids with $n_H=1$~cm$^{-3}$ (consistently with shock grids usually shown in other works; e.g., \citealt{allen08}).
Also, at sub-solar metallicities, we show the grids with (C/O)/(C/O)$_\odot = 0.26$, which are also more consistent with the properties of the CLASSY galaxies.

\section{Results: Optical diagnostics sensitive to the ionization source} \label{sec:results-diagnostics-opt}
In this Section we present the most frequently used classification methods in the optical, showing and discussing where the CLASSY galaxies are located in different diagnostic diagrams. 
{In particular, we compared different criteria to discriminate between ionization processes, sometimes finding that there is not a one-to-one agreement in classification, especially for low-mass and low-metallicity galaxies (e.g., \citealt{groves06,shirazi12,sartori15,polimera22,nakajima22}). In Table~\ref{tab:classification}, we summarize the explored optical diagnostics and the corresponding classifications for each galaxy. 
Overall, a combination of criteria represents the best way to provide a robust determination of the main ionization source.}

{Before going into the details listed in this Section, here we disclose our main results about optical diagnostics. In this work,} we find that CLASSY galaxies are characterized by \oiii/\hb\ narrow and broad components typical of star-forming galaxies. 
Furthermore, their broad components --- if present --- have enhanced \nii/\ha\ (and slightly enhanced \sii/\ha\ and \oi/\ha) and \oi/\oiii\ (possible shock evidence), but lower \oiii/\oii\ (i.e., lower level of ionization) and do not show a trend with \heii$\lambda$4686 (that could be enhanced by shocks/AGN).
We do not preclude that a few of them can have shocks (and possibly hidden AGN activity in the trans-solar galaxy J0808+3948), but, in general, the main ionizing mechanism is clearly star formation. 
This conclusion is not surprising since the sample was selected excluding targets with a classification other than star-forming, but here we further confirmed it by exploring different diagnostic diagrams. 

Performing this optical-diagnostic test is important since in Sec.~\ref{sec:results-diagnostics-uv} we test how UV diagnostic diagrams compare to the well-known and reliable optical ones presented here.
In the following plots, we highlight with a black thick edge the objects with at least one UV line detection (i.e., $S/N$(\ciii~$\lambda\lambda$1907,9)~$>3)$, since only a sub-sample of low-metallicity objects shows UV emission lines. 
% Overall, this section is important to demonstrate the star-forming nature of CLASSY galaxies, highlighting which are the objects that we test and interpret the UV diagnostic diagrams in Sec.~\ref{sec:results-diagnostics-uv}.

\begin{table*}
\begin{center}
\caption{{CLASSY sample (name and alternative name shown in (1) and (2)) classification according to the optical diagnostic diagrams taken into account: \nii-BPT (3), \sii-BPT (4), \oi-BPT (5), the [OI]-[OII]-[OIII] discussed in Sec.~\ref{sec:oi-shocks} (6), \heii-diagram discussed in Sec.~\ref{sec:shirazi} (7). The dominant ionization mechanism is classified as star formation (SF), composite (Comp), Active Galactic Nuclei (AGN), or, if one or more of the involved lines have $S/N<3$, not classified (NC). The '...' indicates that some of the involved emission lines are out of the observed wavelength range. Columns (3), (4), and (5) report the classification using the standard separators shown in black in Fig.~\ref{fig:optbpt}. According to the updated BPT separators from \citetalias{xiao18} (shaded green in Fig.~\ref{fig:optbpt}) all the galaxies are classified as SF dominated. None of the CLASSY galaxies are classified as a low-ionization (nuclear) region or LI(N)ER. For the optical diagnostics, we specify the classification of both the narrow (N) and (B) components - if present. {The 28 galaxies in bold are those which show a $3\sigma$ detection in \ciii$\lambda\lambda$1907,9} (12+log(O/H)~$\lesssim8.3$, $Z\lesssim 50$~\% Z$_\odot$).}}
\label{tab:classification}
\begin{tabular}{ccccccc} 
\hline
\hline   
Object  &  & [NII]-BPT & [SII]-BPT & [OI]-BPT &  [OI]-[OII]-[OIII] & HeII\\
        & & N/B & N/B & N/B & N/B & N/B \\
\hline
J0021+0052 & & SF/SF  &   SF/SF   & SF/SF         &  SF/SF & SF      \\
J0036-3333 &  Haro 11 knot C & SF/SF  &   SF/SF   & SF/SF               & SF/SF & SF   \\
{\bf J0127-0619} & Mrk~996& SF/SF  &   SF/NC   & SF/NC               & SF/NC & NC       \\
{\bf J0144+0453} &UM~133& SF/NC  & {\it AGN}/NC & {\it AGN}/NC &  SF/NC & SF        \\
{\bf J0337-0502} &SBS~0335-052~E& SF/NC     &   SF/NC      & SF/NC            &  SF/NC    & SF/NC           \\
J0405-3648 && SF     &   SF      & SF                 & SF & SF       \\
J0808+3948 && {\it Comp}/SF&   SF/SF   & SF/SF          & {\it AGN-shocks}/NC  & {\it AGN}        \\
{\bf J0823+2806} & LARS9& SF/{\it Comp}&   SF/SF   & SF/SF            & SF/SF  & SF        \\
{\bf J0926+4427} & LARS14 &SF/{\it Comp}&{\it AGN}/SF& {\it AGN}/SF  & SF/SF &   NC       \\
{\bf J0934+5514} &Izw~18& ...    &   ...     & ...                 & ...  & SF      \\
J0938+5428 & &SF/SF  &   SF/SF   & SF/SF          &SF/SF & SF        \\
J0940+2935 & & SF     &   SF      & SF                  &SF & NC          \\
{\bf J0942+3547} & CG-274, SB 110 & SF     &   SF      & SF               &SF     & SF          \\
J0944+3442 & &SF     &   SF      & SF                 & {\it AGN-SF}    & NC         \\
{\bf J0944-0038} & CGCG007-025 &SF     &   SF      & SF               &SF    & SF           \\
{\bf J1016+3754} & 1427-52996-221 &SF     &   SF      & SF               &SF    & SF            \\
{\bf J1024+0524} & SB~36&SF     &   SF      & SF                  &SF  & SF           \\
J1025+3622 & & SF/SF  &   SF/SF   & {\it AGN}/SF    &SF/SF  & SF        \\
{\bf J1044+0353} & & SF     &   SF      & SF                &SF  & SF         \\
{\bf J1105+4444} & 1363-53053-510& SF/SF  &   SF/SF   & SF/{\it AGN}  &SF & SF        \\
J1112+5503 & &{\it Comp}/SF&   SF/SF   & SF/SF       &SF/SF   & {\it Comp}     \\
{\bf J1119+5130} & &SF     &   SF      & SF               &SF    & SF        \\
{\bf J1129+2034} & SB~179 &SF     &   SF      & SF            &SF      & SF              \\
{\bf J1132+1411} & SB~125  &SF     &   SF      & SF             &SF     & SF            \\
J1132+5722 & SBSG1129+576 &SF     &   SF      & SF             &SF     & SF           \\
J1144+4012 & & SF/SF  &   SF/SF   & SF/SF         &SF & NC       \\
{\bf J1148+2546} & SB~182& SF     &   SF      & SF                 &SF  & SF          \\
{\bf J1150+1501} & SB~126, Mrk~0750 &SF     &   SF      & SF              &SF       & SF            \\
J1157+3220 & 1991-53446-584 & SF     &   SF      & SF                 &SF    & SF          \\
{\bf J1200+1343} & & {\it Comp}/{\it Comp} & SF/SF& {\it AGN}/NC  &SF/NC & SF        \\
{\bf J1225+6109} & 0955-52409-608 & SF/NC  &   SF/NC   & SF/NC           &SF/NC   & SF         \\
{\bf J1253-0312} & SHOC391 & ...  &   ...   & ...            & SF/SF & NC        \\
{\bf J1314+3452} & SB~153 & SF     &   SF      & SF               &SF     & SF          \\
{\bf J1323-0132} & & SF     &   SF      & SF                 &SF   & SF          \\
{\bf J1359+5726} & Ly~52, Mrk~1486 & SF/SF  &   SF/SF   & SF/SF           &SF/SF & SF        \\
J1416+1223 & & {\it Comp}/SF&   SF/SF   & SF/SF         &SF/SF     & NC       \\
{\bf J1418+2102} & & SF     &   SF      & SF            &SF     & SF        \\
J1428+1653 & & SF/SF  &   SF/SF   & SF/SF          &SF/SF & NC   \\
{\bf J1429+0643} & & SF/SF & {\it AGN}/SF & {\it AGN}/{\it AGN}  &SF/SF & SF     \\
{\bf J1444+4237} & HS1442+4250 &SF     &   SF      & SF              &SF   & NC           \\
{\bf J1448-0110} & SB~61 &SF  &   SF & SF   &SF  & SF       \\
J1521+0759 && SF/SF  &   SF/SF   & {\it AGN}/SF   &SF/SF & NC         \\
J1525+0757 && SF/SF  &   SF/SF   & SF/SF          &SF/SF & NC        \\
{\bf J1545+0858} & 1725-54266-068 & SF/{\it Comp} &  SF/SF   & SF/{\it AGN}  & SF/SF  & SF       \\
J1612+0817 && {\it Comp}/{\it Comp} &SF/SF   & SF/SF          & SF/SF  & SF     \\
\hline
(1)&(2)&(3)&(4)&(5)&(6)&(7)\\
\hline
%\multicolumn{5}{l}{Input Parameters}\\ 
\end{tabular}
\end{center}
% \tablecomments{}
\end{table*}

\subsection{The classical optical BPT diagrams}\label{sec:results-bpt}
\begin{figure*}
\includegraphics[width=0.5\textwidth]{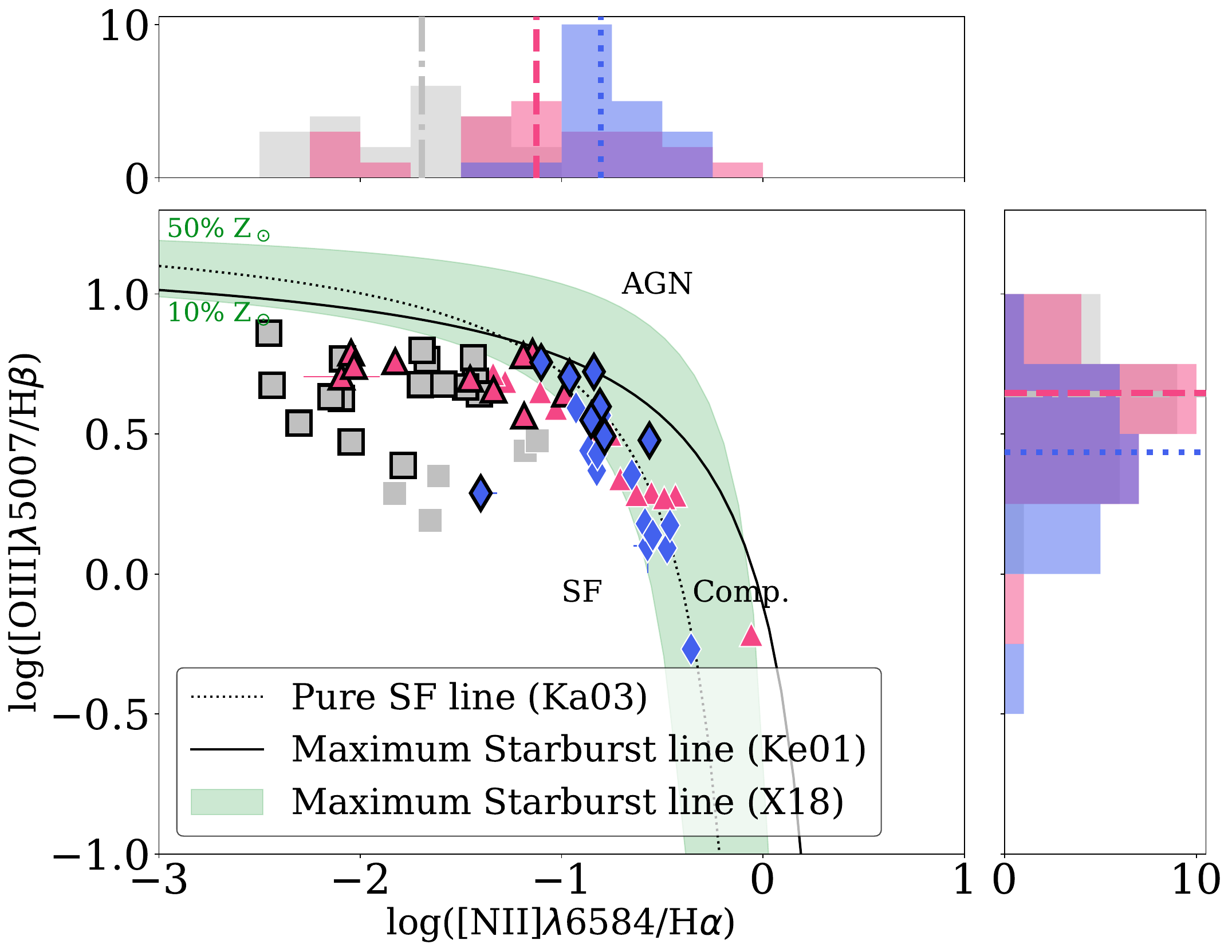}
\includegraphics[width=0.5\textwidth]{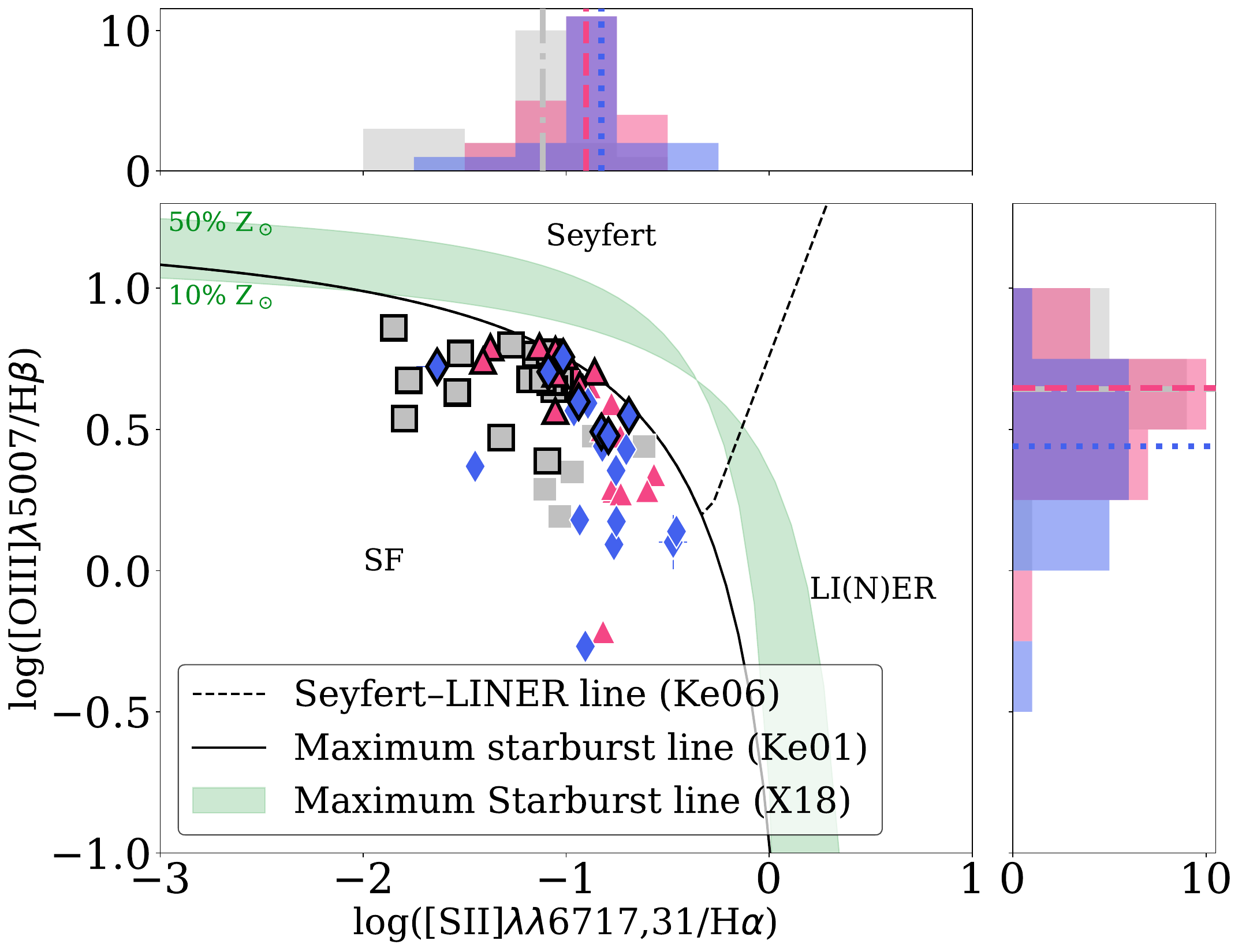}
\begin{center}
\includegraphics[width=0.51\textwidth]{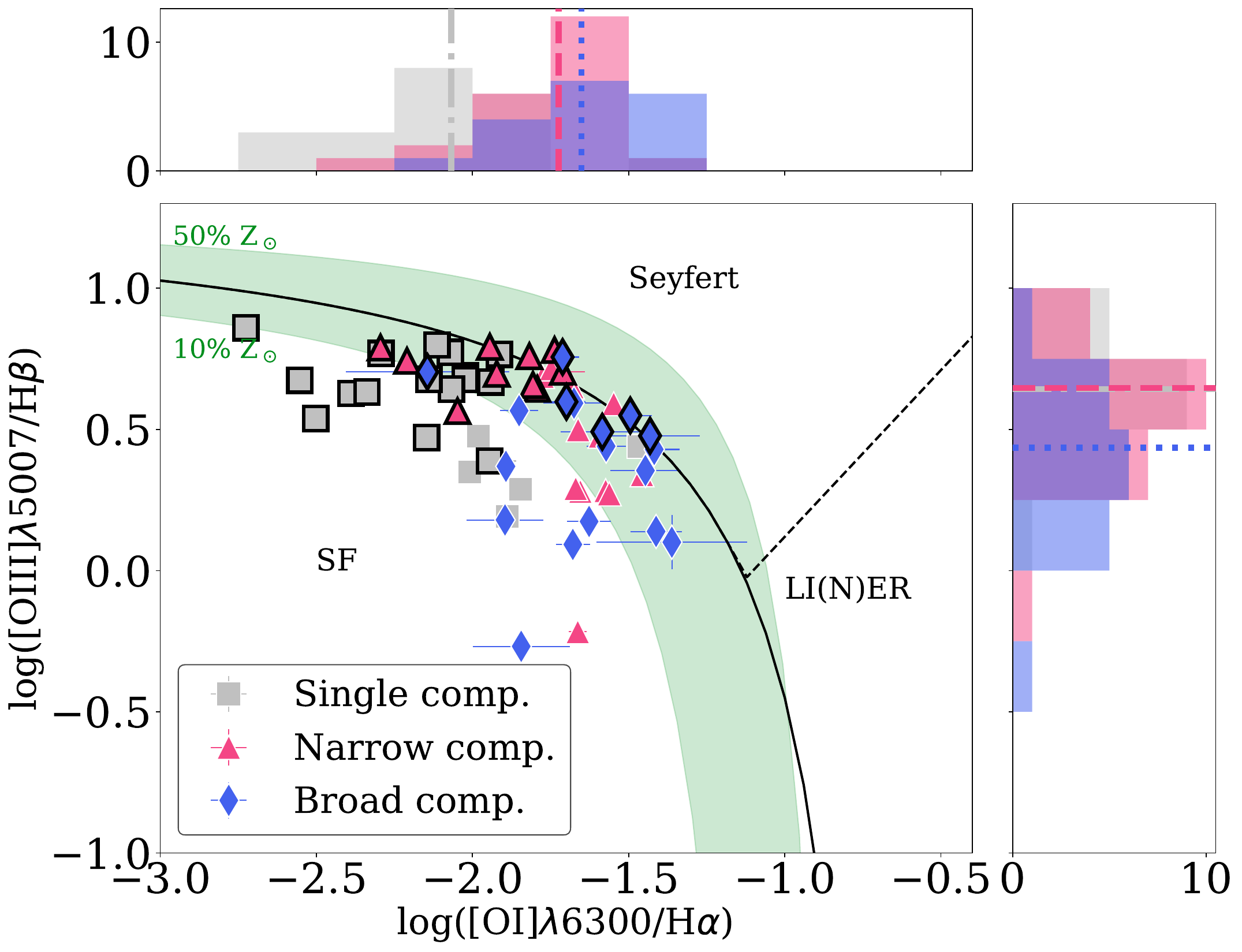}
\end{center}
\caption{\nii\ (top left), \sii\ (top right), and \oi\ (bottom) BPT diagrams for the {single,} narrow and broad components, shown as {gray squares, red triangles} and blue diamonds, respectively. All the measurements have $S/N>3$ for all the lines involved.  
The black dotted curve is the empirical relation to divide pure star-forming from Seyfert–\hii\ composite objects by \citet{kauffmann03}, while the black solid curves are the maximum starburst line derived by \citet{kewley01}. The black dashed lines are the \citet{kewley06} boundary between Seyferts and low-ionization (nuclear) emission-line regions (LI(N)ERs; see also \citealt{heckman80,belfiore16}). 
Histograms showing the distribution of \nii/\ha, \sii/\ha, \oi/\ha, and \oiii/\hb\ line ratios in the {single}, narrow and broad components are shown in {gray}, red and blue, respectively, on each axis, with median values shown by the {dash-dotted}, dashed and dotted black lines. The broad component clearly shows higher x-axis line ratios. The green-shaded regions represent the maximal starburst predictions at metallicity between $10$\%~Z$_\odot$ and $50$\%~Z$_\odot$ values, using pure stellar photoionization including massive star binaries through the BPASS stellar evolution models \citep{xiao18}, to show the extent to which the maximal starburst line can change in the metallicity range in which we observe UV emission lines. {According to the updated BPT separators from \citetalias{xiao18}, thus taking into account the gas-phase metallicity of each target, all the galaxies are classified as SF dominated.} The galaxies with at least one UV emission line (i.e., \ciii~$\lambda\lambda$1907,9) are highlighted by a black thick edge. }
\label{fig:optbpt}
\end{figure*}
The three panels of Fig.~\ref{fig:optbpt} illustrate the \nii, \sii\ and \oi-BPT diagrams for the single {(grey squares),} narrow (red {triangles}) and broad (blue diamonds) components of the CLASSY galaxies. The SF models are not shown for clarity reasons but can completely cover the classical SF locus, while the \citetalias{feltre16} models predict higher \oiii/\hb, consistently with the classical AGN locus.
According to these diagnostics, the majority of the CLASSY galaxies are dominated by SF, even though the classification is not always consistent among different diagnostics, with many objects at the edge of the SF locus (indicated by the black solid maximum starburst line). 
In particular, at these edges, we mainly find the CLASSY objects characterized by UV emission lines (highlighted by thick black edges), that have generally higher \oiii/\hb\ than the others and low gas-phase metallicity (12+log(O/H)~$\lesssim8.3$, corresponding to $Z\lesssim 50$\% Z$_\odot$).
However, as we discuss in the following paragraphs, two main factors can affect the definition of the SF locus and need to be taken into account: the metallicity and the hardness of the radiation field.
% For instance, the broad \nii/\ha\ component is classified as composite in J0823+2806, J0926+4427, J1200+1343, J1545+0858 and J1612+0817, while their broad \sii/\ha\ and \oi/\ha\ components as star-forming. Also, the broad components of J1105+4444, J1429+0643 and J1545+0858 fall in the AGN locus of the \oi-BPT. % % The broad component of J1448-0110 is classified as AGN according to the \nii- and \sii-BPT diagram, while the broad components of J1105+4444, J1429+0643 and J1545+0858 falls in the AGN locus of the \oi-BPT. % No narrow components are AGN-dominated according to the \nii-BPT, while those of J0808+3948, J1112+5503, J1200+1343, J1416+1223 and J1612+0817 fall in the composite locus. 
% Finally, the narrow components of three and eight galaxies are in the AGN locus according to \sii\ (J0144+0453, J0926+4427, J1429+0643) or \oi-BPTs (J0144+0453, J0926+4427, J1025+3622, J1200+1343, J1429+0643, J1521+0759), respectively. We stress that there are no CLASSY galaxies in the LI(N)ER locus. 

On the one hand, \nii/\ha\ is particularly sensitive to metallicity \citep{kewley02, denicolo02, pettini04, kewley08}, with lower \nii/\ha\ at decreasing metallicity (see e.g., the CLASSY galaxies J1323-0132 and J0808+3948 in \citetalias{arellano-cordova22} Fig.~2). Since higher \nii/\ha\ line-ratios are required to classify galaxies as AGN, the \nii-BPT by itself is not sufficient for identifying AGN in typical low-metallicity star-forming $z\sim0$ galaxies (e.g., \citealt{groves06,reines20,molina21,polimera22}). 
%In this context, \citet{reines20,molina21} %suggested using the \oi-BPT for the AGN classification, finding 
%found that galaxies characterized by compact radio sources in their dwarf galaxy subsample and classified as star-forming according to the \nii-BPT had systematically higher \oi/\ha\ line-ratios. %
For instance, \citet{polimera22} showed that the \oi-BPT\footnote{\citet{polimera22} also tested the \sii-BPT diagram, which resulted in being less sensitive to SF dilution compared to the \oi-BPT.} can identify a theoretical dwarf AGN with a spectrum characterized by $\sim90$\% SF contribution and classified as star-forming according to the \nii-BPT (see also \citealt{hogarth20}). 
Indeed, \oi/\ha\ is highly sensitive to hard radiation fields, particularly shock emission in the neutral ISM (e.g., \citealt{osterbrock06,allen08}) and is not very dependent on metallicity. 
Hence, an \oi/\ha\ enhancement could suggest that there is an additional excitation mechanism other than stellar photoionization, missed by the \nii-BPT diagram. %, is required to explain the nature of the broad emission. 
Looking at Table~\ref{tab:classification}, some galaxies are classified as SF-dominated according to the \nii-BPT and as ``AGN" according to the \oi-BPT, because they are located slightly beyond the \citet{kewley06} maximal starburst line. However, here we stress that the \oi$\lambda6300$ line is fainter than \nii\ and \sii, and thus the \oi/\ha\ line ratio has larger error bars, as shown in Fig.~\ref{fig:optbpt}. %\oi$\lambda6300$ lines, however, can be very faint and also difficult to reproduce with models.
% A similar result was found by \citet{hogarth20}, who analyzed the ionized gas kinematics, physical properties, and chemical abundances of a `Green Pea' galaxy at $z\sim 0.17$, the high-$z$ analogue J142947. 

On the other hand, low metallicity stellar populations produce harder radiation fields and thus higher \oiii/\hb\ (and lower low-ionization line ratios) without the need to invoke other mechanisms than SF (e.g., \citealt{feltre16,byler18,xiao18}). 
In this context, \citetalias{xiao18} explored the variation of the {\it maximal starburst line} (i.e., the highest line ratios reproduced with SF models) in BPT diagrams as a function of the metallicity, using pure stellar photoionisation including massive star binaries through the BPASS (Binary Population and
Spectral Synthesis; \citealt{eldridge09}) stellar evolution models\footnote{Binary-star evolution pathways produce harder radiation up to older ages than single stars mainly because of the gas accretion onto compact objects (e.g., HMXBs).}.  
The green shaded regions in Fig.~\ref{fig:optbpt} show the predictions of the \citetalias{xiao18} {\it maximal starburst lines} between $Z\sim10$\%~Z$_\odot$ and $Z\sim50$\%~Z$_\odot$ to underline how difficult the separation between different ionization mechanisms can be for galaxies that lie very close to the edges of the classical BPT diagrams in this range of metallicity (at which we observe also the UV emission lines we discuss in Sec.~\ref{sec:results-diagnostics-uv}). 
% In particular, at decreasing metallicities %($Z\sim0.1-0.5Z_\odot$ or 12+log(O/H)~$\sim 7.64-8.55$) %or at increasing gas density ($n_e\sim10^3$~cm$^{-3}$), 
% the {\it maximal starburst line} can move to lower \nii/\ha, \sii/\ha, \oi/\ha\ and \oiii/\hb\ consistently with observed rare local metal-poor \hii\ regions and metal-poor star-forming galaxies at high redshift (e.g., \citealt{steidel16}). 
{Overall, we stress that taking into account the criteria of \citetalias{xiao18} and the gas-phase metallicity measured for each object of our sample, all the galaxies are classified as SF dominated.}

Another interesting implication from Fig.~\ref{fig:optbpt} is that the broad component of our galaxies is usually characterized by higher \nii/\ha\ and \oi/\ha, slightly higher \sii/\ha, and slightly lower \oiii/\hb\ line ratios. {This difference is shown by the grey, red and blue histograms on each axis of Fig.~\ref{fig:optbpt}, and is more enhanced when comparing results of the single-Gaussian fit with the broad component of the two-Gaussian fit (up to $\sim1$~dex in log(\nii/\ha) and $\sim0.5$~dex in log(\oi/\ha)).}
In particular, strong low-ionization lines could be a signature of shocks \citep{osterbrock06}.
% The median value of velocity dispersion of the broad component is $\sigma\sim193$~km/s, while is $\sigma\sim61$~km/s for the narrow component.
Also, the broad component has (by definition) a much larger velocity dispersion than the narrow one, as shown in Fig.~\ref{fig:fitcomps}.
This $\sigma$ enhancement at increasing \nii/\ha, \sii/\ha, and \oi/\ha\ line ratios suggests that the kinematics and ionization state could be coupled, implying the same physical origin, which could be different from the one producing the narrow components.
This correlation has been already found in normal star-forming galaxies and AGN (e.g., \citealt{ho14,mingozzi19,hogarth20}), and it was attributed to a shocked gas component, as stellar/AGN photoionization is not expected to cause such a trend (see e.g. \citealt{dopita95,rich10,mcelroy15,kewley19}. 
We note that the $\sigma$ enhancement is more evident in \nii/\ha\ and so could also be related to a nitrogen enhancement of the more perturbed gas (e.g., \citealt{james09}). 
We will explore this further in Arellano-Córdova et al. 2023 in prep.

Finally, in Appendix~\ref{app:other-diag} we also comment about an alternative optical BPT diagnostic diagram that exploits only line ratios in the blue part of the optical spectrum (i.e., \oii$~\lambda\lambda$3727/\hb\ versus \oiii$~\lambda5007$/\hb, Fig.~\ref{fig:lamareille}; \citealt{lamareille04,lamareille10})\footnote{This diagram is useful to check two galaxies of our sample, the famous blue compact dwarf I\,Zw~18 (J0934+5514) and J1253-0312, which are not shown in Fig.~\ref{fig:optbpt} because we do not have \ha\ coverage (see \citetalias{mingozzi22} Sec.~3). These two objects are classified as star-forming in Fig.~\ref{fig:lamareille}.}. 

To summarize, since \oiii/\hb\ is in the star formation locus in the BPT diagrams (Fig~\ref{fig:optbpt}) and is not enhanced in the broad component, we think that is unlikely that AGN are present in our sample. 
Here we acknowledge that recently \citet{hatano23} claimed that SBS~0335-052~E hosts an AGN on the basis of a recent NIR variability and the broad \ha\ component. 
We also reveal a broad \ha\ component ($\sigma\sim1000$km/s) not visible in the other emission lines (this could be due to low S/N and low metallicity), but the narrow component line ratios analyzed in this work are all classified as SF-dominated.
Finally, the \oi-excess that we observe in some galaxies of the sample, as well as the increase of \nii/\ha, \sii/\ha, and \oi/\ha\ line ratios with velocity dispersion, indicate the possible presence of additional mechanisms such as shocks, which we further discuss in the following section.

\subsection{[OI]$~\lambda$6300 as a shock indicator}\label{sec:oi-shocks}
\begin{figure*}
\begin{center}
    \includegraphics[width=0.7\textwidth]{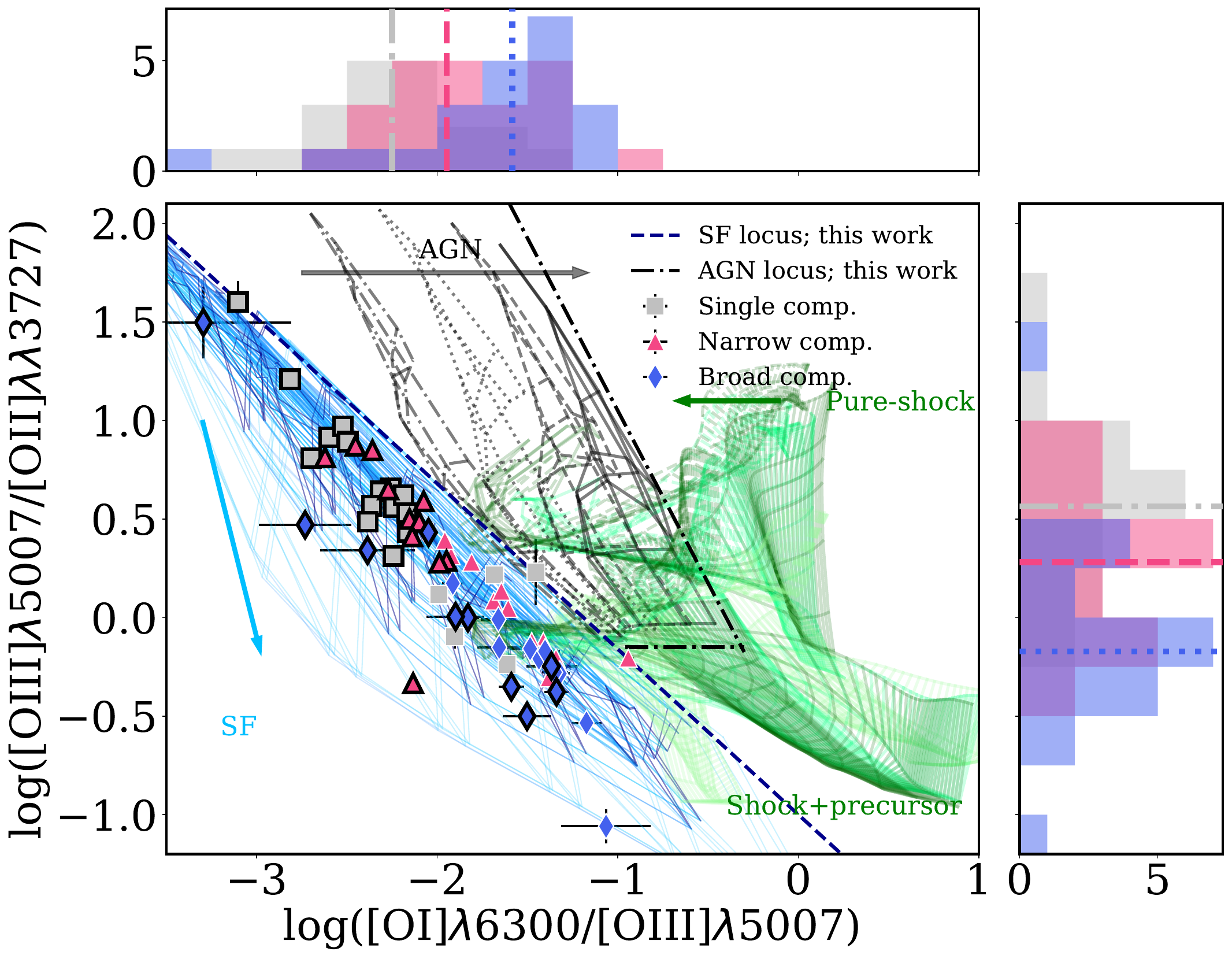}
\end{center}
\caption{Location of the CLASSY galaxies in the log(\oiii/\oii)\ versus log(\oi/\oiii)\ diagnostic diagram from \citet{kewley19} (see also \citealt{stasinska15}). The {single,} narrow and broad components are shown as {grey squares, red triangles,} and blue diamonds and the histograms on the axes show their distributions and median value as in Fig.~\ref{fig:optbpt}. The blue grids show the \citetalias{gutkin16} single-burst SF models, with ages of 3~Myr (darker blue) and 5~Myr (lighter blue). The \citetalias{gutkin16} SF grids indicatively follow the blue arrow at increasing ages. The \citetalias{gutkin16} constant SF models are not shown for clarity, because they completely overlap the single-burst grids. The black grids show the \citetalias{feltre16} AGN models, with the parameter $\alpha$ increasing from -2 (dashed-dotted grid) to -1.2 (solid grid) - and thus harder AGN ionizing radiation - in the direction of the black arrow. \citetalias{alarie19} pure-shock (dashed) and shock+precursor (solid) model grids at varying velocities and magnetic fields are shown at different metallicities, with increasing values from lighter to darker green (following the green arrow). Shock models are more sensitive to the metallicity and have log(\oiii/\oii)~$\gtrsim 0$, but lower log(\oi/\oiii) ($\lesssim 0$) than shock+precursor ones, that can reach log(\oi/\oiii)~$\lesssim1$. The dashed blue line qualitatively separates the SF from AGN/shock models, while the dash-dotted black line indicates the AGN locus.
This plot represents a good tool to separate SF, AGN and shock models for our sample, which includes sub-solar to solar metallicity targets, underlining the importance of \oi$~\lambda$6300 as a good shock diagnostic at optical wavelengths. Only two CLASSY galaxies are at the edge of the star-forming locus. The galaxies with at least one UV emission line (i.e., \ciii$\lambda\lambda$1907,9; 12+log(O/H)~$\lesssim8.3$) are highlighted by a black thick edge.}
\label{fig:opt-shocks}
\end{figure*}
Fig.~\ref{fig:opt-shocks} shows the \oiii$\lambda5007$/\oii$\lambda\lambda3727$ versus \oi$\lambda6300$/\oiii$\lambda5007$ diagram ([OI]-[OII]-[OIII] diagram) for the CLASSY galaxies ({single,} narrow and broad components shown as in Fig.~\ref{fig:optbpt}), presented as a diagnostic plot for the first time in \citet{heckman80} to separate Seyferts from LINERs (see also \citealt{stasinska15}). This diagram is not very dependent on gas-phase metallicity \citep{stasinska15}, avoiding the drawbacks discussed for the classical \nii-BPT diagram. 
Clearly, Fig.~\ref{fig:opt-shocks} represents a good tool to discriminate between SF, AGN, and shocks for our sample, as also shown by the good separation of the SF, AGN, and shock model grids, with just a slight overlap between the grids at the highest \oi/\oiii\ line ratios\footnote{\citet{kewley19} present different shock diagnostics (see their Fig.~11), including one comparing \oiii/\oii\ versus \oi/\ha, that we tested but excluded because the AGN and SF grids of models could not be separated as well as in Fig.~\ref{fig:opt-shocks}.}. 
We highlight that increasing the preshock density $n_H$ from $n_H\sim1$~cm$^{-3}$ to $n_H\sim1000$~cm$^{-3}$ would lead to higher log(\oiii/\oii), that can increase up to $\sim 0.5-1.5$ for shock+precursor and pure-shock grids, respectively. Still, shock models are generally characterized by higher log(\oi/\oiii) than AGN models, reaching values up to $\sim1$.  
The dashed blue line 
\begin{equation}\label{eq:1}
    y=-0.84\,x -1.0
\end{equation}
qualitatively separates the SF from AGN/shock models, while the dash-dotted black line indicates the AGN locus:
\begin{equation}\label{eq:2}
\begin{aligned}
    & y=-0.15 \,\,\,\,\,\,  and \,\,\,\,\,\, -0.96< x <-0.3 \\
    & y=-1.75\,x-0.70 \,\,\,\,\,\, and \,\,\,\,\,\, -3.5< x <-0.3
\end{aligned}
\end{equation}
with $y=$~log(\oiii~$\lambda5007$/\oii~$\lambda\lambda3727$) and $x=$~log(\oi~$\lambda6300$/\oiii~$\lambda$5007).

Overall, Fig.~\ref{fig:opt-shocks} shows that the vast majority of the CLASSY galaxies can be reproduced by SF models, further confirming the classification obtained with the classical BPT diagrams (Fig.~\ref{fig:optbpt}). 
In particular, the optical line broad velocity components show a significant enhancement in terms of the \oi/\oiii\ line ratio (with the broad showing a larger median value by $\sim 0.5$~dex {and $\sim 1$~dex than the narrow and single components, respectively}; except for J0926+4427, J1148+2546 and J1253-0312) and a lower \oiii/\oii\ (in all galaxies except for J1148+2546). 
Shocks can naturally explain high \oi/\oiii\ line ratio with high \oiii/\oii, while photoionization models that produce high \oi/\oiii\ show, in turn, low \oiii/\oii\ (e.g., \citealt{stasinska15,plat19}).
This suggests that also the more turbulent ionized gas has a stellar source of ionization. 
The narrow components of only two CLASSY galaxies are at the edge of the SF locus defined by our Eq.~\ref{eq:1} in this diagram: J0944+3442 and J0808+3948, which could be both consistent with AGN and shock models. 
J0944+3442 is classified as star-forming according to the optical BPT diagrams, while only the narrow component of J0808+3948 is classified as composite according to the \nii-BPT diagram. J0808+3948 has also an extremely high nitrogen enhancement and high N/O (Arellano-Córdova et al. 2023 in prep.; see also \citealt{stephenson23}). 
We stress that these galaxies have no detections of UV lines, so are not shown in the UV diagnostic diagrams discussed in Sec.~\ref{sec:results-diagnostics-uv} (J0808+3948 is the only trans-solar metallicity galaxy of the sample (12+log(O/H)~$\sim8.77$); J0944+3442 has instead low-metallicity (12+log(O/H)~$7.83$).

Even though the \oiii/\oii\ versus \oi/\oiii\ diagram can be a very good tool to understand the role of shocks and separate SF from harder mechanisms, it can also have drawbacks.
\oi~$\lambda$6300 is produced in the warm transition region between the fully ionized gas and neutral gas \citep{osterbrock06,draine11}. As such, it traces the external parts of \hii\ regions, close to the ionization front, where shocks and non-equilibrium heating are important \citep{dopita03,dopita13}, which makes it difficult to predict \oi~$\lambda$6300 in photoionization models \citep{dopita13}. Furthermore, \oi~$\lambda$6300 is generally a faint emission line (this makes it even more tricky to disentangle a broad second component). 
% These implications affect the reliability of diagnostic plots that take \oi~$\lambda$6300 into account. 
Nevertheless, an observed \oi~$\lambda$6300 enhancement could still represent a sign of ionizing mechanisms other than SF. {Overall, the values of the discussed oxygen line ratios are mostly consistent with SF models, further confirming the lack of strong ionizing shocks and/or AGN activity in the CLASSY galaxies.}

\subsection{The Shirazi-Brinchmann diagram}\label{sec:shirazi} 
\begin{figure}
\begin{center}
    \includegraphics[width=0.5\textwidth]{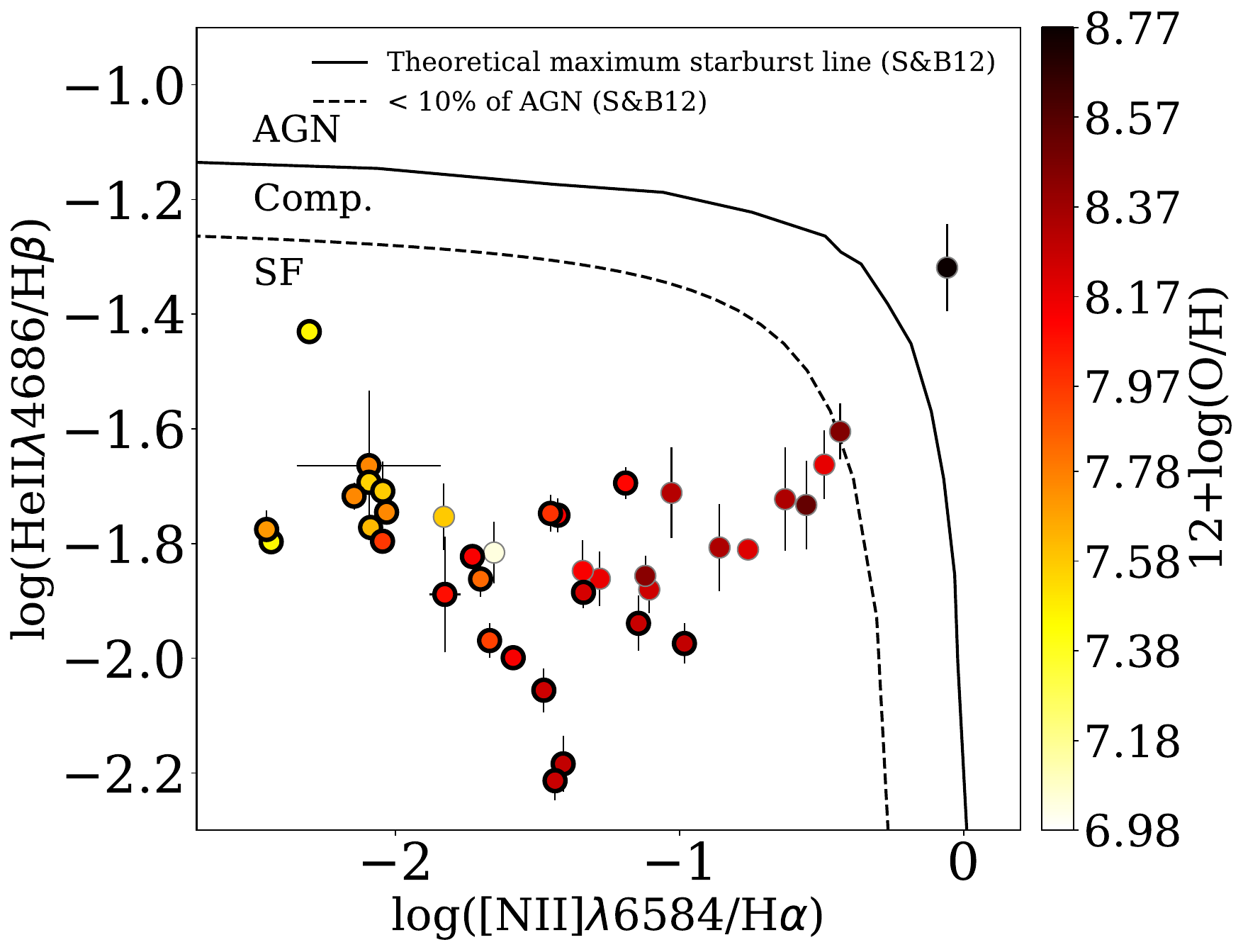}
\end{center}
\caption{{\heii$~\lambda$4686/\hb\ versus \nii/\ha}\ diagnostic diagram for the CLASSY galaxies, shown as dots, color-coded as a function of the gas-phase metallicity. Only the measurements with $S/N>3$ for all the lines involved are reported. The dotted and solid black lines represent the locus where $\sim 10$\% of \heii$~\lambda$4686 comes from an AGN and the theoretical maximum starburst line, respectively, as defined by \citet{shirazi12}. Only two CLASSY galaxies are outside the star-forming locus, but we note that their \heii$~\lambda$4686\ lines are particularly faint ($S/N\lesssim5$), hence this classification is not reliable by itself. The galaxies with at least one UV emission line (i.e., \ciii$\lambda\lambda$1907,9; 12+log(O/H)~$\lesssim8.3$) are highlighted by a black thick edge.}
\label{fig:shirazi}
\end{figure}

To further confirm the SF classification of CLASSY galaxies we also consider high-ionization lines such as \heii$~\lambda4686$. Indeed, as introduced in Sec.~\ref{sec:intro}, \citet{shirazi12} proposed to use {\heii$~\lambda$4686/\hb\ versus \nii/\ha}\ to better constrain the ionization source.
%Exploiting the large statistics of the SDSS DR7 \citep{abazajian09}, \citet{shirazi12} proposed a new diagnostic diagram to discriminate the ionization source: \nii/\ha\ versus \heii$~\lambda$4686/\hb. 
% \citet{garnett91} and \citet{stasinska03} first documented the systematic presence of \heii\ lines in \hii\ galaxies. Moreover, \citet{thuan05} revealed nebular \heii\ emission also in metal-poor blue compact dwarf (BCD) galaxies that were lacking WR features (see also \citealt{guseva00,brinchmann08,shirazi12}), proposing that fast radiative shocks could be the ionizing source for this emission (see also \citealt{dopita96}). 
% This lack of correlation between nebular \heii\ emission and WRs has been found also in spatially spectroscopic studies (e.g., \citealt{kehrig08,kehrig11,neugent11}), further confirming that at low metallicity \heii\ emission can have different origins than WR stars.
The two major advantages of this diagram (HeII diagram) are that (i) it is more sensitive to the hardness of the ionizing source than \oiii/\hb, thus allowing a cleaner separation between star-forming and composite galaxies; and (ii) it is not particularly sensitive to the metallicity \citep{shirazi12}. %However, one major drawback is the weak nature of the \heii$~\lambda$4686 line. 
%This line  can represent a crucial diagnostic, since it indicates a very hard ionizing radiation field (above 54.4~eV). 
% Indeed, the presence of \heii$~\lambda$4686 (as well as \heii$~\lambda1640$) in a galaxy spectrum indicates the existence of sources of hard ionizing radiation, since the ionization energy for He$^+$ is 54.4~eV. 
% Such a high radiation field can be produced by young stellar populations, including Wolf–Rayet (WR) stars (e.g., \citealt{schaerer98}), but also other mechanisms such as AGN \citep{shirazi12}, X-ray binaries \citep{garnett91} or shocks \citep{thuan05,dopita96,stasinska15} can play a role. \citet{garnett91} and \citet{stasinska03} first documented the systematic presence of \heii\ lines in \hii\ galaxies. 
% Moreover, \citet{thuan05} revealed nebular \heii\ emission also in metal-poor blue compact dwarf (BCD) galaxies that were lacking WR features (see also \citealt{guseva00,brinchmann08,shirazi12}), proposing that fast radiative shocks could be the ionizing source for this emission (see also \citealt{dopita96}). 
% This lack of correlation between nebular \heii\ emission and WRs has been found also in spatially spectroscopic studies (e.g., \citealt{kehrig08,kehrig11,neugent11}), further confirming that at low metallicity \heii\ emission can have different origins. 
% Exploiting the large statistics of the SDSS DR7 \citep{abazajian09}, \citet{shirazi12} proposed a new diagnostic diagram to discriminate the ionization source: \nii/\ha\ versus \heii$~\lambda$4686/\hb.
The HeII diagram for the CLASSY sample is shown in Fig.~\ref{fig:shirazi} and is color-coded as a function of 12+log(O/H) (the broad component fit of \heii$~\lambda$4686 line is not detected because of the low S/N). 
Similarly to the standard BPT diagrams described in Sec.~\ref{sec:results-bpt}, SFGs are expected to lie below the maximum starburst (black solid) line, while AGN are located above.
The \citetalias{gutkin16} models are not shown for clarity reasons but can completely cover the star formation locus.
% The two major advantages of the \heii/\hb\ line-ratio are that (i) it is more sensitive to the hardness of the ionizing source than \oiii/\hb, thus allowing a cleaner separation between star-forming and composite galaxies; and (ii) it is not particularly sensitive to the metallicity \citep{shirazi12}. However, one major drawback is the weak nature of the \heii$~\lambda$4686 line. 
Almost all the CLASSY galaxies shown in the HeII diagram are classified as star-forming. 
The only targets with a different classification are J0808+3948 (classified as AGN) and J1112+5503 (at the edge between the SF and composite locii). 
However, we emphasize that their \heii\ lines are particularly faint (S/N$\sim3-5$), hence this classification is not reliable by itself. 
Given that J0808+3948 is not located in the SF locus also in Fig.~\ref{fig:opt-shocks} and in the \nii-BPT (Fig.~\ref{fig:optbpt}), we conclude that it is possible that this galaxy has a further ionizing mechanism than pure SF. 
% Targets with log(\nii/\ha)~$>-1$ and log(\heii/\hb)~$\gtrsim-1.8$ have also a slightly larger velocity dispersion ($\sigma \gtrsim 80$~km/s), with respect to objects characterized by lower line ratios ($\sigma \gtrsim 60$~km/s). This could be a hint of the presence of shocks, even though we stress that the difference is not very significant. 
In Sec.~\ref{sec:results-bpt} and \ref{sec:oi-shocks} we could not exclude the presence of shocks in our targets, but, if present, shocks should easily enhance the \heii$~\lambda$4686 flux (e.g., \citealt{shirazi12,stasinska15}). However, here we stress that we do not see a particular enhancement of \heii\ at increasing \oi/\oiii, which we consider a good shock diagnostic (Fig.~\ref{fig:opt-shocks}) or in galaxies where we identify a second broader component characterized by higher \nii/\ha, \sii/\ha, and/or \oi/\ha, as shown and discussed in Fig.~\ref{fig:optbpt}. This is a further confirmation of the SF classification of our systems.

Finally, we highlight that almost all the \ciii~$\lambda\lambda1907,9$-emitters are shown in Fig.~\ref{fig:shirazi}. 
Those missing are: Izw~18 and J1253-0312, because without \ha\ measurement in our sample, but with log(\heii$~\lambda4686$/\hb)~$\sim -1.35, - 1.86$ and 12+log(O/H)~$\sim6.98, 8.06$, respectively, and thus still lying in the SF locus; Mrk~996 (J0127-0619) with an unreliable \heii~$\lambda4686$ flux measurement, due to the contamination from the WR red clump, but still classified as star-forming (see \citealt{james09} for further details); J1444+4237, which does not show \heii~$\lambda$4686 emission (upper limit of log(\heii$~\lambda4686$/\hb)~$\lesssim -1.8$), even though it has bright UV emission lines. 
Overall, all the galaxies that we will show in Sec.~\ref{sec:results-diagnostics-uv} are classified as SF-dominated also according to the \citet{shirazi12} criterion.

% Four of them (i.e., J0940+2935, J0944+3442, J1144+4012 and J1444+4237) have $S/N$(\heii$~\lambda4686$)$<3$. Interestingly, J0944+3442 and J0808+3948 are the only two galaxies out from the star formation locus in Fig.~\ref{fig:opt-shocks}.
% The \heii$~\lambda4686$ measured flux of Mrk~996 is not reliable due to the contamination from the Wolf-Rayet red clump (see \citealt{james09} for further details). 
% Finally, as explained in Sec.~\ref{sec:results-bpt}, we cannot calculate \nii/\ha\ for I\,Zw\,18 (J0934+5514) and J1253-0312.
% The famous blue compact dwarf I\,Zw\,18 has a bright 
% \heii$~\lambda4686$ with log(\heii$~\lambda4686$/\hb)\~$\sim-1.35$. Given its low-metallicity (12+log(O/H)~$\sim6.98$; 
% \citetalias{berg22},\citetalias{mingozzi22}), this galaxy would 
% probably be located at the lowest \nii/\ha\ line-ratios, close to the 
% low-metallicity and blue compact dwarf SBS~0335-052~E 
% (log(\heii$~\lambda4686$/\hb)~$\sim-1.43$). J1253-0312 has  
% log(\heii$~\lambda4686$/\hb)~$\sim-1.86$ and with its metallicity of 
% 12+log(O/H)~$\sim8.06$ would probably be located at low \nii/\ha\ as well.

\subsection{The importance of optical coronal lines}\label{sec:discussion-3}
As a concluding note on optical diagnostics, we discuss the so-called coronal lines as proof of the presence of AGN activity. 
These are high-ionization~($\geq70$~eV) forbidden transitions excited by collisions similarly to \oiii$~\lambda$5007 and are usually considered undeniable evidence of AGN ionization (e.g., \citealt{korista89}).
Their presence in low-massive metal-poor galaxies optical (and IR) spectra is a current {intensively discussed issue} in the hunt to reveal intermediate-mass black holes (e.g., \citealt{cann20,cann21,molina21}). 

After a careful analysis of our spectra, we exclude the presence of these lines (i.e., \fevi~$\lambda$5146, \fexiv~$\lambda$5303, \fevii~$\lambda$5721, \fevii~$\lambda$6087, \fex~$\lambda$6374) in the optical spectra we analyzed\footnote{J0944-0038 MUSE data recently presented in \citet{delValle-Espinosa22} reveal \fevi~$\lambda$5146 line in the brightest star-formation knot, only partly covered by the SDSS and LBT data that we took into account.}. 
 In the upper panel of Fig.~\ref{fig:fitcomps}, next to \oi$~\lambda$6364 there is a ``bump" that could be interpreted as a tentative \fex~$\lambda$6374. 
However, this feature is more consistent with the \silii~$~\lambda$6371 (actually a multiplet, \silii~$\lambda\lambda$6348,71), previously identified in some CLASSY galaxies in high-resolution optical spectra (e.g., VLT/GIRAFFE SBS~0335-052~E spectrum, \citealt{izotov06b}). 
These permitted \silii\ lines may have a fluorescent origin and be produced by absorption of the intense UV radiation \citep{grandi76,izotov01}.
The detection of very high-ionization emission lines could be an alternative channel to identify hidden AGN activity in dwarf galaxies, especially in the JWST era with rest-frame IR coverage (e.g., \citealt{cann18}). However, here we want to highlight that it can be easy to misidentify \fex~$~\lambda$6374 and \silii~$~\lambda$6371 (see also \citealt{herenz23}).

\vspace{0.5cm}
{In conclusion, we stress that none of our galaxies are uniformly classified as non-SF dominated by \textit{all} the optical diagnostics considered here. % and that all the galaxies characterized by UV emission lines discussed in the following section are SF dominated. 
According to the classical BPT diagrams with the standard separators, few galaxies are  classified as Comp/AGN as shown in Tab.~\ref{tab:classification}, but never consistently, while they are all classified as SF dominated using the separators from \citetalias{xiao18}, thus taking into account the gas-phase metallicity of each object (Sec.~\ref{sec:results-bpt} and Fig.~\ref{fig:optbpt}). 
Only two objects (J0944+3442, classified as SF by all the other diagrams, and again J0808+3948; both without UV lines) fall in the non-SF locus according to the oxygen line ratios used as shock diagnostics (Sec.~\ref{sec:oi-shocks} and Fig.~\ref{fig:opt-shocks}), ruling out shock dominated emission.
Finally, the absence or very weak He~II~$\lambda$4686 lines is another strong indication of the lack of a harder mechanism than SF, with all the galaxies but J0808+3948 (with trans-solar metallicity and no UV lines) classified as SF dominated according to the Shirazi\&Brinchmann diagram (Sec.~\ref{sec:shirazi} and Fig.~\ref{fig:shirazi}).}

\section{UV diagnostic diagrams}\label{sec:results-diagnostics-uv} 
In the previous section, we showed how all the CLASSY galaxies characterized by UV emission lines are dominated by star formation, using different optical criteria, highlighting all their caveats. 
Further confirmation of their SF-dominated nature lies in the lack of \nv~$\lambda\lambda$1239,1243 doublet in emission. 
Indeed in the UV, an unequivocal signature of AGN activity is generally represented by this very high-ionization emission doublet ($E>77.5$~eV; \citealt{feltre16}). We do not observe the ISM \nv\ doublet in emission in any of the CLASSY galaxies, but only in P-Cygni shape - strongly indicating the presence of massive stars (e.g., \citealt{chisholm19}). 
Having said that, here we want to explore if UV diagnostic diagram classification can be consistent with the optical and thus identify the best-observed UV diagnostics to discriminate SF from AGN and shock ionization. Of course, without AGN within the sample, we are unable to assess the ability of UV diagnostics to correctly classify AGN, but we are able to test whether SF galaxies are incorrectly classified as AGN. 
We do this by comparing our sample of CLASSY galaxies with \citetalias{gutkin16} constant and {\citetalias{xiao18} BPASS bursty SF} models, \citetalias{alarie19} shock models and \citetalias{feltre16} AGN models presented in Sec.~\ref{sec:methods} and Tab.~\ref{tab:models}.
{Furthermore, in App.~\ref{app:uv-diag-models}, we show and discuss \citetalias{gutkin16} single-bursts and \citetalias{byler17} models, also introduced in Sec.~\ref{sec:methods} and Tab.~\ref{tab:models}.}
As we introduced in Sec.~\ref{sec:intro}, UV diagnostics can be crucial to investigate ISM properties with JWST observations in objects above $z\sim6$, and thus it is important to understand in detail their strengths and drawbacks. 

Specifically, in the following subsections, we show the UV diagnostic plots that we found more promising among those proposed in the literature on the basis of models and simulations \citep{jaskot16,feltre16,nakajima18,hirschmann19,hirschmann22}:
\begin{itemize}
    \item C3He2-C4He2 (Fig.~\ref{fig:diagnostic-jr16}): \ciii~$~\lambda\lambda$1907,9/\heii~$~\lambda$1640\ vs \civ~$~\lambda\lambda$1549,51/\heii~$~\lambda$1640;
    \item C3He2-O3He2 (Fig.~\ref{fig:diagnostic-f16}): 
    \ciii~$~\lambda\lambda$1907,9/\heii~$~\lambda$1640 versus \oiiiuv~$~\lambda$1666/\heii~$~\lambda$1640;
    \item C4C3He2-C4C3 (Fig.~\ref{fig:diagnostic-n18}): \civ~$~\lambda\lambda$1548,51/\ciii~$~\lambda\lambda$1907,9 versus (\civ+\ciii)/\heii~$~\lambda$1640;
    \item EWC4 and EWC3 (Fig.~\ref{fig:diagnostic-n18-1}): EW(\civ) and EW(\ciii) versus \ciii/\heii$~\lambda$1640.
\end{itemize}
We also report in App.~\ref{app:uv-diag} the diagnostics diagrams \ciii/\heii~$~\lambda$1640 versus \oiiiuv~$~\lambda$1666/\ciii~$~\lambda\lambda$1907,9 (C3He2-O3C3) and \ciii/\heii~$~\lambda$1640 versus \civ~$~\lambda\lambda$1548,51/\ciii~$~\lambda\lambda$1907,9 (C3He2-C4C3). 
{Here we highlight that \civ\ and \heii\ lines are broadly used in these diagrams, but they can have both a nebular and stellar contribution. We discuss this caveat in detail in Sec.~\ref{sec:civheii-discussion}.}

As a general note about the \citetalias{alarie19} shock grids, in all the following plots they are often completely overlapped in terms of velocity and magnetic field and gas-phase metallicity, but we decided to show the entire range of the $v$ and $B$ parameter space at three representative values of metallicity ($Z=0.05,0.5,1$~Z$_\odot$) for completeness. These grids also cover very narrow regions in C4He2 space that correspond to the lowest velocities in the parameter space ($v<200$~km/s, seen as spike-shaped features in grid coverage).

\subsection{C3He2-C4He2}\label{sec:C3He2-C4He2} %CIII]~$~\lambda\lambda$1907,9/He~II$~\lambda$1640 versus CIV~$~\lambda\lambda$1548,51/He~II$~\lambda$1640}
\begin{figure*}
\begin{center} 
    \includegraphics[width=0.9\textwidth]{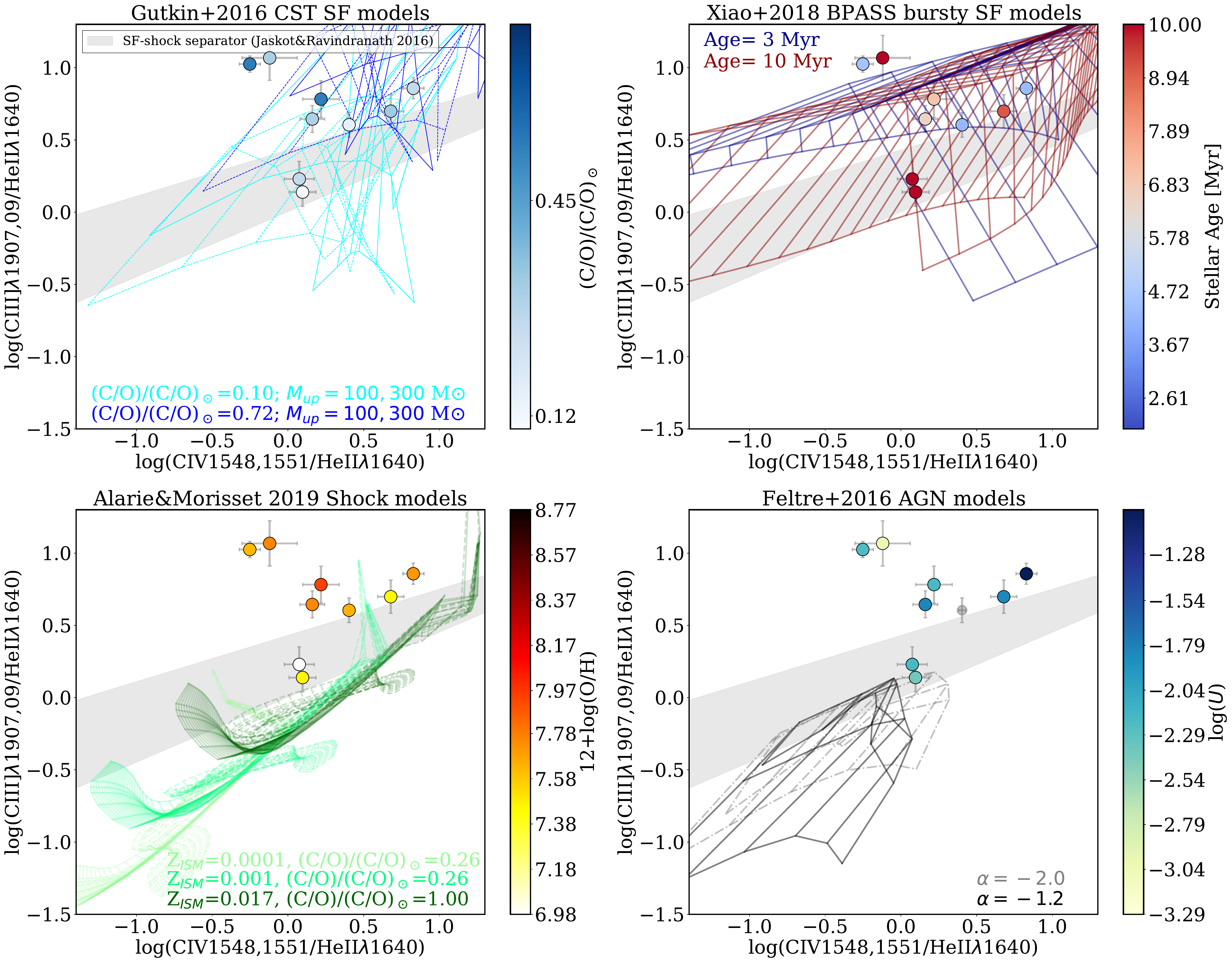}   
\end{center}
\caption{C3He2-C4He2 diagram: \ciii~$~\lambda\lambda$1907,9/\heii~$~\lambda$1640 vs \civ~$~\lambda\lambda$1549,51/\heii~$~\lambda$1640 diagnostic plot (\citetalias{feltre16}, \citealt{jaskot16}) to separate SF and AGN activities. {Here we show only the CLASSY galaxies with \civ\ detected in pure emission.} 
The gray shaded region represents the separator proposed by \citet{jaskot16} between pure-SF models (located above) and models with $\geq 50$\% shock contribution (located below).
{Upper left panel:} \citetalias{gutkin16} models with continuous SF are shown in light ((C/O)/(C/O)$_\odot$=0.1) and dark blue ((C/O)/(C/O)$_\odot$=0.72), with the scatter plot color-coded as a function of the (C/O)/(C/O)$_\odot$ measured for the sample. The solid and dashed grids refer to different IMF cut ($M_{up}=100$~M$_\odot$ and $M_{up}=300$~M$_\odot$, respectively). 
{Upper right panel:} {\citetalias{xiao18} models with burst of SF at 3~Myr and 10~Myr are shown in blue and red, respectively,} with the scatter plot color-coded as a function of the stellar age derived from the UV stellar component fitting. 
{Lower left panel:}  \citetalias{alarie19} shock models with increasing metallicities from light to dark green, as shown in the legend, and (C/O)/(C/O)$_\odot=0.26$ are shown on a scatter plot color-coded as a function of the gas-phase metallicity 12+log(O/H) (darker green means higher metallicity, up to the trans-solar value of $Z_{ISM} = 0.017$). {Lower right panel:}  \citetalias{feltre16} AGN models with $\alpha = -2$ (solid - softer ionization) and $\alpha = -1.2$ (dash-dotted - harder ionization) overplotted on a scatter plot color-coded as a function of the ionization parameter log($U$). 
The two galaxies located in the gray-shaded region are I\,Zw\,18 and SBS~0335-052~E, which have the lowest 12+log(O/H) and C/O abundance in the sample. Overall, the C3He2-C4He2 diagram can be used to discriminate SF from AGN and shocks, with some caveats. Shocks and AGN are harder to discriminate, but still the former are expected to have higher carbon-to-helium line-ratios, especially at increasing metallicity.}
\label{fig:diagnostic-jr16}
\end{figure*}

Fig.~\ref{fig:diagnostic-jr16} shows the 
\ciii~$~\lambda\lambda$1907,9/\heii~$~\lambda$1640 versus the 
\civ~$~\lambda\lambda$1548,51/\heii~$~\lambda$1640 diagnostic diagram (C3He2-C4He2) with previously described SF, shock and AGN grid models overlaid (see caption for details).
This diagnostic diagram was first proposed as a good AGN and star-forming galaxies separator in \citetalias{feltre16}. \citet{jaskot16} further supported this as a good diagnostic plot to discriminate between shocks and pure SF, where their pure photoionization models lie above the shaded gray region in Fig.~\ref{fig:diagnostic-jr16}, while most of their models with a shock contribution of $\geq 10$\% and $\geq 50$\% to the shown line ratios fall within and below the gray shaded region, respectively. 
Indeed, the \heii$~\lambda$1640 line can be enhanced by hard ionizing radiation with high-energy photons ($>54.4$~eV), which can be produced by either AGN or shocks, causing a decrement of \ciii~$~\lambda\lambda$1907,9/\heii~$~\lambda$1640 and \civ~$~\lambda\lambda$1548,51/\heii~$~\lambda$1640. 
This also implies that the C3He2-C4He2 diagram cannot clearly distinguish between AGN and shocks. 
Indeed, looking at the bottom panels of Fig.~\ref{fig:diagnostic-jr16}, shock and AGN grids cover a similar \ciii/\heii\ vs \civ/\heii\ parameter space, with the most metal-poor shock grids reaching \ciii/\heii\ and \civ/\heii\ down to $\sim-1.75$ and $\sim-2.5$, respectively, as the $\alpha=-1.2$ AGN models (hardest AGN radiation) at the lowest values of metallicity and ionization parameter. 
% This could be because shocks are expected to enhance carbon emission \ciii$\lambda\lambda$1907,9 (e.g., \citealt{jaskot16}, Fig.~12). 
% However, at decreasing metallicity, the shock models also move to lower line ratios, with the lowest metallicity grid down to log(\ciii/\heii)~$\sim-1.5$ and log(\civ/\heii)~$\sim-1.7$, respectively. 
Indeed, the strength of \ciii\ and \civ\ is known to drop at decreasing metallicity (12+log(O/H)~$\lesssim7.5$; e.g., \citealt{jaskot16}; see also \citetalias{mingozzi22}) and decreasing C/O.
We also notice that the \citetalias{alarie19} shock models shown in Fig.~\ref{fig:diagnostic-jr16} can reach higher values than \citetalias{feltre16} AGN models in terms of \ciii/\heii\ line ratios, going beyond the gray shaded SF-shock separator region at low values of velocities ($v<200$~km/s, spike-shaped feature in the shown grids; see also \citealt{jaskot16} Sec.~4.4).

The scatter plots in Fig.~\ref{fig:diagnostic-jr16}, color-coded as reported in the color bar labels, show the 9 galaxies of our sample with \civ~$~\lambda\lambda$1548,51 in pure emission (the other galaxies show pure absorption or P-Cygni profiles).
%: UM133 (J0144+0453), SBS~0335-052~E (J0337-0502), I\,Zw\,18 (J0934+5514), J0944-0038, J1044+0353, J1148+2546, J1323-0132, J1418+2102 and J1444+4237. 
In particular, SBS~0335-052~E (J0337-0502) and I\,Zw\,18 (J0934+5514) are located in the gray shaded region that separates the pure-SF and shock locii according to \citet{jaskot16}. Moreover, the two objects partially overlap with the \citetalias{feltre16} AGN models. However, from the optical diagnostics, we know that these galaxies are SF-dominated (Sec.~\ref{sec:results-diagnostics-opt}).
Interestingly, these objects are the two lowest metallicity galaxies of the sample showing UV emission lines (12+log(O/H)~$\sim 7.46$ and $\sim 6.98$, respectively) with also the lowest C/O abundances ((C/O)/(C/O)$_\odot$~$\sim 0.26$ and $\sim 0.12$, respectively). 
This means that the \citet{jaskot16} criterion can fail in classifying extremely metal-poor star-forming objects possibly characterized by very hard ionization and low C/O, which could mimic AGN or shock ionization (see also \citealt{maseda17}). 
In agreement with this, \citetalias{gutkin16} SF models (both CST and SSP) can fall beyond the \citet{jaskot16} pure-SF line when considering low C/O (cyan grids have (C/O)/(C/O)$_\odot$=0.1), which would reduce \ciii\ and \civ\ emission. 
Another factor that pushes the \citetalias{gutkin16} SF models below the \citet{jaskot16} pure-SF locus is a higher IMF upper-cut ($M_{up}=300$~M$_\odot$ instead of $M_{up}=100$~M$_\odot$, dashed and solid lines, respectively), which would imply more massive stars, hence a harder ionization field (i.e., higher \heii\ emission). 
Concerning \citetalias{gutkin16} single-burst SF grids {(that we do not show for clarity reasons but see Fig.~\ref{fig:modelscomp})}, only ages younger than $\sim4$~Myr can reproduce the line ratios observed for the CLASSY galaxies. 
Since the \citetalias{gutkin16} models do not take into account either stellar rotation or binary interactions (which would amplify the galaxies ionizing fluxes for several Myr after a SF burst), the ionizing photons drop at older ages than 4~Myr, with subsequent reduction of \heii\ (but also \ciii\ and \civ), inducing a shift of the grids to higher \ciii/\heii\ and \civ/\heii\ than the range shown in Fig.~\ref{fig:diagnostic-jr16} ({see App.~\ref{app:uv-diag-models} and} e.g., \citealt{jaskot16} Sec.~3.1 for a detailed comparison between different model prescriptions). 
{In BPASS models instead the stellar multiplicity impacts the duration of the ionizing photon production (as well as in \citetalias{byler17} models that take into account stellar rotation; see Fig.~\ref{fig:modelscomp}). Indeed, both the BPASS 3~Myr and 10~Myr star-forming grids of \citetalias{xiao18} can fully cover the CLASSY data and the SF locus in Fig.~\ref{fig:diagnostic-jr16}, only slightly overlapping with the \citetalias{alarie19} shock models.}

Overall, to explain the locations of galaxies such as SBS~0335-052~E and I\,Zw\,18, lying in the region separating SF from harder mechanisms, the most likely explanation is their low O/H and C/O abundances (see also \citealt{wofford21} for SBS~0335-052~E). 
Hence, before using the C3He2-C4He2 diagram the user should be mindful of the metallicity and C/O abundance of the targets, in that ionization sources in low-metallicity objects are less easily distinguishable for this diagnostic.
% Finally, we note that \citetalias{gutkin16} SSP SF models cover the entirety of the CLASSY galaxies and the region above and below the \citet{jaskot16} shock locus, without showing particular trends with the stellar age. {BLJ: This suggests that under the assumption of a single-burst stellar population, there may be difficulties in using this line ratio to distinguish between photoionization, shocks and AGN for log(\ciii/\heii)$<0$.}
% Overall, we conclude that the C3He2-C4He2 diagram can be used to discriminate SF from AGN and shocks. However, the user should be mindful of the metallicity and C/O abundance of the targets, in that ionization sources in low-metallicity objects are less easily distinguishable for this diagnostic. Shocks and AGN are harder to discriminate, but still the former are expected to have higher carbon-to-helium line-ratios, especially at increasing metallicity.

\subsection{C3He2-O3He2}\label{sec:O3He2-C3He2}
\begin{figure*}
\begin{center} 
    \includegraphics[width=0.9\textwidth]{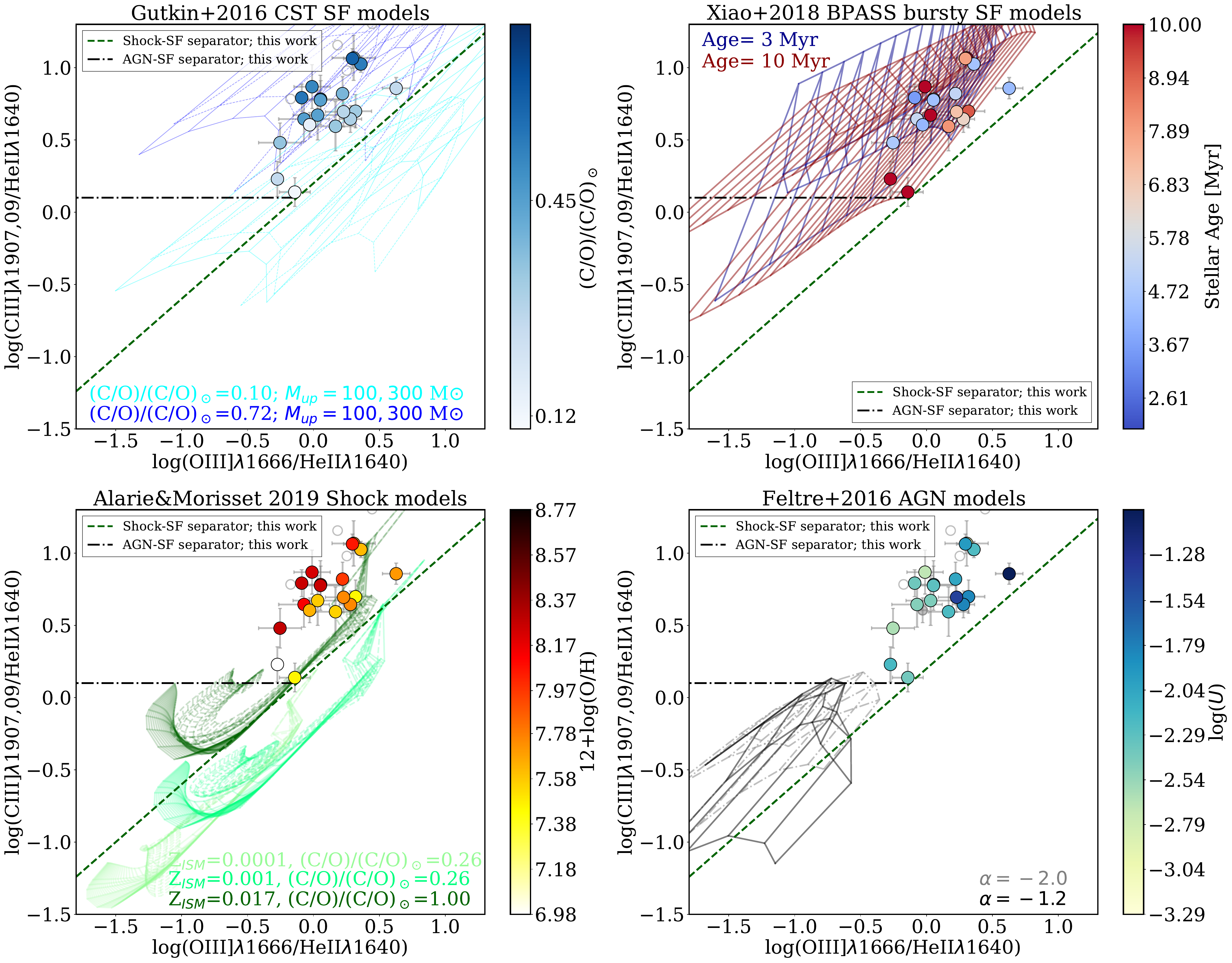}   
\end{center}
\caption{C3He2-O3He2 diagram: \ciii~$~\lambda\lambda$1907,9/\heii~$~\lambda$1640 versus \oiiiuv~$~\lambda$1666/\heii~$~\lambda$1640 diagnostic diagram, proposed by \citetalias{feltre16}. {The filled and open dots show $S/N>3$ and $S/N<3$ fluxes for all the emission lines taken into account, respectively.}
The models superimposed are as explained in Fig.~\ref{fig:diagnostic-jr16}. Overall, this represents the best diagram to discriminate SF, AGN and shocks at sub-solar metallicities (i.e., excluding the dark-green shock grid), due to the minimal overlap between each set of grids and the fact that all CLASSY galaxies are correctly classified as star-forming, as shown by the dashed green and dash-dotted lines to separate shock and AGN models from the SF locus (Eq.~\ref{eq:3} and Eq.~\ref{eq:4}), respectively.}
\label{fig:diagnostic-f16}
\end{figure*}

In Fig.~\ref{fig:diagnostic-f16} we show the \ciii~$~\lambda\lambda$1907,9/\heii~$~\lambda$1640 versus \oiiiuv~$~\lambda$1666/\heii~$~\lambda$1640 diagnostic diagram, proposed by \citetalias{feltre16} and further explored by \citet{hirschmann22}. % (who note a slight overlap between AGN and shocks within their simulated parameter space). 
In contrast to the C3He2-C4He2 diagram shown in Fig.~\ref{fig:diagnostic-jr16}, C3He2-O3He2 in Fig.~\ref{fig:diagnostic-f16} can separate all of the three contributions of SF, AGN, and shocks quite well. 
This is with the exception of the \citetalias{gutkin16} SF models with either low C/O or high $M_{up}$ {(both with constant and bursty SF)} which, again, cover the entirety of the parameter space (see Sec.~\ref{sec:C3He2-C4He2}), as we further comment below.
Most importantly, AGN and shock grids are only slightly overlapped, considering sub-solar metallicities (considering (C/O)/(C/O)$_\odot = 0.26$), with shock models showing higher \oiiiuv/\heii\ line ratios than AGN grids. Instead, the dark-green grid at trans-solar metallicity ($Z_{ISM}=0.017$; (C/O)/(C/O)$_\odot = 1$) shifts to higher \ciii/\heii, overlapping more with the AGN and SF grids.
In particular, \heii\ emission is expected to increase in both AGN and shock models, as already discussed in Sec.~\ref{sec:C3He2-C4He2}, explaining the shift of AGN and shocks to lower \ciii/\heii\ and \oiiiuv/\heii\ than SF grids (but $v<200$~km/s shock models - spike-shaped in the shown grids - can reach \ciii/\heii\ comparable to SF grids as in Fig.~\ref{fig:diagnostic-jr16}). 
However, shocks are expected also to have an increase of \ciii, which generally tightly correlates with temperature-sensitive lines such as \oiiiuv~$\lambda$1666 (see e.g., \citealt{jaskot16} Fig.~12 and 9). This could explain the higher \oiii/\heii\ for shocks than AGN models. %Finally, \ciii\ emission reduces at decreasing C/O, where log(\oiiiuv/\heii) reaches values $\sim 1$, as it can be seen in the upper two panels of Fig.~\ref{fig:diagnostic-f16}. 
As a reference, in Fig.~\ref{fig:diagnostic-f16} we added a dashed green line to separate shock models from the SF locus
\begin{equation}\label{eq:3}
    y=0.8\,x+0.2
\end{equation}
and a horizontal dash-dotted line to separate AGN models from the SF locus
\begin{equation}\label{eq:4}
    y=0.1 \,\,\,\,\,\, and \,\,\,\,\,\, -1.8<x<-0.15
\end{equation}
with $x = $~log(\oiiiuv~$~\lambda$1666/\heii~$~\lambda$1640) and $y=$~log(\ciii~$~\lambda\lambda$1907,9/\heii~$~\lambda$1640).
The AGN grids again extend beyond the shown x- and y-ranges, with log(\oiiiuv/\heii) and log(\ciii/\heii) down to $\sim-1.5$ and $\sim-2.5$, respectively, at the highest values of metallicity and lowest of ionization parameter. %In particular, the ionization parameter increases at higher log(\oiiiuv/\heii) values, while the grids are overlapped in metallicity. %, with their lowest values at the lowest log(\ciii/\heii).
% The trends with log($U$) and 12+log(O/H) also hold true for \citetalias{gutkin16} SF models (top left and right panels), even though \citetalias{byler17} grids show less degeneracy. 
% While the shock grids instead are completely folded in terms of velocity and magnetic field and gas-phase metallicity, we decided to show the entire range of their values for completeness.

The scatter plot in Fig.~\ref{fig:diagnostic-f16} shows the 19 CLASSY galaxies with all the involved UV emission lines with $S/N>3$. %, but we also show upper-limits ($S/N\lesssim 2$) as gray dots.
Interestingly, the parameter space covered by the CLASSY galaxies in this diagnostic diagram is very similar to that of the $z\sim$2--4 galaxies of \citet{nanayakkara19}, which {further} supports the classification of these systems as analogues to high-$z$ SF-galaxies.
As in Fig.~\ref{fig:diagnostic-jr16}, \citetalias{gutkin16} SF models cover the majority of the C3He2-O3He2 plane. We notice that grids with low C/O abundances (solid and dashed cyan \citetalias{gutkin16} grids in the upper left panel) are shifted to lower \ciii/\heii\ and higher \oiiiuv/\heii\ (trend in agreement with the C/O color-coding shown in Fig.~\ref{fig:diagnostic-f16} upper left-hand panel). 
This is probably due to the fact that carbon is a key coolant in the nebula, and thus a lower C/O ratio raises the electron temperature -- as commented in \citet{jaskot16}, and demonstrated by the decrease in metallicity (which is color-coded in the bottom left-hand panel) as O3He2 increases. Hence, the increased collisional excitation rate partially compensates for the reduced C abundance, preventing an extreme drop of the \ciii/\heii\ ratio, and possibly enhancing \oiii~$~\lambda$1666. As a result, the low C/O \citetalias{gutkin16} grids (solid and dashed cyan, according to the different IMF cut) manage to cover the lowest \ciii/\heii\ and \oiiiuv/\heii SBS~0335-052~E and I\,Zw\,18. %SBS~0335-052~E and I\,Zw\,18 are still the galaxies with the lowest \ciii/\heii\ and \oiiiuv/\heii, and can be covered by \citetalias{gutkin16} SF models only taking into account either low C/O or high $M_{up}$, that can extend to lower \ciii\heii\ line ratios, as commented for Fig.~\ref{fig:diagnostic-jr16}. 
{\citetalias{xiao18} BPASS models instead are more confined in the SF locus, with the lowest metallicity and ionization parameter points of the grid slightly covering the region that we define as the AGN locus (below the black dash-dot line and above the dashed black line in Fig.~\ref{fig:diagnostic-f16}).}

Also \citet{hirschmann19} proposed line separators to classify the main ionization mechanisms for the C3He2-O3He2 diagnostic plot, obtained from synthetic emission line ratios computed by coupling \citetalias{gutkin16} and \citetalias{feltre16} models with high-resolution cosmological zoom-in simulations of massive galaxies. 
However, according to \citet{hirschmann19} AGN-Composite and Composite-SF separators (that we do not report in Fig.~\ref{fig:diagnostic-f16} for clarity reasons), {all the CLASSY galaxies} would be classified as composite.
\citet{hirschmann19} also compared their simulated galaxies with observations of AGN and star-forming galaxies, finding generally good agreement with their classification. {However,} the local and high-$z$ low-metallicity galaxies that they {overplotted to their diagnostic diagrams ended up in the composite region as well}. 
Hence, we conclude that the \citet{hirschmann19} {composite region also hosts star-forming galaxies with more ``extreme'' high-$z$ characteristics that are typical of our sample}.

Overall, we conclude that C3He2-O3He2 represents a better diagnostic diagram than C3He2-C4He2, since it is able to clearly separate (up to solar metallicities) {the different mechanisms, with AGN showing lower \oiiiuv/\heii\ than shocks, while SF dominated emission shows higher \ciii/\heii}. Moreover, all the CLASSY galaxies, even the most metal-poor SBS~0335-052~E and I\,Zw\,18, are covered by the SF grids and do not overlap with the AGN or shock grids.

\subsection{C4C3He2-C4C3, EWC4, and EWC3}\label{sec:C4C3He2-C4C3}
%CIV~$~\lambda\lambda$1548,51/CIII]~$~\lambda\lambda$1907,9 versus (CIV+CIII])/He~II$~\lambda$1640}
\begin{figure*}
\begin{center} 
    \includegraphics[width=0.9\textwidth]{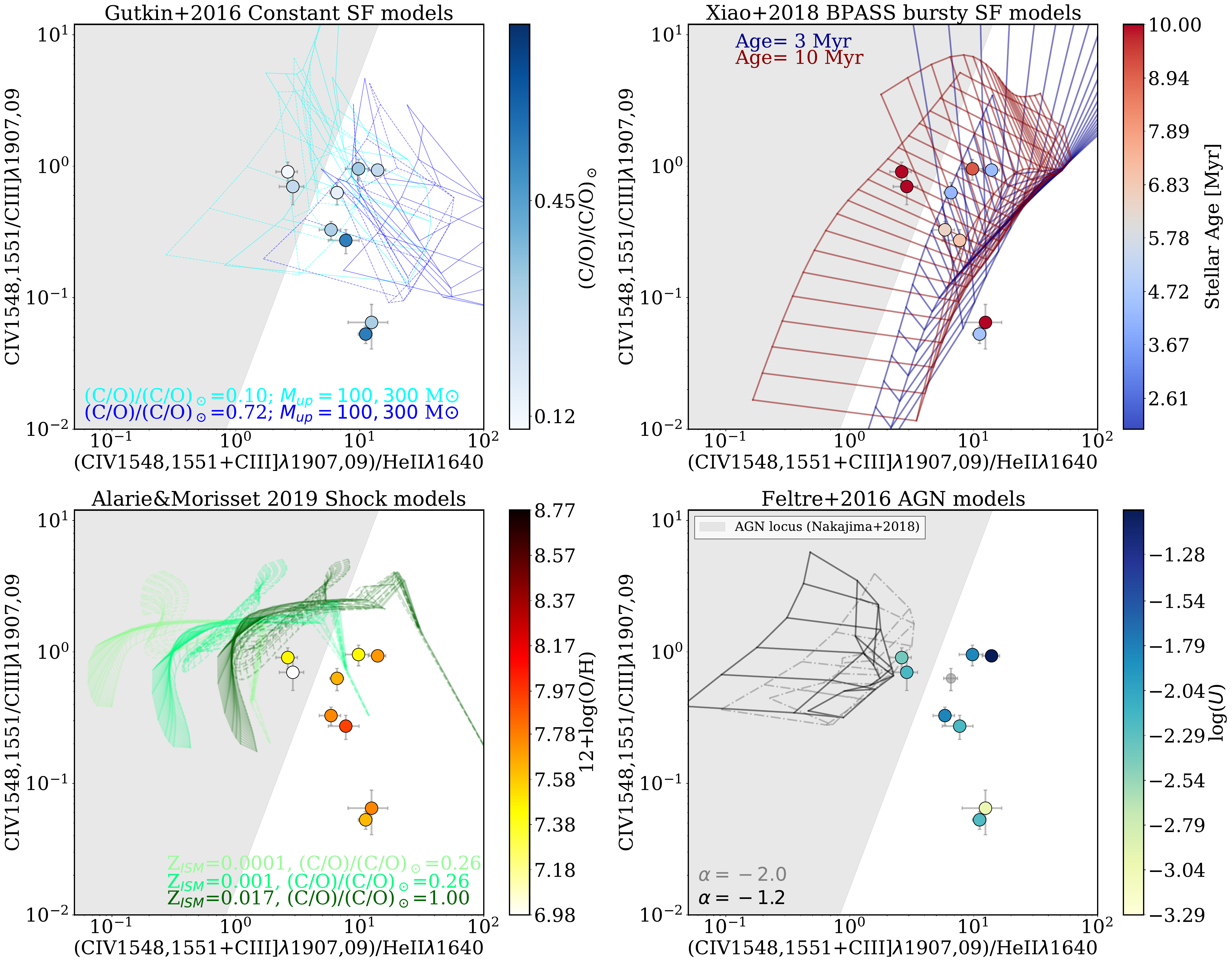}   
\end{center}
\caption{C4C3He2-C4C3 diagram: (\civ~$~\lambda\lambda$1548,51+\ciii~$~\lambda\lambda$1907,9)/\heii~$~\lambda$1640 vs \civ/\ciii\ diagnostic plot proposed by \citet{nakajima18} to separate SF and AGN activities, with their proposed AGN-dominated region highlighted in shaded gray. The models superimposed are as explained in Fig.~\ref{fig:diagnostic-jr16}. Similarly to Fig.~\ref{fig:diagnostic-jr16}, this plot can be used to separate SF from AGN with some caveat. Indeed, the two galaxies which end up in the AGN locus are again I\,Zw\,18 and SBS~0335-052~E, the two lowest 12+log(O/H) and C/O abundance galaxies of the CLASSY sample, and can be reproduced by the low C/O SF grids (in cyan). }
\label{fig:diagnostic-n18}
\end{figure*}
\begin{figure*}
\begin{center} 
    \includegraphics[width=0.8\textwidth]{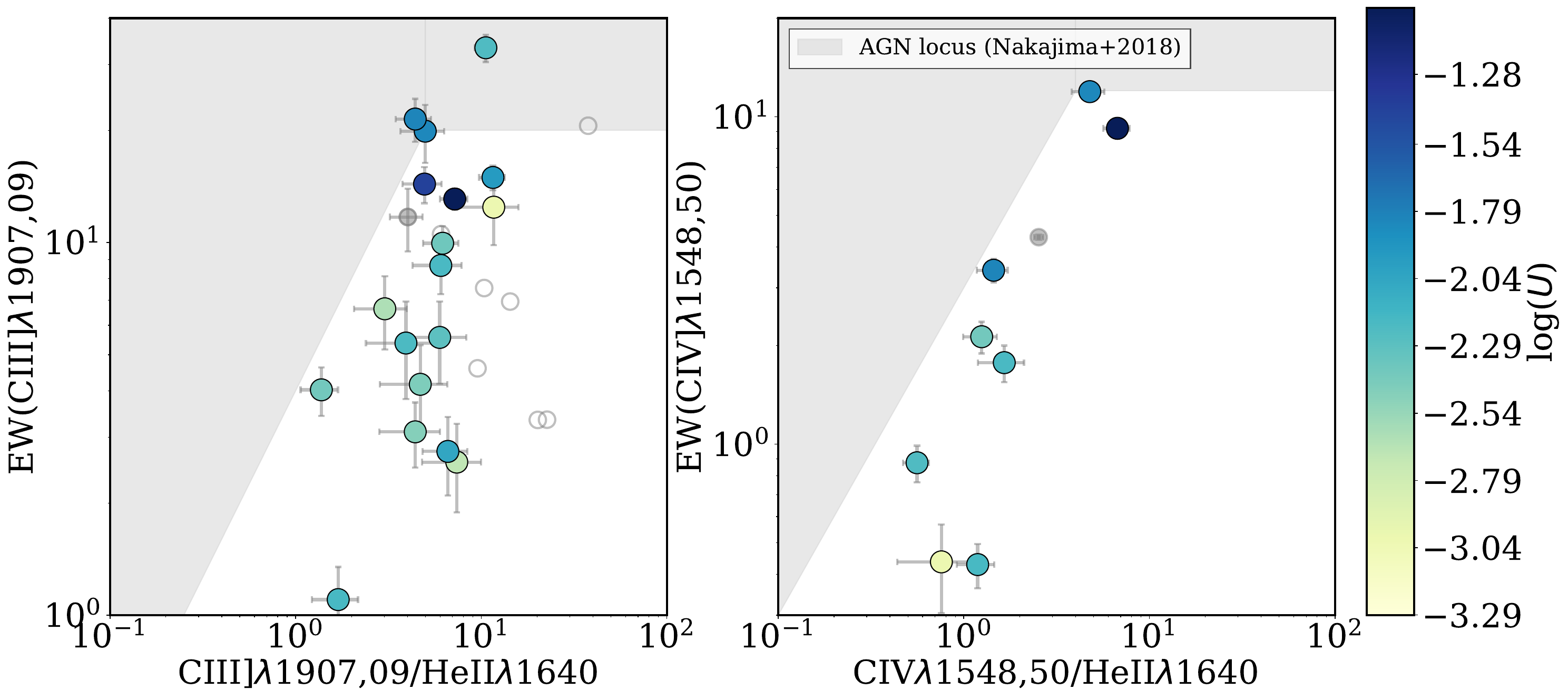}   
\end{center}
\caption{EW(\ciii) and EW(\civ) diagnostic plots, where the gray shaded regions represent the \citet{nakajima18} AGN-locus, separated from the star formation locus. {The filled and open dots show $S/N>3$ and $S/N<3$ fluxes for all the emission lines taken into account, respectively.} Interestingly, this classification scheme works well for the CLASSY galaxies, even for the most metal-poor systems.} %Our targets would be all classified as composite according to the \citet{hirschmann19} separators proposed for this diagram.}
\label{fig:diagnostic-n18-1}
\end{figure*}

Fig.~\ref{fig:diagnostic-n18} shows the (\civ+\ciii)/\heii\, vs \civ/\ciii\ diagnostic plot proposed by \citet{nakajima18}, with their proposed AGN locus shown in shaded gray. %The dashed black line indicates the separation from AGN and SF, on the left- and right-side of it, respectively \citep{nakajima18}. 
As in Fig.~\ref{fig:diagnostic-jr16}, SBS~0335-052~E and I\,Zw\,18 end up in the AGN locus,  overlapping slightly with the \citetalias{feltre16} AGN models.  
Analogously to what we commented in Sec.~\ref{sec:C3He2-C4He2}, their shift can be explained in terms of their low 12+log(O/H) and low C/O abundance. 
In this figure, it is even clearer that only the \citetalias{gutkin16} SF models (both CST and SSP) with low C/O (cyan grids) can cover this sub-set of CLASSY galaxies, excluding two outliers. A higher C/O abundance grid with a high-cut in the IMF (dashed blue grids) lies close but does not cover the lowest (\civ+\ciii)/\heii\ line ratios, while the higher C/O abundance grid with M$_{up}=100$~M$_\odot$ (solid blue grids) does not cover the CLASSY galaxies. 
Overall, an increase of C/O quickly pushes the \citetalias{gutkin16} grids {(also the bursty SF grids at ages below 4~Myr; see Fig.~\ref{fig:modelscomp})} toward values of (\civ+\ciii)/\heii\ {higher} than what we typically observe in our sample.
{\citetalias{xiao18} BPASS models (as well as \citetalias{byler17}; see Fig.~\ref{fig:modelscomp}) can well reproduce the observed CLASSY line ratios, with the 3~Myr grids confined in the \citet{nakajima18} SF locus and the 10~Myr grids slightly shifted on the grey shaded area of Fig.~\ref{fig:diagnostic-n18}, reaching the lowest values of \civ+\ciii)/\heii\ at decreasing metallicity and ionization parameter.}
Moreover, AGN and shock grids are almost entirely overlapped (apart from the most metal-poor shock models, characterized by the lowest (\civ+\ciii)/\heii).

\citet{nakajima18} proposed other two additional diagnostic plots to separate SF and AGN activities, shown in Fig.~\ref{fig:diagnostic-n18-1}, with the CLASSY data color-coded as a function of the ionization parameter log($U$). 
In particular, we notice that EW(\ciii) and EW(\civ) increase at higher log($U$), and also the \civ\ and \ciii\ emissions are higher with respect to \heii. This is clearer in the right panel since EW(\civ) provides a valid log($U$) tracer as demonstrated in \citetalias{mingozzi22}.
As in Fig.~\ref{fig:diagnostic-n18}, the AGN regions according to \citet{nakajima18} criteria are shown in shaded gray.
\citet{nakajima18} computed AGN photoionization models, assuming a continuum underneath the line emission dominated by the accretion disk, commenting that this is an overestimate since it does not take into account the effects of dust attenuation (i.e., presence of the torus). % and thus,  their modeled \ciii\ and \civ\ EWs could be only lower limits.
In general, evaluating the continuum emission to compute EWs in shock and AGN models implies many arbitrary assumptions, since the continuum can have a composite nature, coming from both the stellar component and the additional ionizing sources as well as their intrinsic characteristics. For this reason \citetalias{feltre16} and \citetalias{alarie19} do not provide EWs for their models. On the other hand, all the displayed CLASSY galaxies are in the SF-locus defined by \citet{nakajima18}, with the exception of J0944-0038 in the left panel, characterized by EW(\ciii)~$\sim 33$~\AA, which still lies in a region where both AGN and SF models could overlap. 
% {ADD COMMENT BPASS HERE, remove gutkin}
{Also, \citetalias{xiao18} BPASS models can predict the observed range of \ciii\ and \civ\ EWs only taking into account very young ages (1~Myr), which are shifted at higher \ciii/\heii\ and \civ/\heii. Thus, it could be that the continuum to compute the EWs in these models is overestimated.}
{Interestingly,} the EW values predicted by \citetalias{gutkin16} single-burst SF models with ages younger than 4~Myr cover the \citet{nakajima18} SF locus, and thus the EW(\ciii) and EW(\civ) values observed for the CLASSY galaxies, while the constant SF models hardly reproduce them.  
% Nevertheless, we want to test these proposed diagnostic diagrams for the CLASSY sample, overplotting our data color-coded as a function of the ionization parameter log($U$). 
As commented also for Fig.~\ref{fig:diagnostic-f16}, CLASSY galaxies fall entirely in the region defined as composite in \citet{hirschmann19}, confirming that star-forming objects with the characteristics of the CLASSY sample (and thus high-$z$ analogs) could risk being misclassified according to their criteria.
Overall, these diagrams could be useful tools to classify even metal-poor star-forming galaxies, but since it is not trivial to evaluate the continuum to compute EWs in shock and AGN models, they cannot be used to separate AGN from shocks.

\subsection{Summary}\label{sec:uv-diagnostics-summary}
All the discussed UV diagnostic diagrams in the previous subsections, with some caveats, can distinguish SF from AGN and shocks, while it is possible to disentangle AGN from shocks only by taking into account \ciii, \heii\ and \oiiiuv\ emission lines.
Here we summarize our main findings:
\begin{itemize}
\item The C3He2-C4He2 diagram (Fig.~\ref{fig:diagnostic-jr16}) from \citet{jaskot16} can be used to discriminate SF from AGN and shocks. However, the users should be mindful of the metallicity and C/O abundance of the targets, in that ionization sources in low-metallicity objects are less easily distinguishable for this diagnostic. Shocks and AGN are harder to discriminate, but still the former are expected to have higher carbon-to-helium line-ratios than the latter, especially at increasing metallicity.
\item C3He2-O3He2, proposed by \citetalias{feltre16}, is the only diagram able to separate all the three ionization mechanisms at sub-solar metallicities (Fig.~\ref{fig:diagnostic-f16} and Eq.~\ref{eq:3}, \ref{eq:4}). In this plot, all the CLASSY galaxies, even the objects characterized by the lowest 12+log(O/H) and C/O (i.e., SBS~0335-052~E and I\,Zw\,18), are covered by the \citetalias{gutkin16} SF models and do not overlap to the AGN or shock grids, that cover similar \ciii/\heii\ but different \oiiiuv/\heii\ line ratios (lower for AGN, higher for shocks). 
Other diagnostic diagrams from \citetalias{feltre16} are C3He2-C3O3 and C3He2-C4C3 (App.~\ref{app:other-diag}), with the former able to separate SF, AGN and shocks as C3He2-O3He2 (Eq.~\ref{eq:5}, \ref{eq:6}), while the latter can only separate SF from AGN and shocks (Eq.~\ref{eq:7}, \ref{eq:8}).
\item The C4C3He2-C4C3 diagram (Fig.~\ref{fig:diagnostic-n18}) from \citet{nakajima18} is a good diagnostic plot to separate SF and AGN, with the caveat of taking into account 12+log(O/H) and C/O, similarly to C3He2-C4He2. However, it is not possible to distinguish between AGN and shocks, for which O3He2-C3He2 and C3He2-C3O3 would be better suited. 
% Also, in this diagnostic plot, the model grids are not folded (apart at the highest 12+log(O/H) and log(U) values) in contrast to those shown in Fig.~\ref{fig:diagnostic-jr16} and Fig.~\ref{fig:diagnostic-f16}. This is because the x-axis of Fig.~\ref{fig:diagnostic-n18} is mainly sensitive to the gas-phase metallicity, while the y-axis mainly to log($U$) \citep{mingozzi22}. As a result, it allows to distinguish among models with different metallicities (including shocks). This means that a combination of C4C3He2-C4C3 and one of the previous diagrams can provide several pieces of information about the ISM gas, while avoiding degeneracies.
\item Finally, EWC4-C3He2 and EWC3-C3He2 (Fig.~\ref{fig:diagnostic-n18-1}) from \citet{nakajima18} provide as well good diagnostic plots to separate star-forming galaxies from objects characterized by non-photoionization mechanisms via the \citet{nakajima18} AGN-SF separator line. This diagnostic diagram holds true even at low metallicity, low C/O abundance and high log($U$), but does not allow the user to discriminate shocks from AGN models, given the difficulties in estimating the corresponding EWs.
\end{itemize}

\section{Discussion}\label{sec:discussion}
In {Sections~4 and 5} we demonstrated that our galaxies are SF-dominated according to optical diagnostics, and then explored the most promising UV diagnostic diagrams, highlighting each of their strengths and weaknesses. 
In the following, we  discuss the \textit{practicality} of the UV diagnostic diagrams, in terms of the observability of UV emission lines shown in this work {and possible caveats when using them as diagnostics, and conclude with a comment on UV emission lines currently observed with JWST at high-$z$}.
We do this for two reasons: (1) to provide guidance on whether or not these lines should be expected in proposed observations of certain targets and (2) to provide context on the properties of the target if the lines \textit{are} detected in their observations.
Indeed, UV emission lines are a powerful (and at increasing redshift also the only) tool to investigate ISM conditions in galaxies. 

\subsection{{Conditions for UV emission in high-$z$ analogues}}
%\subsection{{Conditions to observe UV emission lines in CLASSY}}
%\footnote{\lya\ usually is the strongest feature in rest-frame UV spectra, and is a resonant line, meaning it is easily scattered by the gas within and surrounding galaxies. This feature deserves a publication by itself (Weida et al. submitted) given its complex nature and interpretation.}). %\mgii\ and \civ\ behave similarly to \lya, with \civ\ tracing the high-ionization gas through which very high energy ionizing photons may escape instead of the neutral phase (e.g., \citealt{berg19}). \mgii\ is not covered by the CLASSY wavelength range, while in this paper we are using \civ\ as a diagnostic only when in pure nebular emission.}).
In general, we confirmed that the main two conditions to have UV emission line detections in CLASSY (i.e., \civ$\lambda\lambda$1548,51, \heii$\lambda$1640, \oiiiuv$\lambda\lambda$1661,6, \ciii$\lambda\lambda$1907,9) are low metallicity and high ionization parameter, traced by the \oiii/\oii\ line ratio (but also \oiii/\hb).
Indeed, we see no \ciii$~\lambda\lambda$1907,9 (and thus no other UV emission lines) in CLASSY galaxies with 12+log(O/H)~$\gtrsim8.3$. 
{Specifically, we do not observe \ciii\ in the eight CLASSY galaxies at metallicity higher than this threshold, where EW(\ciii)~$\lesssim 2$~\AA\ and log($U$)~$\sim -2.9$.}
This is also the reason why \ciii-emitting galaxies (highlighted by thick black edges in all the figures) are systematically shifted to lower log(\nii/\ha) ($\lesssim -1$) in Fig.~\ref{fig:optbpt} and Fig.~\ref{fig:shirazi}), with the lowest metallicity targets (12+log(O/H)~$<7.8$), having also higher log(\heii$\lambda4686$/\hb) line ratios ($\sim-1.4$ to $-1.8$).
However, the trend between the EW of UV lines and metallicity is not linear because carbon lines in our galaxies can be suppressed at 12+log(O/H)~$<7.5$ (see e.g., Fig.~15 in \citetalias{mingozzi22}) and at decreasing C/O (Berg et al. in preparation). 
{As such, low-metallicity is not a sufficient condition to ensure the presence of UV emission lines. In fact, there are nine CLASSY galaxies at 12+log(O/H)~$<8.3$ without UV emission lines. These objects also have median values of EW(\ciii)~$\lesssim2$~\AA\ and log($U$)~$\sim-2.8$.}
Indeed, the EW of UV lines slightly correlates with log($U$) as well, with a significant decrease of detected UV lines at log($U$)~$\lesssim -2.5$ for our sample. %: for instance, S/N(\ciii) decreases from $\sim10$ to $\sim3$ at log($U$) in the range [-1,-2.5].
This is also visible from Fig.~\ref{fig:optbpt} and Fig.~\ref{fig:opt-shocks}, in which \ciii-emitters are clearly shifted to higher log(\oiii/\hb) ($\sim0.5-1$) and higher log(\oiii/\oii) ($\sim0.5-1.5$). {Overall, this finding is consistent to previous works in the nearby Universe (e.g., \citealt{heckman98,ravindranath20}), as well as at higher redshift (e.g., \citealt{shapley03,rigby15,du17}).}

The physical reason why UV metal lines are {mostly} visible only at low metallicity is well-explained in \citet{jaskot16} (see their Sec.~3.2). In particular, a decrease in metallicity implies a harder stellar ionizing radiation and a higher temperature nebula, and thus a larger amount of ionizing photons and a larger collisional excitation rate, respectively. 
Indeed, the UV metal lines are collisionally excited with energy levels above the ground state of E/k~$\sim 6.5-8.3$~eV~$\sim 75,000- 97,000$~K, much higher than optical collisional lines that typically have E/k~$<<70,000$~K (e.g., $\sim 7.5$~eV~$\sim87,000$~K for \oiiiuv$\lambda\lambda$1661,6 versus $\sim2.5$~eV~$\sim30,000$~K for \oiii$\lambda$5007 and $\sim5.4$~eV~$\sim62,000$~K for \oiii$\lambda$4363; \citealt{osterbrock06,draine11}). 
Specifically, \ciii$\lambda\lambda$1907,9 (the brightest and most common) has the lowest E/k~$\sim 75,000$, while the less common \civ$\lambda\lambda$1548,51 and 
\niv$\lambda\lambda$ have the highest E/k~$\sim 93,000-97,000$~K.
Hence, the UV lines are extremely temperature sensitive in photoionized gas that can reach T~$ \sim 2\times10^4$~K in our systems (see also \citetalias{mingozzi22}), compensating for the lower ionic relative abundances. 
%, while T~$ \sim 10^4$~K at solar metallicities. 
% From T. Heckman: This might seem suprising that metal lines are only seen when the metallicty is low. You could explain that physically, this is because the UV lines are collisionally-excited with levels having energies above the ground-state of E/k ~ 75,000 to 90,000 K. This makes them extremely temperature sensitive in photoionized gas with T ~ 10^4 K. At low metallicity, the reduced cooling by metal lines means that the equilibrium T is more like 20,000K vs. 10,000 K at solar metallicity. This higher T more than compensates for the lower relative abundances of O, C etc.
As discussed in previous works (e.g., \citealt{jaskot16}), dust attenuation, density, SFR and stellar population characteristics (stellar age and metallicity) can also play a role in the presence or absence of these UV lines, but we find no clear correlations comparing the derived quantities for the CLASSY galaxies with their UV emission strengths.
% Our results are consistent with %statistically large 
% rest-UV spectroscopic studies pre-JWST typically targeting $2<z<4$ star-forming galaxies (e.g., MEGaSaURA, \citealt{rigby18a}; VANDELS, \citealt{mclure18,llerena21}; VUDS, \citealt{lefevre15}; and MUSE samples, \citealt{maseda17,nanayakkara19,feltre20,schmidt21}). 

\subsection{{Caveats in Using CIV and HeII}}\label{sec:civheii-discussion}
UV lines such as \heii$~\lambda$1640 and \civ~$~\lambda\lambda$1548,51 can have both a stellar and nebular origin, with \civ\ being also a resonant line affected by radiation transfer effects (e.g., \citealt{steidel16}). 
For many CLASSY galaxies, we can confirm \heii\ and \civ\ nebular origin thanks to the general consistency of their intrinsic velocity dispersion values with those of \oiiiuv~$~\lambda\lambda$1661,6 ($\sigma\sim 50$~km/s; Fig.~\ref{fig:fitcomps} bottom panel). 

\begin{figure}
\begin{center}
    \includegraphics[width=0.5\textwidth]{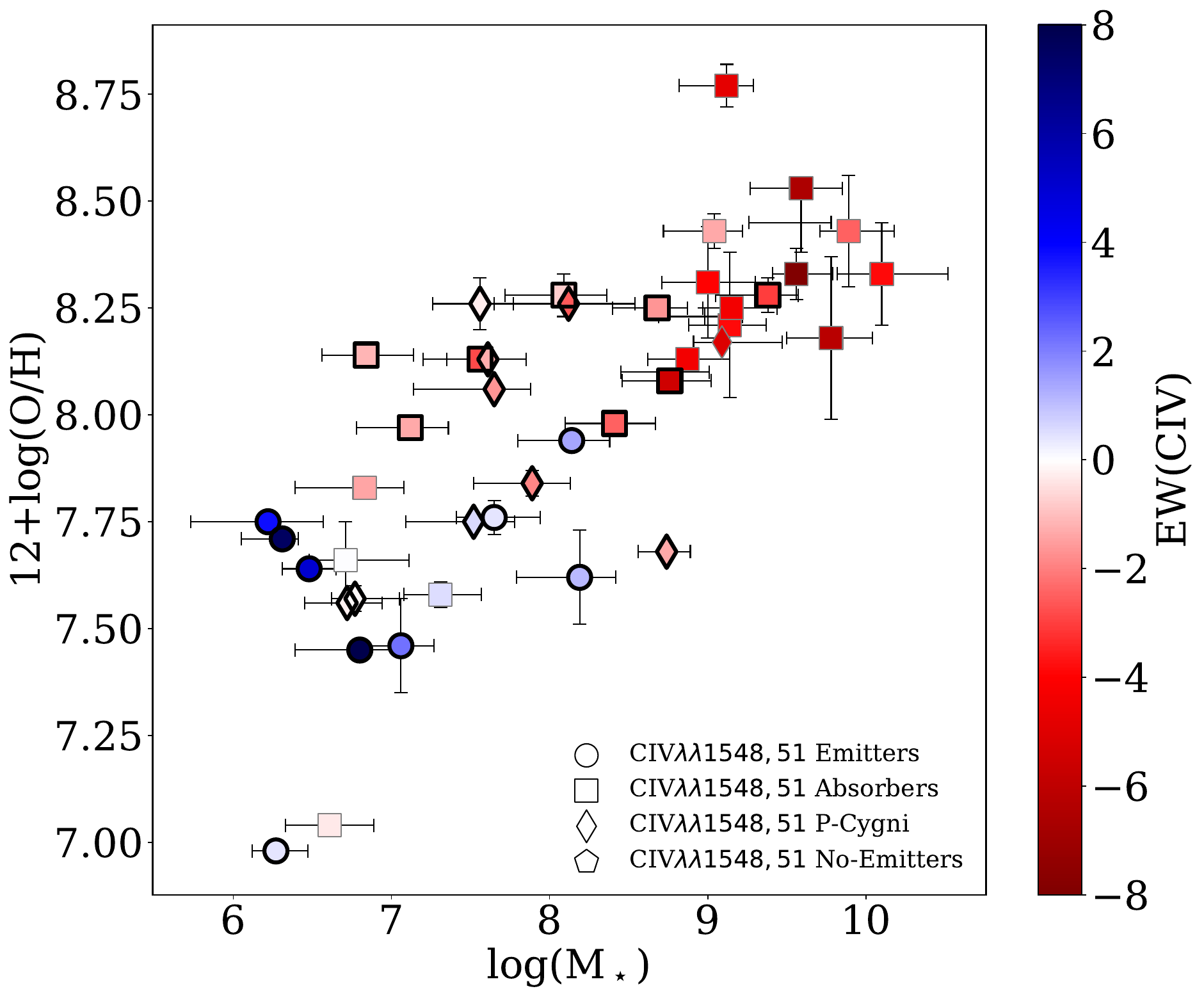}
\end{center}
\caption{Mass-metallicity relation for CLASSY galaxies with \civ\ in emission (dots), absorption (squares) and P-Cygni (diamonds), color-coded by EW(\civ), with red and blue values indicating negative (lines in absorption) and positive (lines in emission) EWs, respectively. The galaxies with at least one UV emission line (i.e., \ciii\ are highlighted by a black thick edge. These plots show that galaxies which are \civ\ emitters, or with narrow P-Cygni non-stellar profiles, tend to have lower metallicities (12+log(O/H)$\lesssim8, 8.2$) and lower stellar masses (log($M\star$/M$_\odot$)~$\lesssim8,9$).} 
\label{fig:whenewciv}
\end{figure}

Concerning \civ, typical galaxies without an AGN show either \civ\ in absorption from the surrounding ISM or circumgalactic medium (CGM) gas or a P-Cygni profile from the stellar winds of massive stars (e.g., \citealt{leitherer01}).
Nebular \civ\ emission is rarely observed in the literature, and comes uniquely from studies of systems with 12+log(O/H)~$\lesssim8$ and generally high ionization parameter log($U$)~$\gtrsim -2.5$ (see Fig.~16 in \citetalias{mingozzi22}), probably tracing a rapid hardening of the ionizing spectrum at low metallicities (see also \citealt{senchyna19a,berg21a,senchyna22,schaerer22}).
From a visual inspection of \civ\ feature, we classified the CLASSY galaxies as being either \civ\ emitters (9 objects), P-Cygni (10), without \civ\ coverage (3), and defining the remaining others as ISM absorbers (23). Specifically, we applied the classification after normalizing for the stellar component, thus we are considering only the ISM contribution of \civ\ doublet. 
We then calculated the EW(\civ) by integrating the \civ\ emission in a $\sim10$~\AA\ window centered on the \civ\ lines, and the continuum in two featureless $\sim5$~\AA\ windows on each side. We show the results of this exercise in Fig.~\ref{fig:whenewciv}, where we color-code the CLASSY mass-metallicity relation as a function of EW(\civ). Also, galaxies with different \civ\ profiles are indicated using different symbols, as listed in the legend.
Hence, within the CLASSY sample, \civ\ emitters (but also P-Cygni) are found to be more metal-poor and low-mass (12+log(O/H)$\lesssim8, 8.25$ and log($M\star$/M$_\odot$)~$\lesssim8, 9$, respectively). It should be noted here, that the P-Cygni profiles here are very narrow and non-stellar (i.e., the stellar component has been subtracted) and should not be confused with the strong stellar P-Cygni profiles seen at high-metallicity.
Also, \civ\ absorbers within the CLASSY sample have increasing EWs at larger stellar masses and metallicities. 
We do not find further particular correlations with other gas and stellar properties (e.g., SFR, E(B-V), density). Of course, these inferences are being drawn from our local sample of high-$z$ analogues and may not hold true at all redshifts.
{Finally, we highlight that \civ\ is a resonant transition, meaning that high-ionization gas can scatter the photons. As discussed in \citet{berg19b}, resonantly scattered \civ\ emission appears broader than non-resonant emission lines (e.g., \oiiiuv~$\lambda$1666) and double-peaked, as it is usually found in \lya\ studies (e.g., \citealt{hayes15, verhamme17}; see also CLASSY \citetalias{hu23}). In particular, \citet{berg19b} confirmed this \civ\ resonant emission in two CLASSY galaxies, J1044+0353 and J1418+2102, which are the two objects that show the largest difference between \civ\ and \oiiiuv\ widths. The other CLASSY galaxies showing \civ\ in emission do not show the clearly double-peaked signatures (see Fig.~2 in \citetalias{mingozzi22}) and have consistent values of $\sigma$(\civ) and $\sigma$(\oiiiuv).}

Overall, given the complex nature of the \civ\ nebular emission (see also Sec.~6.4.1. of \citetalias{mingozzi22}), diagnostic diagrams involving the \civ\ lines (i.e., \civ/\heii\ and \civ/\ciii\ line ratios, as well as EW(\civ)) should be used with caution.
The complexity of this line is also evident in the velocity offset that we find for \civ\ lines, in that they are systematically redshifted by $\sim 25-100$~km/s with respect to the other UV lines (see Fig.~\ref{fig:uv-opt-kins} in App.~\ref{app:uv-opt-kins}), also seen in \citet{berg19b,wofford21,senchyna22}, which could be consistent with resonance lines in an outflowing medium. Indeed, we found that to properly fit the \civ\ lines, we have to employ a different value of their velocity\footnote{\civ\ lines were fit separately from the other UV lines, as explained in \citetalias{mingozzi22} Sec.~3.2.} (and thus line center) with respect to \heii\ and \oiiiuv. 
%The velocity shifts between \civ\ and the other lines are in the range $\sim 25-100$~km/s, and they could imply that the emission is not coming from the same regions of the ISM.
However, in order to fully explore and compare in detail the different line kinematics, we would need integral-field spectroscopy data in the UV to see how the spectra change as a function of the location within the galaxy.
% \begin{figure}[!h]
% \begin{center}
%     \includegraphics[width=0.45\textwidth]{vel_comparison_ciii_civ.pdf}
% \end{center}
% \caption{Offset in velocity between \civ$~\lambda\lambda$1548,51 and \ciii$~\lambda\lambda$1907,9 emission lines as a function of the galaxy stellar mass. The shift can imply that the two emission lines are not tracing the same regions of the ISM, coming from distinct kinematical components of the gas. \heii\ and \oiiiuv\ lines have instead consistent velocities with \ciii. Hence, diagnostic diagrams not involving the \civ\ lines should be preferred when exploring the gas ionizing conditions.}
% \label{fig:c_velocity_offset}
% \end{figure}

\begin{figure}
\begin{center}
    \includegraphics[width=0.4\textwidth]{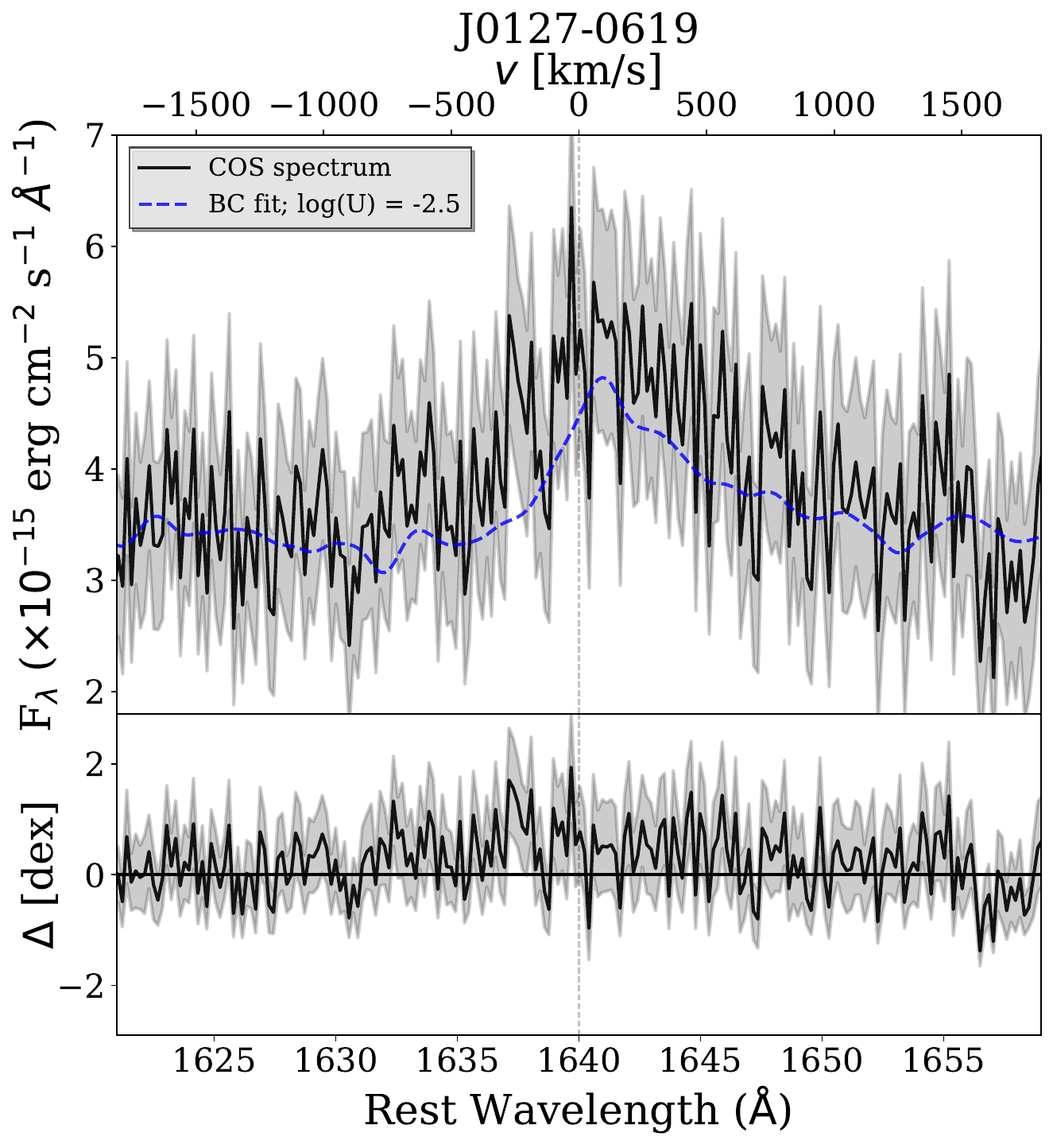}
\end{center}
\caption{The \heii$\lambda$1640 feature of Mrk~996 (J0127-0619) superimposed with the stellar component best-fit made with C\&B (blue dashed line) described in detail in Sec.~3.1.1 of \citetalias{mingozzi22}. The bottom panel shows the difference between the observed spectrum and the C\&B model. The very broad \heii\ component ($\sigma > 100$~km/s) has a stellar origin, in contrast to the nebular \heii\ emission that we generally observe in other CLASSY galaxies ($\sigma\sim 50$~km/s; see {Fig~\ref{fig:fitcomps} of this paper and also} Fig.~22 of \citetalias{mingozzi22}). This figure aims to illustrate the broad stellar \heii\ profile and guide the user in its identification.}
\label{fig:heiimrk996}
\end{figure}

Moving to \heii$\lambda$1640 instead, its stellar contribution can be due to massive and very massive stars, including O-type stars and WRs, as commented in Sec.~\ref{sec:intro}. A typical signature of stellar \heii\ is a very broad width (e.g., \citealt{sixtos23,wofford23,smith23}), that can be distinguished from nebular emission via a  comparison with close emission lines not affected by stellar contamination (i.e., \oiiiuv~$\lambda\lambda$1661,6). The contribution of WR stars is more evident in optical spectra, where the \heii$~\lambda$4686 line can be blended with several metal lines in the so-called ``blue-bump'' \citep{brinchmann08}. 
As discussed in Sec.~3.1.1 of \citetalias{mingozzi22}, our stellar component fitting accounts for different prescriptions for the evolution and atmospheres of massive stars, better accounting for \heii\ stellar contribution than other modeling prescriptions (e.g., \citealt{senchyna21}). 
The best example is Mrk~996 (J0127-0619) shown in Fig.~\ref{fig:heiimrk996}, which hosts a significant WR population \citep{james09} and a broad \heii$\lambda$1640 feature ($\sigma > 100$~km/s), that is completely stellar according to the C\&B stellar component fit (blue dashed line), without signs of nebular emission.
Only four CLASSY galaxies (i.e., J1129+2034, J1253-0312, J1200+1343, J1314+3452; see Fig.~22 of \citetalias{mingozzi22}) have $S/N$(\heii)~$>3$ and $\sigma$(\heii)~$>100$~km/s (up to $200$~km/s {and broader than $\sigma$(\oiiiuv1666})), indicating that a stellar contribution can still be present despite our subtraction of the stellar-component emission. 
After a visual inspection of their optical spectra, all these four galaxies show the presence of the WR blue-bump.

Overall, we acknowledge that \heii$\lambda$1640 (both for its stellar/nebular origin and high ionization potential) can be also problematic to reproduce by photoionization models (e.g., \citealt{gutkin16,  steidel16,nanayakkara19}). 
This led some authors (e.g., \citealt{hirschmann19, feltre16}) to propose UV diagnostic diagrams without \heii, comparing \civ~$\lambda\lambda$1549,51 and \ciii~$\lambda\lambda$1907,9 with other UV lines, such as \nv~$\lambda\lambda1239,1243$, \niv~$\lambda\lambda$1483,87 and \niii~$\lambda\lambda$1747–54, or optical lines like \oiii~$\lambda5007$ and \oi~$\lambda6300$ (that can be a good low metallicity AGN and shock diagnostic, Sec.~\ref{sec:results-bpt},~\ref{sec:oi-shocks}). 
However, as shown in \citetalias{mingozzi22}, these UV lines, as well as \siiii~$\lambda\lambda$1893,92, are very faint and observable in a handful of objects, while it is difficult to have extended coverage from the UV up to the very faint \oi~$\lambda6300$. 
% Hence, we are not discussing here further diagnostics diagrams.
This is the reason why we have not explored further UV diagnostic diagrams in this paper. %, while in our next work, we will explore possible combinations of mixed optical-UV diagnostics, taking into account the observable lines with the current facilities such as JWST as a function of $z$ (and especially at $z>6$).

{To conclude, we want to highlight the advantages of detecting \oiiiuv$\lambda\lambda$1661,6 in UV spectra. As discussed in \citetalias{mingozzi22}, \oiiiuv\ is an auroral line like \oiii$\lambda$4363 and is thus sensitive to the gas temperature. It can be used to estimate C/O (see Sec.~\ref{sec:thesample}) and is more sensitive to shocks than carbon lines (Fig.~\ref{fig:diagnostic-f16}). Finally, the proximity of \oiiiuv\ to \civ\ and \heii\ lines in UV spectra is useful for comparisons of the corresponding line widths, to understand if the \civ\ and \heii\ emission is purely nebular, as discussed in this subsection.}

\subsection{{Current Observations of UV Emission Lines at High-$z$}}
% {In this work we want to present UV lines to study at high-z BUT there not UV lines at high-z, why? We see nitrogen lines etc etc, case of gnz11 (but less text), say it's open problem, we can see them in stack (priv. comm.)}
{The UV diagnostics presented in this work and in CLASSY~\citetalias{mingozzi22} can provide informative guidelines for interpreting high-$z$ spectra. Such guidelines are especially important now that JWST has pushed the UV spectroscopic frontier to higher redshifts than ever before. 
Currently, large and deep JWST surveys, such as JADES, CEERS, and GLASS (e.g., \citealt{bagley23,bunker23b,treu22}), have started to provide statistically significant galaxy samples for studying galaxy evolution from $z\sim 1$ till the epoch of reionization (up to $z\sim13$).}
Up to now, the furthest spectroscopically confirmed galaxy is among the four young, metal-poor (stellar metallicity $\log(Z/{\rm Z_\odot}) \sim -1.91 - -0.18$), low-mass ($M\sim10^7-10^8$~M$_\odot$) systems at $10.3<z<13.2$ presented in \citet{curtis-lake23}. 
Their redshifts were estimated using the Lyman-break as a reference, because of the complete lack of both UV and optical emission lines.
{Indeed, so far, there has been a deficit in the amount of UV emission lines measured in the epoch of reionization. Despite becoming visible when stacking NIRSpec/prism spectra (using 38 JADES targets at $5<z<8$, priv. comm.), UV lines are detected in $\lesssim4$\% objects in the latest JADES data release \citep{bunker23b}.
This is puzzling given that pre-JWST ground-based rest-frame optical observations proved that galaxies at $z\sim5-7$ show prominent high-ionization nebular emission UV lines due to their extreme radiation fields and low metallicity (e.g., \citealt{stark15,mainali17,mainali18}).
The lack of observed UV emission lines in the current JWST observations could be due to a combination of several factors, such as the low spectral resolution ($R\sim100$) of the NIRSpec prism (the most commonly used mode so far for wide-field MSA observations currently published), the reduced sensitivity of NIRSpec detector at shorter wavelengths, the depth of the exposures, or the physical conditions of the galaxies themselves. With regards to the galaxy conditions, metallicity in particular may not be the issue - 
%Having said that, the currently revealed 
current observations are finding that high-$z$ galaxies are generally metal-poor but not metal \textit{deficient}, with metallicities of Z$\sim 0.03-0.6$~Z$_\odot$ (median $Z\sim 0.1$~Z$_\odot$ at $z\sim3-10$; \citealt{curti23}), a range comparable to the CLASSY sample. This suggests that UV lines are being produced in these systems, but not observed due to observational limitations.}
%This suggests that UV lines could potentially be observed and thus that the current JWST observations may be just not deep enough.}

The current furthest ($z\sim10.603$) galaxy spectrum with clear UV and optical emission lines (from \lya\ to \hg) is GN-z11, the most luminous candidate $z > 10$ Lyman break galaxy in the GOODS-North field \citep{bunker23}. 
% This galaxy is compact and metal-poor ($\log(Z_{\rm neb}/{\rm Z_\odot}) \sim −0.92$; 12+log(O/H)~$\sim7.5$), more massive ($\sim$$10^{9}~M_{\odot}$) and luminous than the objects analyzed in \citet{curtis-lake23} \citep{bunker23,tacchella23}.
{Surprisingly, GN-z11 shows UV emission lines that we typically do \textit{not} observe in CLASSY. Indeed, its NIRSpec/prism spectrum } clearly shows both \niv$\lambda\lambda$1483,7 and \niii~$~\lambda$1750 \citep{bunker23}, very rarely observed in the local Universe (see \citetalias{mingozzi22}). 
{This has opened an intense debate in the literature, with authors implying an extremely elevated nitrogen enrichment (at only 440~Myr from the Big Bang) and proposing different scenarios to explain it, including signatures of globular cluster precursors, massive star winds, runaway stellar collisions or tidal disruption events \citep{senchyna23,cameron23,charbonnel23}. On the other hand, \citet{maiolino23} claimed that this object hosts AGN activity which would be the origin of the peculiar nitrogen emission lines, thus solving the issue of the presence of the nitrogen lines and the puzzling N enrichment.}

In particular, \niv$\lambda\lambda$1483,7 lines (ionization potential $E>47.5$~eV) are rarely seen in emission in SFGs \citep{fosbury03,raiter10,vanzella10,mingozzi22,harikane23}, trace very dense gas ($n_e>10^4-10^5$~cm$^{-3}$) and, when present, they can also be - but not unambiguously - a signature of AGN ionization. 
\niii~$~\lambda\lambda$1750 instead is a multiplet that could have a nebular origin and be prominent in SFGs, but has been rarely observed (e.g., in the Sunburst Arc, a lensed $z\sim2.4$ galaxy hosting a $\sim 3-4$~Myr super star cluster; \citealt{pascale23}) and is visible in only two galaxies of the CLASSY sample: Mrk~996 (i.e., J0127-0619) and J1253-0312 (Fig.~20 of \citetalias{mingozzi22}; log(\niii$\lambda$1750/\heii$\lambda$1640)~$\sim1.34, 0.05$, respectively), both also characterized by WR features in their optical spectra, as commented above.
WR stars represent the best tracer of a young ongoing starburst in galaxies, since they are uniquely characterized by very fast and strong winds (especially at high-metallicity), and thus a short lifetime ($\sim10^5$~yr; \citealt{crowther07}), that can lead to the production of prominent nitrogen features both at optical and UV wavelengths (e.g., \citealt{crowther97}). 
In WR-dominated spectra, it is expected also to observe \niv~$\lambda$1719 resonance lines with a clear P-Cygni profile, which can help to discriminate between the nebular and stellar origin of the \niii\ and \niv\ emission. 
The CLASSY galaxy Mrk~996 shows a \niv~$\lambda$1719 P-Cygni feature, a broad \heii~$\lambda1640$ (Fig.~\ref{fig:heiimrk996}), clear WR features in the optical spectra \citep{james09} as well as the \niii~$\lambda$1750 multiplet, so a (partly) stellar origin of the UV N emission lines is plausible. 
% Different is the case of GN-z11, which 
In summary, {the rare nature of nitrogen lines in UV spectra of nearby SFGs such as Mrk~996 could be due to the limited time frame of the starburst events that are causing them. To shed light on this and understand high-$z$ galaxies such as GN-z11 }
we need further studies on local samples {(including AGN galaxies)} where we can broadly observe these UV nitrogen features to ultimately clarify their origin and diagnostic power. Taking into account all these considerations, at the moment we consider N lines not reliable diagnostics for the so-called {\it UV BPT diagrams}.

\section{Conclusions} \label{sec:conclusions}
In this work, we investigated the optical and UV diagnostics tracing different ionization mechanisms (i.e., SF, AGN, shocks), using the CLASSY survey \citep{berg22,james22}, that collects for the first time high-quality, high-resolution, and broad-wavelength-range FUV ($\sim1200-2000$~\AA) spectra for 45 nearby ($0.002 < z < 0.182$) star-forming galaxies, including analogs of the high-$z$ universe thanks to the broad ISM properties parameter space covered.
This paper builds on our results shown in CLASSY~\citetalias{mingozzi22} in which we measured CLASSY UV (from \niv$\lambda\lambda$1483,7 to \ciii$\lambda\lambda1907,9$) and optical (from \oii$\lambda\lambda$3727 to \siii$\lambda$9069) emission lines, and several ISM properties (i.e., $n_e$, $T_e$, $E(B-V)$, 12+log(O/H), log($U$)). 

As a first result of this paper, we confirmed the star-forming nature of our systems using different well-known optical diagnostics sensitive to the dominant ionization source, also taking into account their limitations due to the low metallicity of some of our galaxies. 
Then, we explored the proposed UV counterparts of these diagnostic diagrams - the so-called ``UV-BPT diagrams'' - from previous works based on photoionization and shock models \citep{gutkin16,jaskot16,hirschmann19,hirschmann22}. 
To do this, we compared our measured line ratios (and their corresponding ISM properties) with a set of state-of-the-art models taken from the literature: constant and single-burst SF models from \citetalias{gutkin16}; AGN models from \citetalias{feltre16}; shock models from the 3MdBs\footnote{\url{http://3mdb.astro.unam.mx/}} database (\citetalias{alarie19}). Finally, we discussed which conditions favor the detection of UV emission lines.
In the following, we summarise our findings:
\begin{enumerate}
    \item In Sec.~\ref{sec:results-diagnostics-opt} we explored the classical BPT diagrams, as well as the use of \oi$\lambda$6300 as a shock diagnostic and the \citet{shirazi12} diagram, that takes into account the \heii$\lambda4686$ emission line. In particular, we showed that CLASSY galaxies have \oiii/\hb\ and \heii$\lambda$4686/\hb\ typical of SFGs. Also, their broad components - if present - are still generally classified as star-forming, and, even though have enhanced \nii/\ha, \sii/\ha, \oi/\ha\ and \oi/\oiii\ and low \oiii/\oii\ (possible evidence of shocks), do not correlate with \heii$\lambda$4686, that could be enhanced by shock ionization.
    Overall, it is not possible to completely exclude the presence of shocks (and possibly AGN activity for the only trans-solar metallicity CLASSY galaxy), but the main ionizing mechanism is clearly SF, especially for the CLASSY subsample showing UV emission lines.
    \item In Sec.~\ref{sec:results-diagnostics-uv}, we presented our list of reliable UV diagnostic plots that we explored for the CLASSY galaxies showing UV emission lines. 
    Many of the discussed UV diagnostic diagrams can separate SF from AGN and shocks, with the caveat that low 12+log(O/H) and carbon abundance C/O objects can fall in the AGN/shock-locus proposed by previous authors (e.g., C3He2-C4He2, Fig.~\ref{fig:diagnostic-jr16}; C4C3He2-C4C3, Fig.~\ref{fig:diagnostic-n18}; C3He2-C4C3 in App.\ref{app:other-diag}), while they usually struggle in separating AGN from shocks. 
    Diagnostics using carbon lines equivalent widths identify the star formation locus, even at low metallicity, low C/O abundance and high log($U$) (e.g., EWC4-C3He2 and EWC3-C3He2; Fig.~\ref{fig:diagnostic-n18-1}), but the modelled EWs are heavily dependent on the model assumptions of the continuum component.
    Also, the \civ\ doublet has a complex nature that still needs to be properly understood and is rarely observed to be purely nebular (see also Sec.~\ref{sec:discussion}), thus diagnostic diagrams to investigate the ionization source or ISM properties involving these lines should be used with caution.
    \item In Sec.~\ref{sec:results-diagnostics-uv}, we also concluded that the combination of \ciii~$\lambda\lambda$1907,9, \heii~$\lambda$1640 and \oiiiuv~$\lambda1666$\ (C3He2-O3He2, Fig.~\ref{fig:diagnostic-f16}, see Eq.~\ref{eq:3},~\ref{eq:4}; C3He2-C3O3 in App.\ref{app:other-diag}, Eq.~\ref{eq:5},~\ref{eq:6}) represents the best diagnostic diagram able to separate all the three ionization mechanisms at sub-solar metallicities. In particular, AGN models show lower \oiiiuv/\heii\ and \oiiiuv/\ciii\ {than shocks}, while SF grids are usually located at higher \ciii/\heii\ {than AGN and shocks}.
    \item In Sec.~\ref{sec:discussion}, we confirmed that UV emission lines observed in CLASSY (with \ciii$\lambda\lambda$1907,9, the most common FUV line, excluding \lya) are mainly favored in ISMs with low metallicity and high ionization parameters, with none of them visible in targets with 12+log(O/H)~$\gtrsim8.3$. 
    In particular, \ciii-emitters have systematically lower log(\nii/\ha) ($\lesssim -1$; see Fig.~\ref{fig:optbpt}), can reach high values log(\heii$\lambda4686$/\hb) ($\sim-1.4 - -1.8$; Fig.~\ref{fig:shirazi}), and are clearly shifted to higher log(\oiii/\hb) ($\sim0.5-1$) and higher log(\oiii/\oii) ($\sim0.5-1.5$; Fig.~\ref{fig:optbpt}), and thus higher log($U$). 
    % \item In Sec.~\ref{sec:discussion}, we also showed that the \civ$\lambda\lambda1548,51$ and \heii$\lambda1640$ emission lines measured in CLASSY are mainly nebular, given the consistency of their velocity dispersion with \oiiiuv$\lambda1666$ ($\sigma\sim50$~km/s; Fig.~\ref{fig:fitcomps}). 
    % However, \civ\ doublet as a complex nature that still needs to be properly understood and is rarely observed to be purely nebular (we see it only in 9 low-mass 12+log(O/H)~$\lesssim8$ targets). Hence, diagnostic diagrams to investigate the ionization source or ISM properties involving these lines should be used with caution.
    We also {describe the caveats involved with using \civ$\lambda\lambda1548,51$ and \heii$\lambda1640$ emission lines} and showed that those measured in CLASSY are mainly nebular, given the consistency of their velocity dispersion with \oiiiuv$\lambda1666$ ($\sigma\sim50$~km/s; Fig.~\ref{fig:fitcomps}). {Finally, while reflecting on the currently puzzling detection of UV emission in high-$z$ systems with JWST,} we underlined the almost total absence of nitrogen UV lines (i.e., \niv$\lambda\lambda$1483,7, \niii$\lambda1750$) in the CLASSY survey. %It is still unclear the nebular origin of these lines, especially for \niii, since they can be the product of WR stars. 
    Hence, we need further studies of samples where we can observe these features before attesting their reliability as diagnostics to investigate the main source of ionization.
\end{enumerate} 

Overall, this paper together with \citetalias{mingozzi22} uses the CLASSY survey to provide the tool-kit to investigate ISM properties using UV emission lines. 
This can be particularly important to explore the $z > 6$ rest-frame UV spectra in the JWST era, in which an unprecedented number of EoR galaxies have already been revealed (e.g., \citealt{curtis-lake23,curtijwst22,arellano-cordova22b,bunker23,cameron23,roberts-borsani22}).

\newpage 
% \begin{acknowledegments}
MM, DAB, KZA-C and XX are grateful for the support for this program, HST-GO-15840, that was provided by NASA through a grant from the Space Telescope Science Institute, which is operated by the Associations of Universities for Research in Astronomy, Incorporated, under NASA contract NAS5-26555. 
MM, BLJ, SH and NK are thankful for support from the European Space Agency (ESA).
Also, MM is grateful to Carlo Cannarozzo, Giovanni Cresci and Alessandro Marconi for inspiring conversations and advice.
AF acknowledges the support from grant PRIN MIUR2017-20173ML3WW\_001.
AW acknowledges the support of UNAM via grant agreement PAPIIT no. IN106922. RA acknowledges support from ANID Fondecyt Regular 1202007.
JB acknowledges support by Fundação para a Ciência e a Tecnologia (FCT) through the research grants UID/FIS/04434/2019, UIDB/04434/2020, UIDP/04434/2020, national funds PTDC/FIS-AST/4862/2020 and work contract 2020.03379.CEECIND.

The CLASSY collaboration extends special gratitude to the Lorentz Center for useful discussions 
during the "Characterizing Galaxies with Spectroscopy with a view for JWST" 2017 workshop that led 
to the formation of the CLASSY collaboration and survey.
The CLASSY collaboration thanks the COS team for all their assistance and advice in the
reduction of the COS data.

Funding for SDSS-III has been provided by the Alfred P. Sloan Foundation, the Participating Institutions, the National Science Foundation, and the U.S. Department of Energy Office of Science. The SDSS-III web site is \href{http://www.sdss3.org/}{http://www.sdss3.org/}.
SDSS-III is managed by the Astrophysical Research Consortium for the Participating Institutions of the SDSS-III Collaboration including the University of Arizona, the Brazilian Participation Group, Brookhaven National Laboratory, Carnegie Mellon University, University of Florida, the French Participation Group, the German Participation Group, Harvard University, the Instituto de Astrofisica de Canarias, the Michigan State/Notre Dame/JINA Participation Group, Johns Hopkins University, Lawrence Berkeley National Laboratory, Max Planck Institute for Astrophysics, Max Planck Institute for Extraterrestrial Physics, New Mexico State University, New York University, Ohio State University, Pennsylvania State University, University of Portsmouth, Princeton University, the Spanish Participation Group, University of Tokyo, University of Utah, Vanderbilt University, University of Virginia, University of Washington, and Yale University.

This work also uses the services of the ESO Science Archive Facility,
observations collected at the European Southern Observatory under 
ESO programmes 096.B-0690, 0103.B-0531, 0103.D-0705, and 0104.D-0503, and observations obtained with the Large Binocular Telescope (LBT).
The LBT is an international collaboration among institutions in the
United States, Italy and Germany. LBT Corporation partners are: The
University of Arizona on behalf of the Arizona Board of Regents;
Istituto Nazionale di Astrofisica, Italy; LBT Beteiligungsgesellschaft,
Germany, representing the Max-Planck Society, The Leibniz Institute for
Astrophysics Potsdam, and Heidelberg University; The Ohio State
University, University of Notre Dame, University of
Minnesota, and University of Virginia.

% Photoionization models were run with {\sc cloudy}, which is currently supported 
% by grants from NSF (AST 1816537 and AST 1910687 National Aeronautics and Space 
% Administration 19-ATP19-0188) and Space Telescope Science Institute (HST-AR-15018). 

This research has made use of the HSLA database, developed and maintained at STScI, Baltimore, USA.

All the {\it HST} data used in this paper can be found in MAST: \dataset[10.17909/m3fq-jj25]{http://dx.doi.org/10.17909/m3fq-jj25}.

% \end{acknowledegments}

\facilities{HST (COS), LBT (MODS), APO (SDSS), KECK (KCWI, ESI), VLT (MUSE, VIMOS)}
\software{
astropy \citep{astropy13,astropy18,astropy22}
BEAGLE \citep{chevellard16}, 
CalCOS (STScI),
dustmaps \citep{green18},
jupyter \citep{kluyver16},
LINMIX \citep{kelly07}, 
MPFIT \citep{markwardt09},
MODS reduction Pipeline,
Photutils \citep{bradley21},
PYNEB \citep{luridiana12,luridiana15},
python,
pysynphot (STScI Development Team),
RASCAS \citep{michel-dansac20},
SALT \citep{scarlata15}, 
STARLIGHT \citep{fernandes05}, 
TLAC \citep{gronke14},
XIDL}

%-----------------------------------------------------------------------------------------
%-----------------------------------------------------------------------------------------
\typeout{} 
\bibliography{CLASSY}

\clearpage

\newpage

\appendix

\section{Comparing UV and optical kinematics} \label{app:uv-opt-kins}
\begin{figure}[!h]
\begin{center}
    \includegraphics[width=0.48\textwidth]{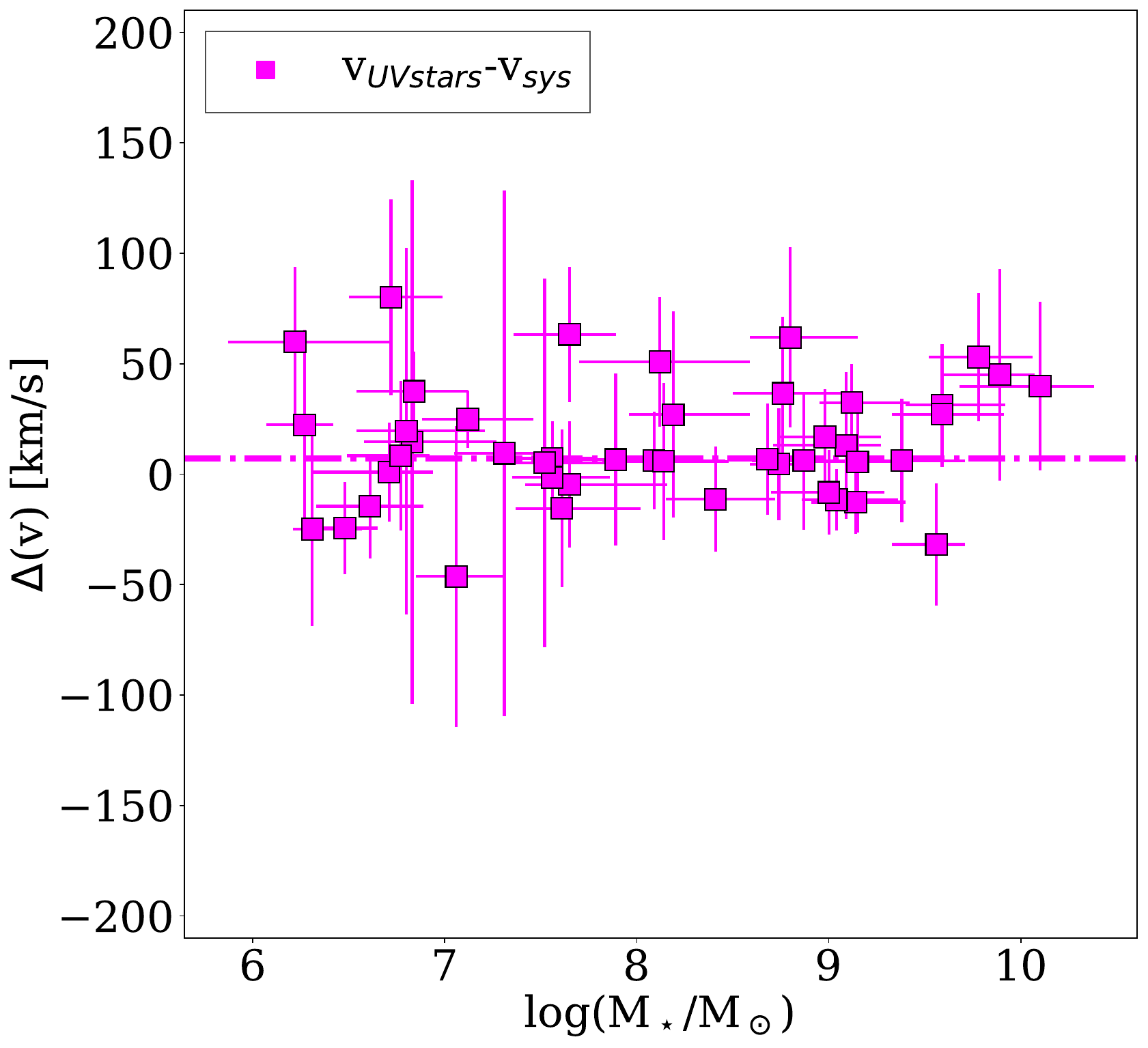}
    \includegraphics[width=0.48\textwidth]{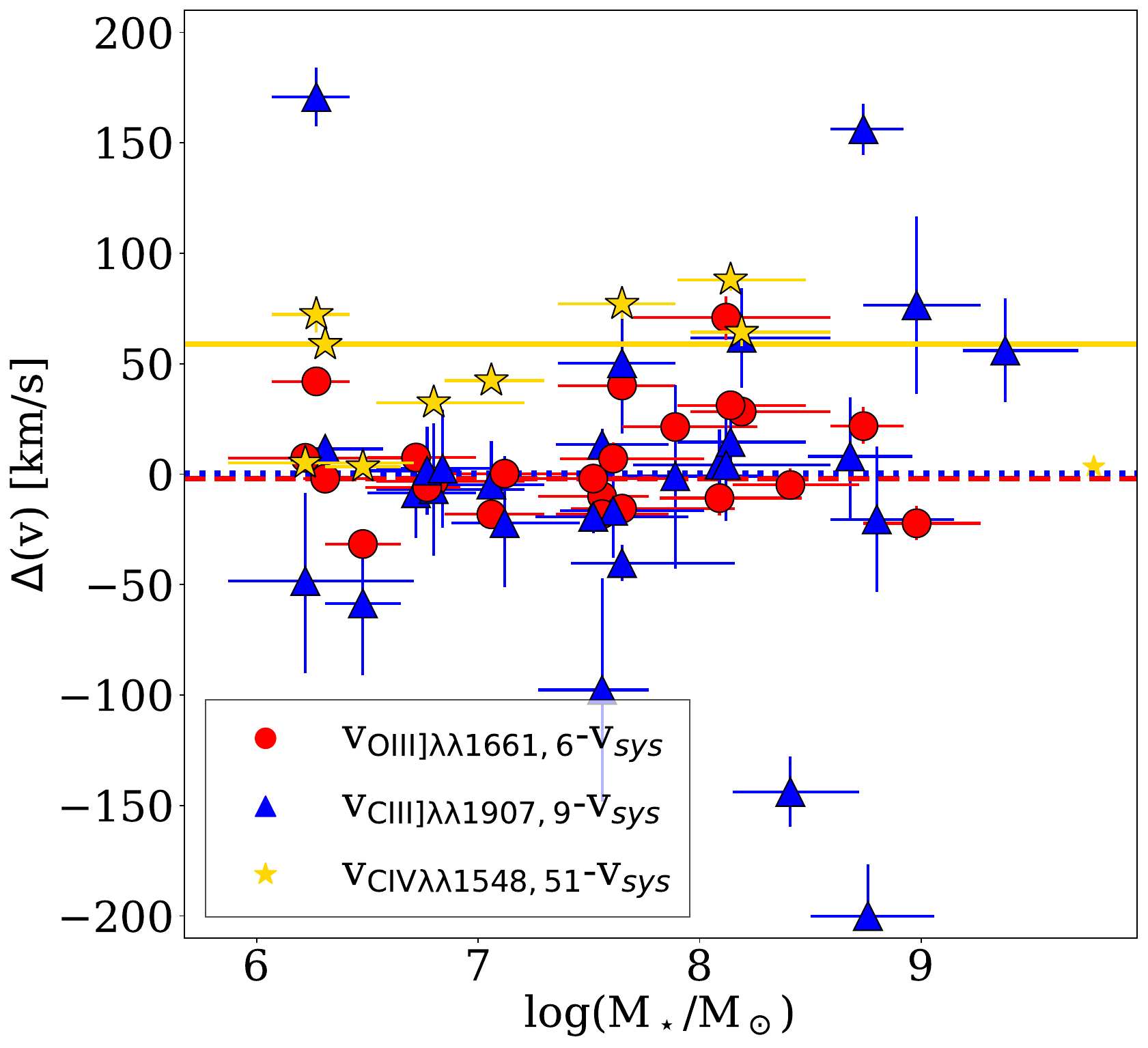}
\end{center}
\caption{Velocity offset of the UV stellar (left panel, magenta squares) and gas kinematics (right panel) with respect to the systemic velocity $v_{sys}$, consistent with the ionized gas velocity measured from optical spectra, and as a function of the galaxy stellar mass. 
In the right panel, we indicate the \oiiiuv$~\lambda\lambda$1661,6, \ciii$~\lambda\lambda$1907,9, and \civ$~\lambda\lambda$1548,51 velocities as red dots, blue triangles and yellow stars, as reported in the legend.}
\label{fig:uv-opt-kins}
\end{figure}

Here we want to further discuss the emission line kinematics, also comparing optical and UV results. 
The left and right panels of Fig.~\ref{fig:uv-opt-kins} show the velocity offset of the UV stellar (left panel) and gas kinematics (right panel) with respect to the systemic velocity $v_{sys}$ (derived from $z_{lit}$), consistent with the ionized gas velocity measured from optical emission lines, as a function of the galaxy stellar mass.
In general, we find no systematic offsets between the ionizing stellar component velocity explored by fitting the UV spectra (magenta squares) and $v_{sys}$, with $\Delta v$ mainly within $\pm \sim 50$~km/s and median value $<\Delta v> \sim 7$~km/s (dash-dotted magenta line). 
One implication is that the narrow components of the optical emission lines are really tracing non-outflowing gas \citep{rigby18a}. 
The right panel of Fig.~\ref{fig:uv-opt-kins} instead shows the \oiiiuv$~\lambda\lambda$1661,6, \ciii$~\lambda\lambda$1907,9, and \civ$~\lambda\lambda$1548,51 velocities as red dots, blue triangles and yellow stars. 
There is a scatter between the UV and systemic velocities mainly within $\pm \sim 50$~km/s, and more enhanced for some \ciii\ measurements, with three galaxies (J1359+5726, J0926+4427 and J0942+3547) showing a clear redshifted offset ($\Delta v\sim +100-200$~km/s) and three (I\,Zw\,18, Mrk~996 and J1025+3622) showing a clear blueshifted offset ($\delta v\sim -150 {-} -200$~km/s). 
% ({explanations? real - emission coming from different regions - or due to issue in NUV wl calibration? add a sentence here referring to \citetalias{james22}}). 
However, we do not see either any systematic offsets, as shown by the dashed red and dotted blue lines indicating the corresponding median values with respect to $v_{sys}$.
Hence, we conclude it is unlikely that they are due to issues in the G185M grating calibration, covering the \ciii, discussed in detail in \citet{james22}.

Interestingly, we find a systematic offset between the $v_{CIV}$ and $v_{sys}$ with a median value of $<\Delta v> \sim +60$~km/s (solid gold line), implying that \civ\ emission is systematically redshifted with respect to the systemic velocity, consistently with resonant scattering/absorption (see also \citealt{berg19b,wofford21}).
For seven out of the nine galaxies for which we measure \civ\ nebular emission, this blueshifted offset increases with the stellar mass, going from $\sim 0$~km/s to $\sim 90$~km/s in the range log(M$_\star$/M$_\odot$)~$6{-}8$. 
The two exceptions are I\,Zw\,18 and J1323-0132, with stellar masses of log(M$_\star$/M$_\odot$)~$\sim 6.3$, showing $\delta v\sim 65$~km/s.

As discussed in Sec.~\ref{sec:data-analysis} and in Fig,~\ref{fig:fitcomps}, optical emission lines of half of the CLASSY sample needed a narrow and a broad Gaussian component to be accurately reproduced.
Looking at the single and narrow component $\sigma$ values shown in the right panel of Fig,~\ref{fig:fitcomps}, we see that $\sigma$ (and also its scatter) increases at larger stellar masses, with only objects with $M_\star \gtrsim 10^7$~M$_\odot$ needing a 2-component fit. 
We highlight here that all the galaxies characterized by a broad component in optical emission lines (apart Mrk~996 and SBS~0335-052~E) were found to have an outflowing component in rest-UV absorption lines analyzed by \citet{xu22,xu23} (CLASSY~\citetalias{xu22}, \citetalias{xu23}). 
Recent studies have shown that low-mass systems ($M_\star<10^9$~M$_\odot$) such as those of the CLASSY survey can have irregular velocity fields in the ionized gas studied through optical emission lines, indicating the presence of non-circular motions (e.g., \citealt{marasco23}). 
However, galactic winds, defined as gas at velocities larger than the galaxy escape speed, have found to be able to account only for a few percent of their observed fluxes both in the local universe \citep{marasco23} and at $z\sim1.2-2.6$ \citep{concas22}. 
Here we highlight that the lack of galactic wind detection in small systems could be also an issue of S/N, but in this work we do not explore this topic any further, leaving it to a future study using optical integral-field spectroscopy data, that can provide both a spatial and kinematical information of our galaxies, and thus a more complete picture that the integrated optical spectra analyzed here cannot provide.

\section{Alternative diagnostic diagrams}\label{app:other-diag}
\begin{figure}[!h]
\begin{center}
    \includegraphics[width=0.5\textwidth]{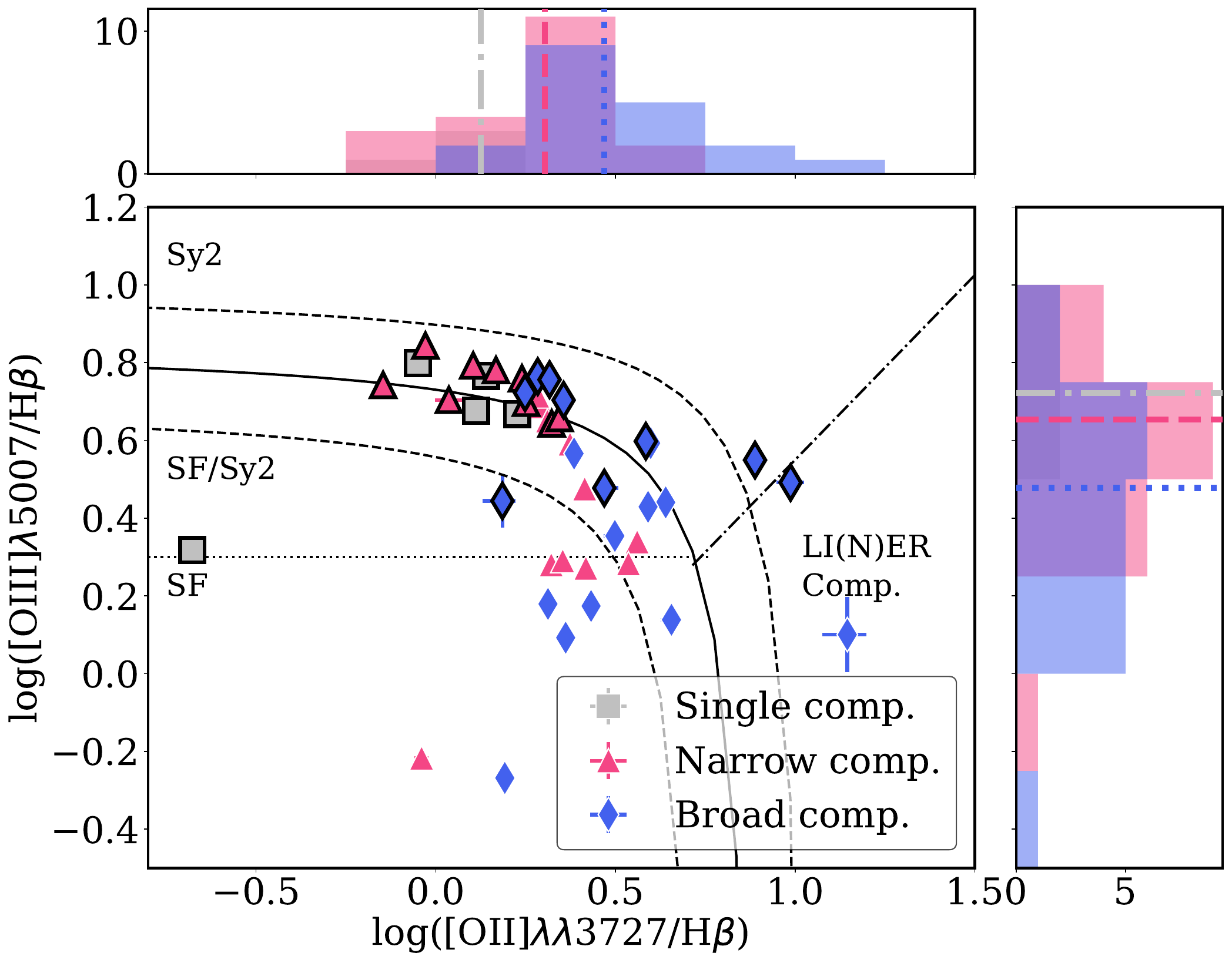}
\end{center}
\caption{Analogously to Fig.~\ref{fig:optbpt}, \oii$~\lambda\lambda$3727/\hb\ versus \oiii$~\lambda5007$/\hb\  diagnostic diagram for the narrow and broad components of the CLASSY galaxies, shown as red dots and blue diamonds, respectively. Only the measurements with $S/N>3$ for all the lines involved are reported. The dotted and solid lines indicate the empirical dividing line between star-forming galaxies and AGN identified by \citet{lamareille04}, the dashed ones show the uncertainty region, while the dash-dotted line identifies the empirical locus of LI(N)ER and Seyfert galaxies, found by \citet{lamareille10}. Unfortunately, this diagnostic plot does not represent a good tool to discriminate among the different ionizing mechanisms of our systems.}
\label{fig:lamareille}
\end{figure}

Fig.~\ref{fig:lamareille} shows an alternative optical diagnostic diagram to discriminate the excitation properties, using \oii$~\lambda\lambda$3727/\hb\ versus \oiii$~\lambda5007$/\hb\ line ratios \citep{lamareille04,lamareille10}. The dotted, solid and dashed lines indicate the empirical dividing line between star-forming galaxies and AGN identified by \citet{lamareille04}, while the dash-dotted line identifies the empirical locus of LI(N)ER and composite galaxies, found by \citet{lamareille10}. 
The advantage of this diagram is that it can be used to explore JWST targets at redshifts up to $z\sim8.5$ in the observed NIR (with NIRISS or NIRspec), for which redder wavelengths are unavailable (e.g., \citealt{curtijwst22}). However, this diagram should be taken with caution since, as shown in Table~\ref{tab:classification}, it does not always disentangle ionizing mechanisms in our sample, with many galaxies (classified as star-forming according to the classical BPT) lying in the boundary region between SF and Seyferts and thus classified as composite. This shows the limit of \oii/\hb\ in distinguishing among ionization mechanisms.

\section{Other UV diagnostic diagrams}
\subsection{The effect of stellar rotation and multiplicity on UV line ratios}\label{app:uv-diag-models}
\begin{figure}
\begin{center} 
    
    \includegraphics[width=0.42\textwidth]{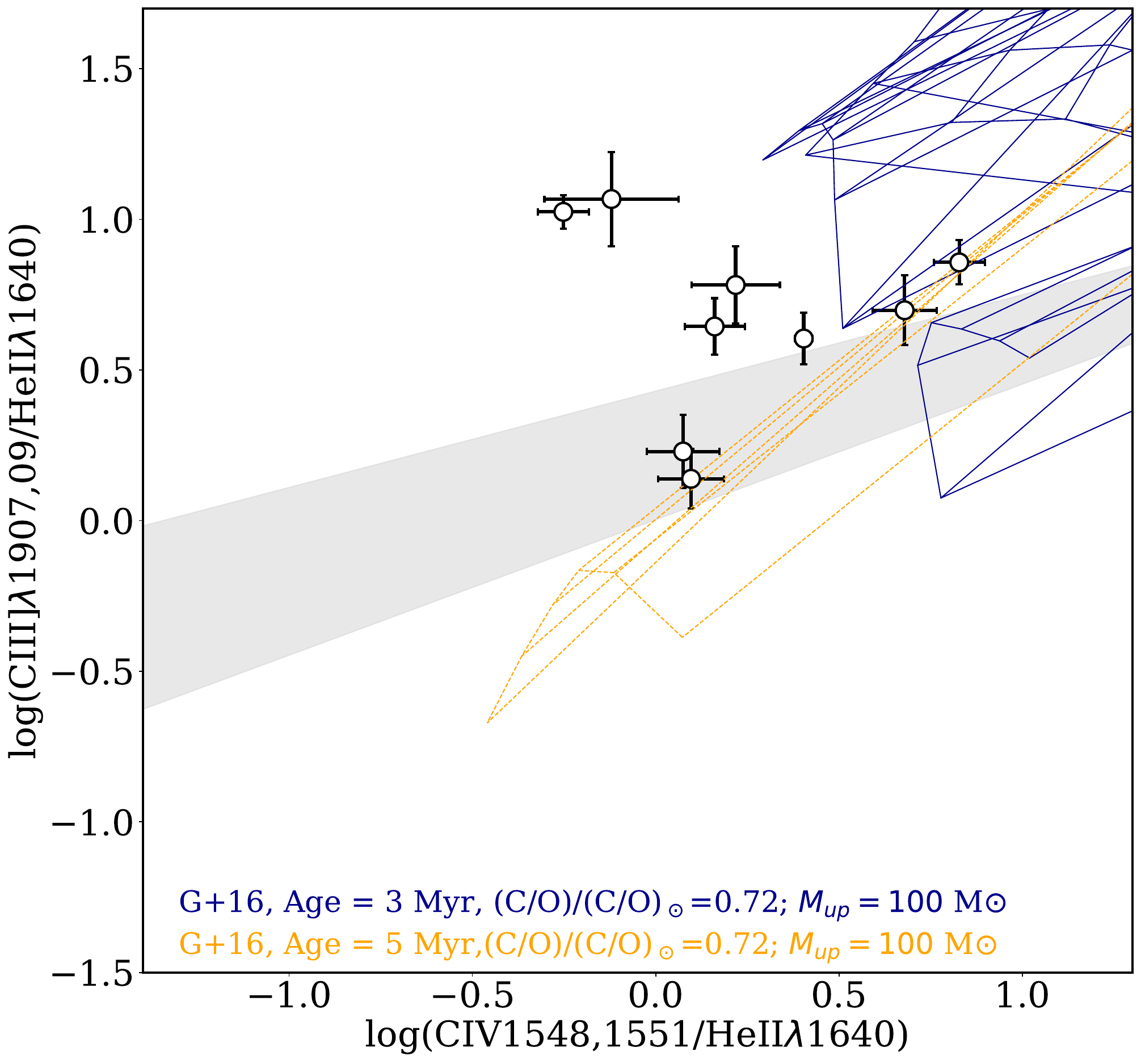}
    \includegraphics[width=0.42\textwidth]{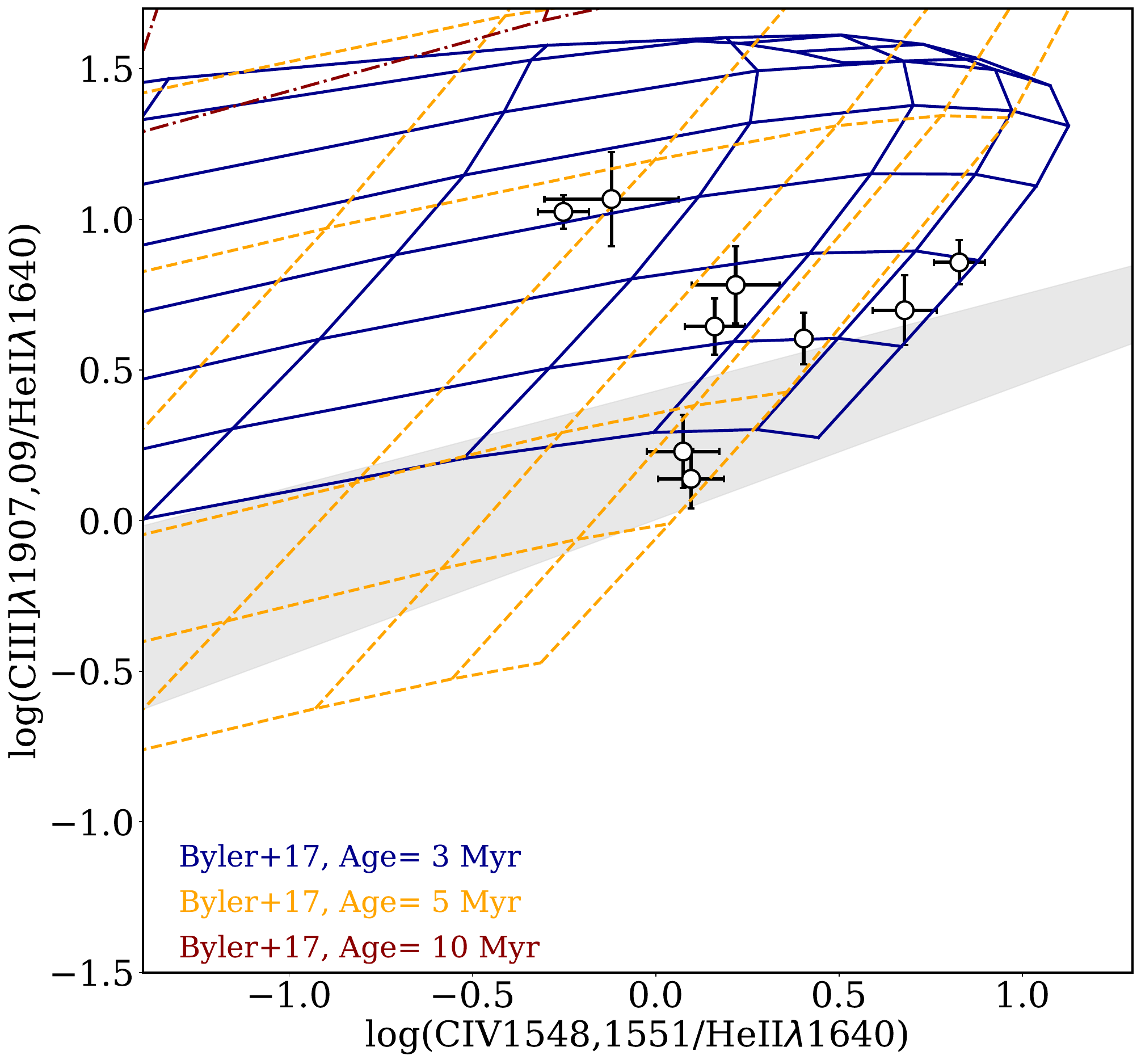}
    
    \includegraphics[width=0.42\textwidth]{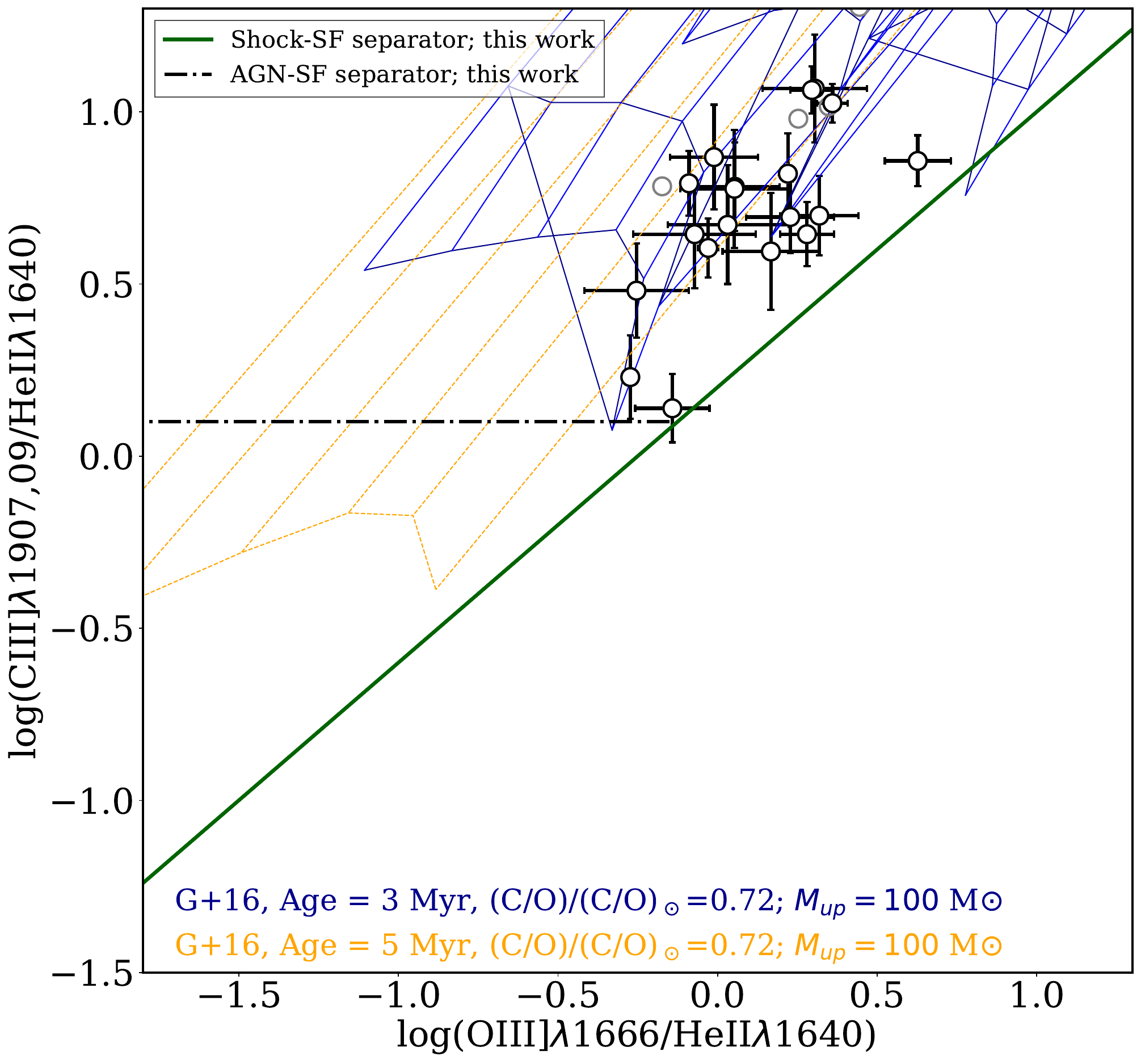}
    \includegraphics[width=0.42\textwidth]{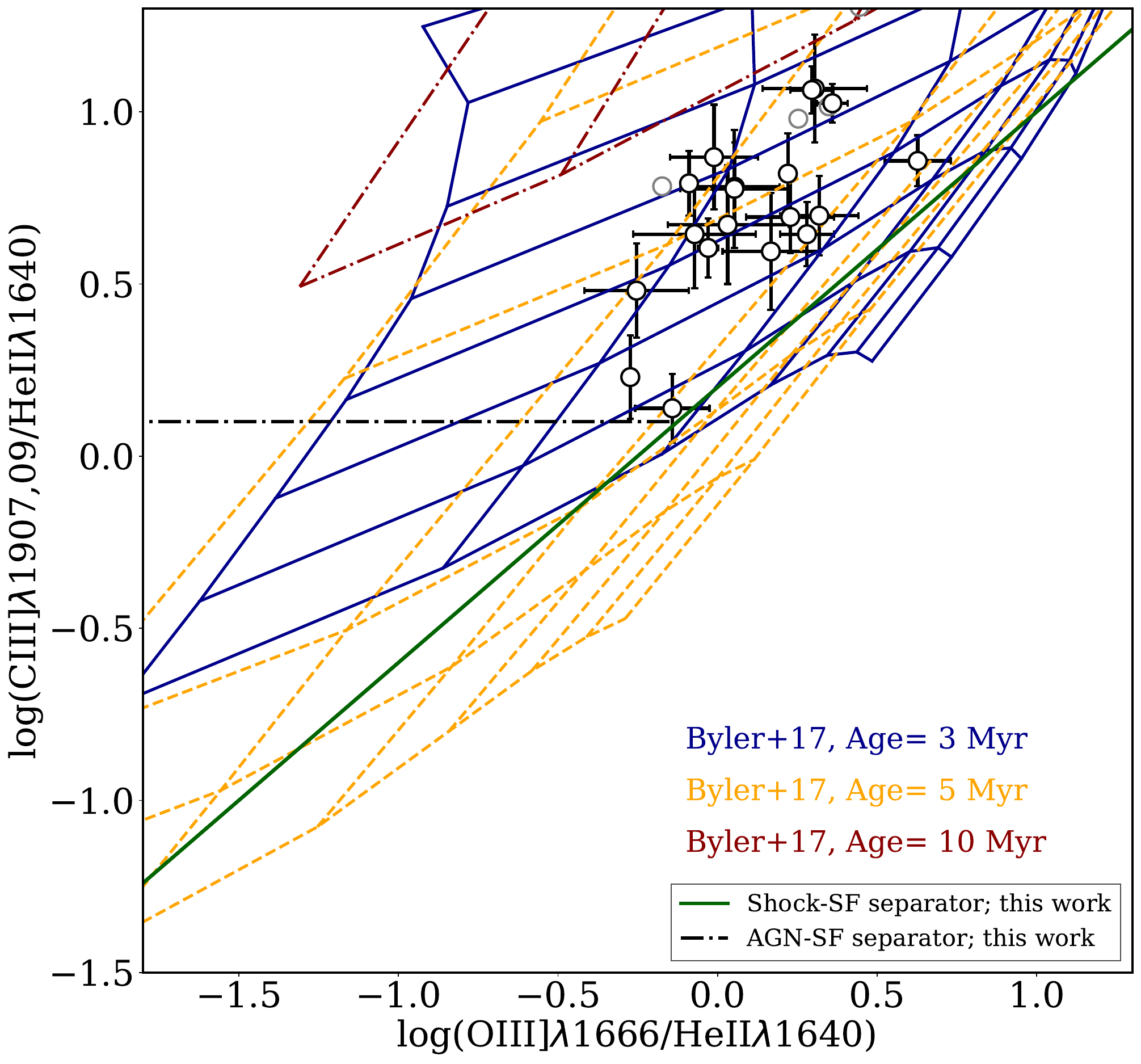}

    \includegraphics[width=0.42\textwidth]{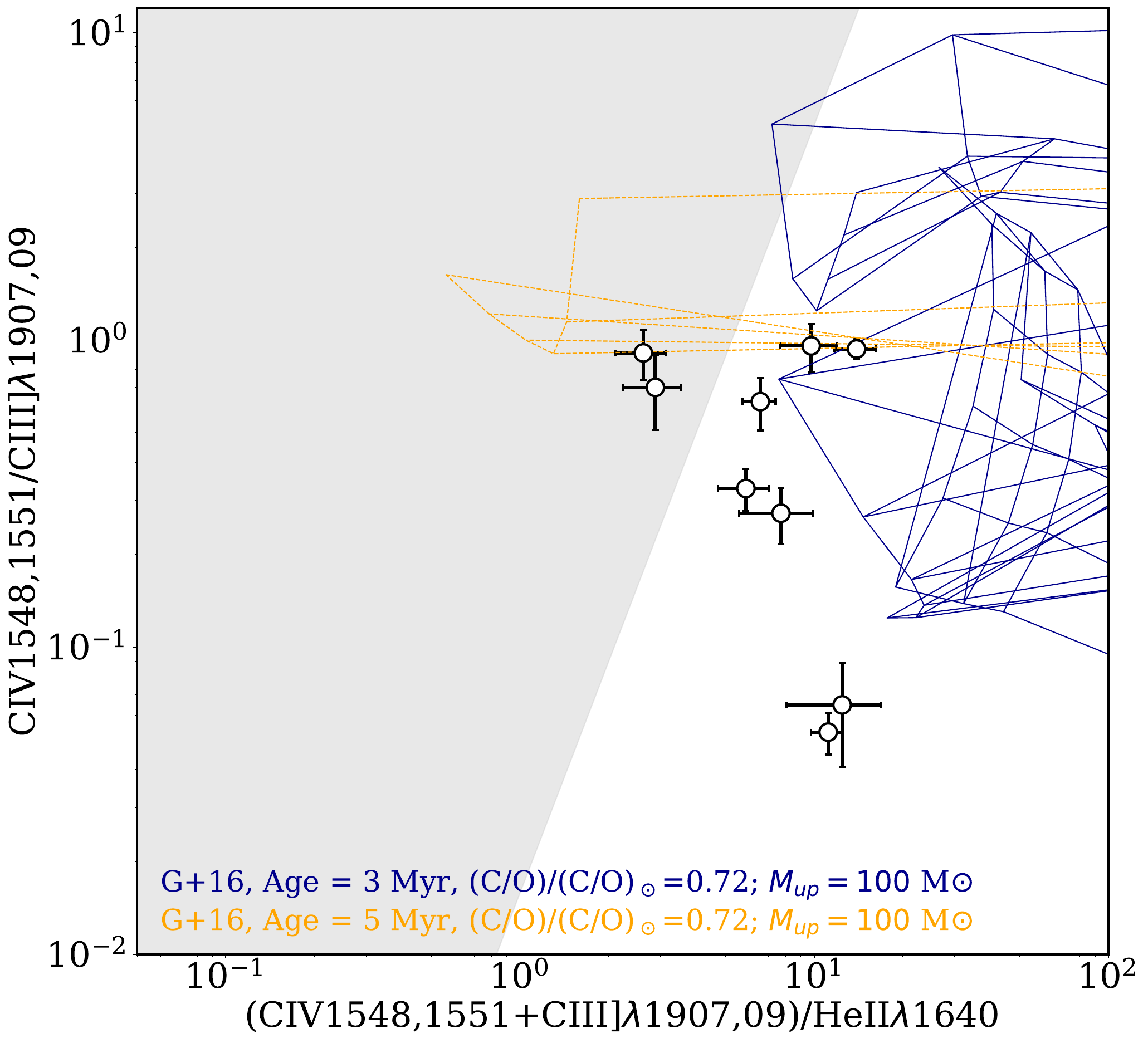}
    \includegraphics[width=0.42\textwidth]{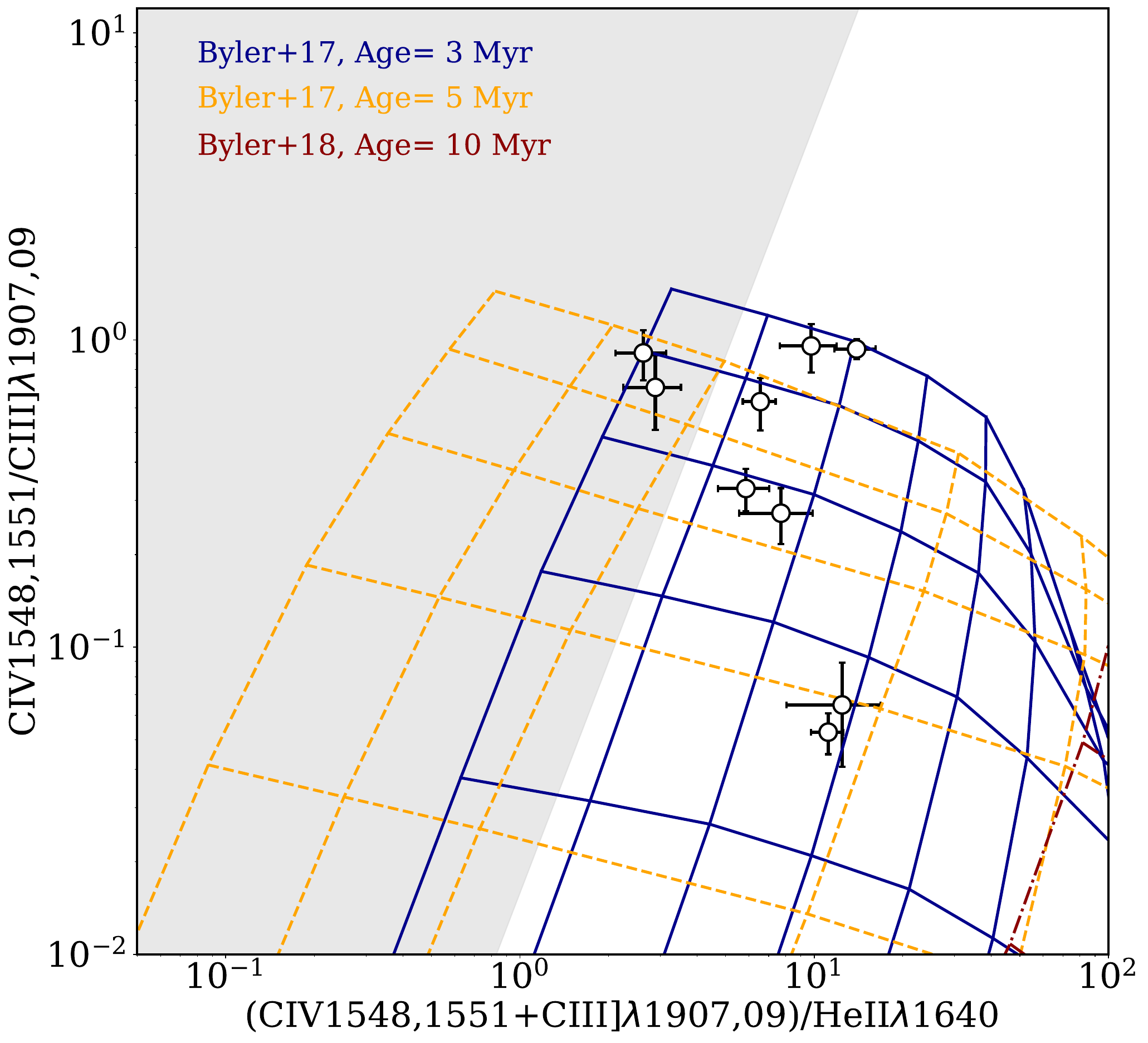}
\end{center}
\caption{C3He2-C4He2, C3He2-C4C3 and C3He2-O3He2 diagrams as shown in Fig.~\ref{fig:diagnostic-jr16}, \ref{fig:diagnostic-f16} and \ref{fig:diagnostic-n18} with overplotted \citetalias{gutkin16} single-burst SF grids (left panels) at 3~Myr and 5~Myr in blue and orange, respectively, and \citetalias{byler17} bursty models (right panels) including stellar rotation at 3, 5 10~Myr in blue, orange and red, respectively, as indicated in the legend. Lower C/O or higher IMF cut (not shown in the plot) has the effect of moving the 3~Myr \citetalias{gutkin16} grids to lower \ciii/\heii\ and (\civ+\ciii)/\heii\ line ratios, reproducing the observations, while at larger ages the stellar ionizing radiation is not enough. Indeed, stellar rotation and multiplicity (both not considered in \citetalias{gutkin16}) have the effect of amplifying the stellar ionizing radiation for several Myr after a star formation burst.}
\label{fig:modelscomp}
\end{figure}

{In this section, we compare \citetalias{gutkin16}, \citetalias{byler17} and \citetalias{xiao18} bursty models presented in Tab.~\ref{tab:models}. 
\citetalias{byler17} models take into account only stellar rotation and \citetalias{xiao18} models only stellar multiplicity, while \citetalias{gutkin16} models neither of the two.
The differences in how \citetalias{gutkin16} and \citetalias{byler17} grids differ with respect to \citetalias{xiao18} models in the UV diagnostic diagrams discussed in Fig.~\ref{fig:diagnostic-jr16}, \ref{fig:diagnostic-f16} and \ref{fig:diagnostic-n18} are shown in Fig.~\ref{fig:modelscomp}. 

In particular, in Fig.~\ref{fig:modelscomp} we show \citetalias{gutkin16} and \citetalias{byler17} grids at 3, 5 and 10~Myr. 
The bursty \citetalias{gutkin16} grids seem unable to reproduce the lowest \civ/\heii, \ciii/\heii\ and (\civ+\ciii)/\heii\ line ratios. 
This is because, for clarity reasons, we have only shown a subset of \citetalias{gutkin16} model parameter space, considering only (C/O)/(C/O)$_\odot$~$=0.72$ and the IMF cut at $M_{up}=100$~M$_\odot$.
Considering either a lower C/O or a higher IMF cut, the 3~Myr grid can reproduce the line ratios observed for the CLASSY galaxies, as shown for the constant SF grids in Fig.~\ref{fig:diagnostic-jr16}, \ref{fig:diagnostic-f16} and \ref{fig:diagnostic-n18}. 
However, at higher ages the ionizing photons drop with a subsequent reduction of \heii\ (but also \ciii\ and \civ), inducing a shift of the grids to higher \ciii/\heii\ and \civ/\heii\ than the range shown in Fig.~\ref{fig:diagnostic-jr16} (see e.g., \citealt{jaskot16} Sec.~3.1 for a detailed comparison between different model prescriptions). 
On the other hand, \citetalias{byler17} models at 3 and 5~Myr behave similarly to \citetalias{xiao18} grids at 3 and 10~Myr\footnote{\citetalias{xiao18} grids at 5~Myr are not shown in this paper, but are consistent with those at 10~Myr}, with the one at 5~Myr going beyond the SF locus defined by the CLASSY galaxies, partly covering the shock/AGN regions. 
The \citetalias{byler17} grid at 10~Myr, instead, is unable to reproduce the observed line ratios, being shifted to higher \ciii/\heii\ and (\civ+\ciii)/\heii\ (in the opposite direction with respect to the AGN/shock regions) because of the reduced ionizing radiation, unable to enhance \heii\ emission.
The main take-away from Fig.~\ref{fig:modelscomp} is that, as discussed in Sec.~\ref{sec:methods}, stellar rotation and multiplicity have the effect of amplifying the stellar ionizing radiation for several Myr after a star formation burst. Overall, the different grids at 3~Myr can reproduce the observed line ratios of star-forming galaxies.
}

\subsection{Other UV diagnostic diagrams}\label{app:uv-diag}
\begin{figure}
\begin{center} 
    \includegraphics[width=0.9\textwidth]{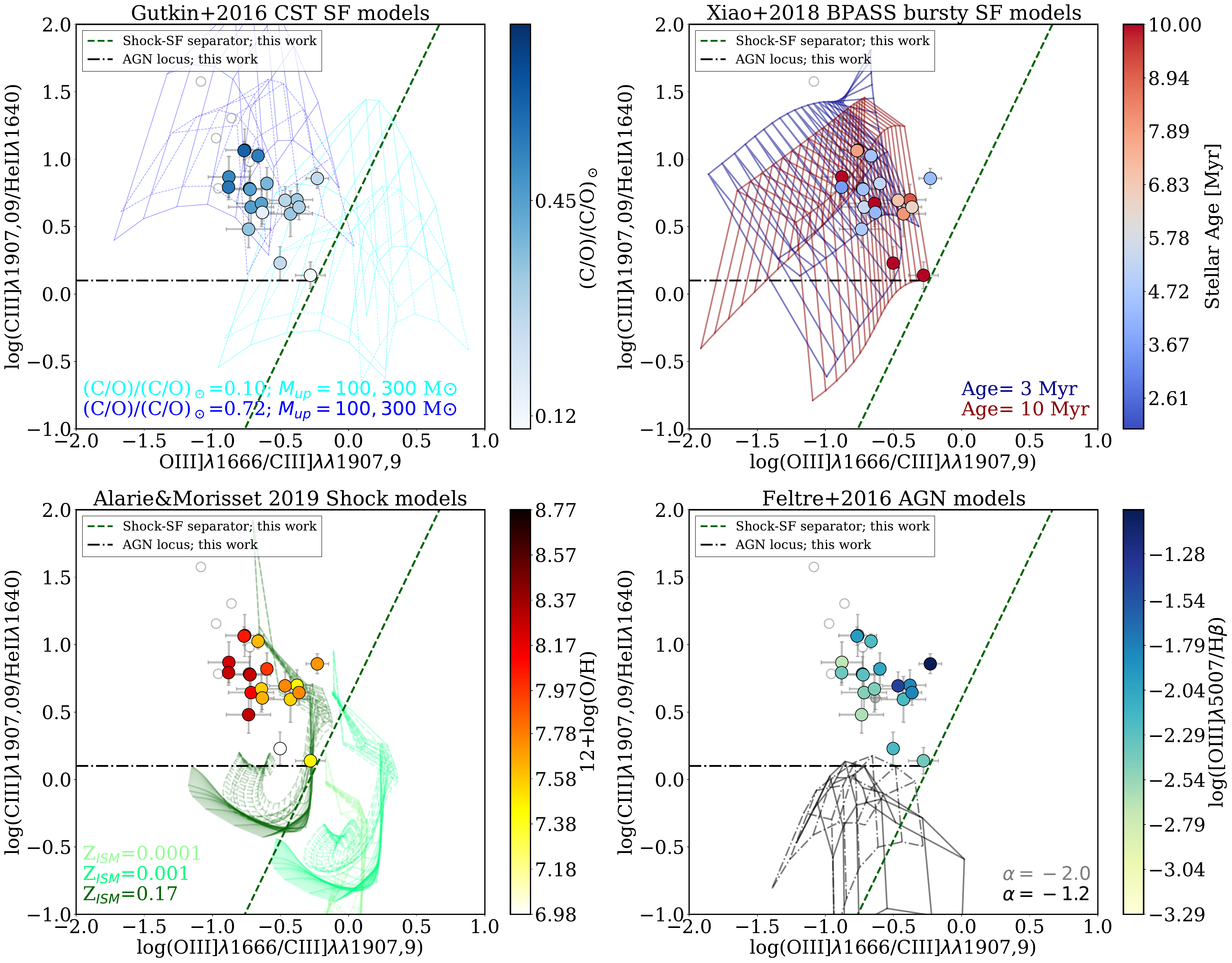}
\end{center}
\caption{C3He2-O3C3 diagram: \ciii~$~\lambda\lambda$1907,9/\heii~$~\lambda$1640 vs \oiiiuv~$~\lambda$1666/\ciii\ diagnostic plot proposed by \citetalias{feltre16} to separate SF and AGN activities. {The filled and open dots show $S/N>3$ and $S/N<3$ fluxes for all the emission lines taken into account, respectively.} The models superimposed are as explained in Fig.~\ref{fig:diagnostic-jr16}. This diagram clearly distinguishes among SF, AGN and shocks at sub-solar
metallicities (i.e., excluding the dark-green shock grid), as C3He2-O3He2 (Fig.~\ref{fig:diagnostic-f16}), as indicated by the dashed dark green and dash-dotted black lines.}
\label{fig:uvbptapp}
\end{figure}

\begin{figure}
\begin{center} 
    \includegraphics[width=0.9\textwidth]{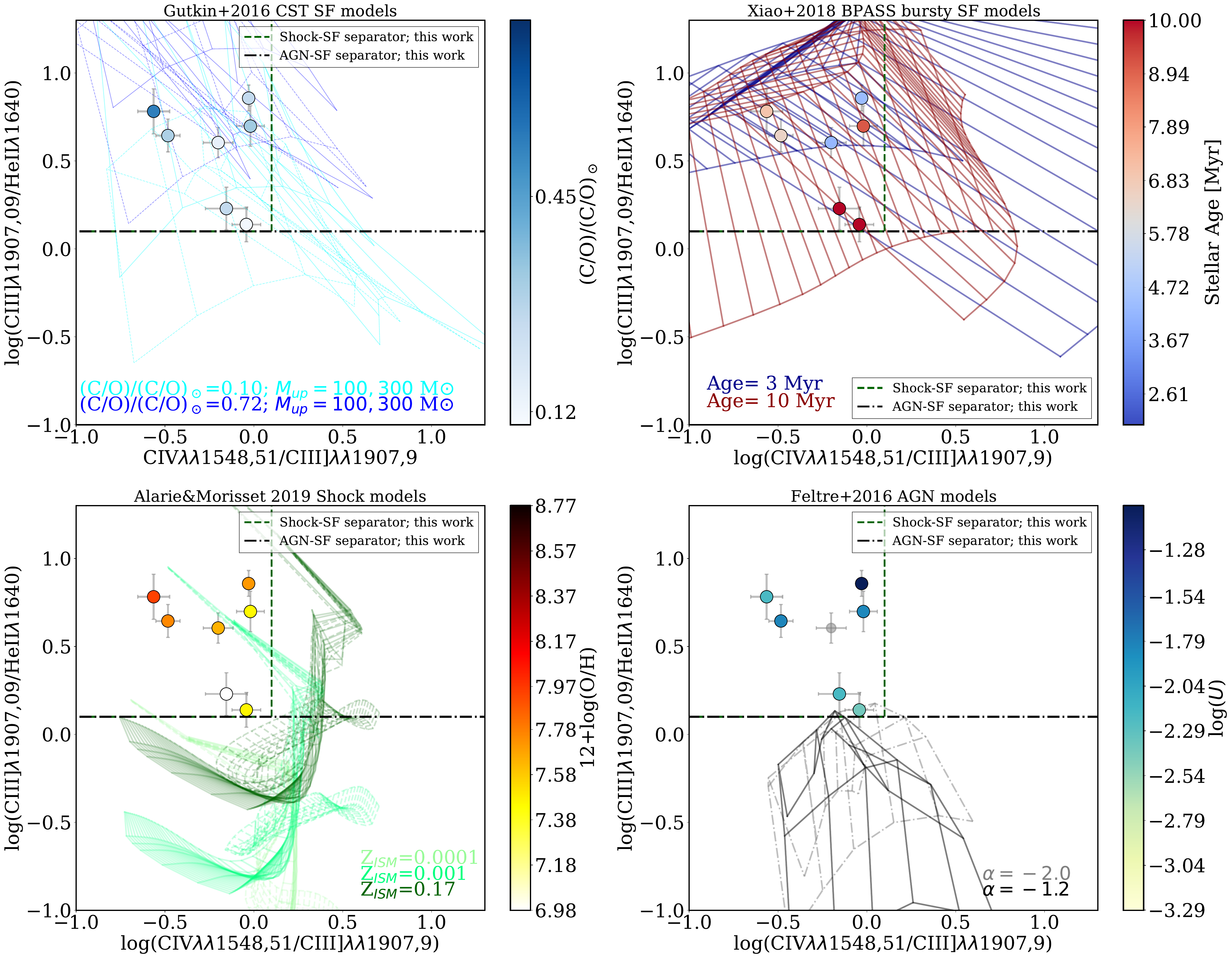}
\end{center}
\caption{C3He2-C4C3 diagram: \ciii~$~\lambda\lambda$1907,9/\heii~$~\lambda$1640 vs \civ~$~\lambda\lambda$1548,51/\ciii\ diagnostic plot proposed by \citetalias{feltre16} to separate SF and AGN activities. {Here we show only the CLASSY galaxies with \civ\ detected in pure emission.} The models superimposed are as explained in Fig.~\ref{fig:diagnostic-jr16}. There is not a clear difference in \civ/\ciii\ line ratios between the different models, and thus the difference between SF and other mechanisms is mainly traced by the different \ciii/\heii\ (higher for SF grids, lower for AGN/shocks), as indicated by the dashed dark green and dash-dotted black lines.}
\label{fig:uvbptapp1}
\end{figure}

Fig.~\ref{fig:uvbptapp} and \ref{fig:uvbptapp1} show the C3He2-C3O3 and C3He2-C4C3 diagrams, respectively, proposed together with the C3He2-O3He2 diagram (Fig.~\ref{fig:diagnostic-f16}) by \citetalias{feltre16}. 
C3He2-C3O3 behaves similarly to C3He2-O3He2 (Fig.~\ref{fig:diagnostic-f16}), and thus, at sub-solar metallicities, it can clearly distinguish among SF, AGN and shocks, as shown by the Shock-SF separator
\begin{equation}\label{eq:5}
    y=2.1\,x+0.6
\end{equation}
and the AGN-SF separator lines
\begin{equation}\label{eq:6}
    y=0.1 \,\,\,\,\,\, and \,\,\,\,\,\, x<-0.25
\end{equation}
with $x = $~log(\oiii$\lambda1666$/\ciii$\lambda\lambda$1907,9) and $y=$~log(\ciii$\lambda\lambda$1907,9/\heii$\lambda$1640).
At sub-solar metallicities (i.e., excluding the dark-green shock grid), shocks can reach higher \oiii/\ciii\ line ratios than AGN models, while SF grids are located at higher \ciii/\heii\ values.

C3He2-C4C3 instead behaves similarly to C3He2-C4He2 diagram (Fig.~\ref{fig:diagnostic-jr16}) and are a very similar version of the C4C3He2-C4C3 diagram (Fig.~\ref{fig:diagnostic-n18}). There is not a clear difference in \civ/\ciii\ line ratios between the different models, and thus the difference between SF and other mechanisms is mainly traced by the different \ciii/\heii\ (higher for SF grids, lower for AGN/shocks), as indicated by the Shock-SF separator
\begin{equation}\label{eq:7}
    y\geq0.1 \,\,\,\,\,\, and \,\,\,\,\,\, x\geq0.1
\end{equation}
and the AGN-SF separator lines
\begin{equation}\label{eq:8}
    y=0.1 \,\,\,\,\,\, and \,\,\,\,\,\, x<-0.25
\end{equation}
with $x = $~log(\civ$\lambda\lambda1548,51$/\ciii$\lambda\lambda$1907,9) and $y=$~log(\ciii$\lambda\lambda$1907,9/\heii$\lambda$1640)
Also, AGN and shocks show similar \civ/\ciii, and thus this diagnostic diagram cannot distinguish them.

\end{document}